\documentclass[a4paper,10pt]{article}
\usepackage{jheppub}
\usepackage{graphicx,floatflt,amssymb,rotate,epsfig}
\usepackage[utf8]{inputenc}
\usepackage[inline]{enumitem}
\usepackage{mathrsfs}
\usepackage{amssymb}
\usepackage{amsmath}
\usepackage{color}
\usepackage{graphicx}
\usepackage{subfig}
\usepackage{multirow}
\usepackage{mathtools}
\usepackage{float}
\usepackage{pstricks}
\usepackage[toc,page]{appendix}
\usepackage{pifont}
\usepackage{braket}
\usepackage{bbold}
\usepackage{hyperref}

\usepackage{relsize}
\def\BaBar{{\mbox{\slshape B\kern-0.1em{\smaller A}\kern-0.1em B\kern-0.1em{\smaller A\kern-0.2em R}}}}

\newcommand{\ba}{\begin{array}}
	\newcommand{\ea}{\end{array}}
\def\beq{\begin{equation}}
	\def\eeq{\end{equation}}
\def\bea{\begin{eqnarray}}
	\def\eea{\end{eqnarray}}
\def\nn{\nonumber}

\def\roughly#1{\mathrel{\raise.3ex\hbox
		{$#1$\kern-.75em\lower1ex\hbox{$\sim$}}}}
\def\lsim{\roughly<}
\def\gsim{\roughly>}
\def\sla#1{\raise.15ex\hbox{$/$}\kern-.57em #1}

\def\bd{B_d^0}

\def\order{\lower 1.8ex \hbox{\LARGE\~{}}}
\def\btopilnu{\bar{B} \to \pi\ell^-\bar{\nu}_{\ell}}

\def\btopi0lnu{B^+ \to \pi^0 \ell\nu}

\def\btorho0nu{B^+ \to \rho^0 \ell\nu}

\def\BtoDstlnu{\bar{B} \to D^* \ell^-\bar{\nu}_{\ell}}

\def\BtoDstmunu{\bar{B} \to D^* \mu^-\bar{\nu}}
\def\BtoDstenu{\bar{B} \to D^* e^-\bar{\nu}}
\def\BtoDmunu{\bar{B} \to D \mu^-\bar{\nu}}
\def\BtoDenu{\bar{B} \to D e^-\bar{\nu}}

\def\BtoDsttaunu{\bar{B} \to D^* \tau^-\bar{\nu}_{\tau}}
\def\BtoDtaunu{\bar{B} \to D \tau^-\bar{\nu}}
\def\BtoDDsttaunu{\bar{B} \to D^{(*)} \tau^-\bar{\nu}}
\def\BtoDDstmunu{\bar{B} \to D^{(*)} \mu^-\bar{\nu}}
\def\BtoDDstenu{\bar{B} \to D^{(*)} e^-\bar{\nu}}
\def\BtoDDstellnu{\bar{B}\to D(D^*) \ell^-\bar{\nu}_{\ell}}

\def\btoclnu{b\to c\ell^-\bar{\nu}_{\ell}}
\def\btoctaunu{b\to c\tau^-\bar{\nu}_{\tau}}
\def\btoulnu{b\to u\ell^-\bar{\nu}_{\ell}}
\def\RDRDST{\frac{\Gamma(\bar{B}\to D^{(*)}\tau^-\bar{\nu}_{\tau})}{\Gamma(\bar{B}\to D^{(*)}\ell^-\bar{\nu}_{\ell})}}
\def\rdrdst{R(D^{(*)})}

\newcommand*{\rom}[1]{\expandafter\@slowromancap\romannumeral #1@}

\def\bd0tau{B\to D \tau\nu_{\tau}}

\def\be {\begin{equation}}
	\def\ee {\end{equation}}

\usepackage[normalem]{ulem}

\definecolor{darkgreen}{cmyk}{1,0,1,0.4}

\definecolor{pink}{cmyk}{0.4,1,0.3,0}

\def\com2#1{\textcolor{red}{\it{#1}}}

		\title{\boldmath Test of new physics effects in $\bar{B} \to (D^{(*)}, \pi) \ell^-\bar{\nu}_{\ell}$ decays with heavy and light leptons  }
			
		\author[a,b]{Ipsita Ray}
		\emailAdd{ipsitaray02@gmail.com }
		
		\author[a]{and Soumitra Nandi}
		\emailAdd{soumitra.nandi@iitg.ac.in}
		
		\affiliation[a]{Indian Institute of Technology, North Guwahati, Guwahati 781039, Assam, India }	
		\affiliation[b]{Indian Institute of Technology Gandhinagar
			Palaj, Gandhinagar - 382055, Gujarat, India }	
		
	\abstract{
We study the $\BtoDDstellnu$ decays based on the up-to-date available inputs from experiments and the lattice. First, we review the standard model (SM) predictions of the different observables associated with these decay channels. In the analyses, we consider new physics (NP) effects in the channels with the heavy ($\tau$), as well as the light leptons ($\mu, e$). We have extracted $|V_{cb}|$ along with the new physics Wilson coefficients (WCs) from the available data on light leptons; the extracted value of $|V_{cb}|$ is $(40.3 \pm 0.5)\times 10^{-3}$. The extracted WCs are consistent with zero, but some could be of order $10^{-2}$. Also, we have done the simultaneous analysis of the data in $\bar{B} \to D^{(*)}(\mu^-,e^-)\bar{\nu}$ alongside the inputs on $\rdrdst = \RDRDST$ and the $D^*$ longitudinal polarisation fraction $F_L^{D^*}$ in different NP scenarios and extracted $|V_{cb}|$ which is consistent with the number mentioned above. Also, the simultaneous explanation of $\rdrdst$ and $F_L^{D^*}$ is not possible in the one-operator scenarios. However, the two operator scenarios with $\mathcal{O}_{S_2}^{\tau} = (\bar{q}_R b_L)(\bar{\tau}_R\nu_{\tau L})$ as one of the operators could explain all these three measurements. Finally, we have given predictions of all the related observables in $\bar{B} \to D^{(*)}(\tau^-,\mu^-,e^-)\bar{\nu}$ decays in the NP scenarios, which could be tested in future experiments. We have repeated this exercise for $\btopilnu$ decays with the light lepton and extracted $|V_{ub}|$ and the new physics WCs. Finally, using all these available data for the light and heavy leptons, we have given bounds on the couplings of the relevant SM effective field theory (SMEFT) operators and the probable NP scale $\Lambda$.   }  
		 
		\arxivnumber{}	

\begin{document}
	\maketitle
	
\section{Introduction}

For the last couple of years, special attention has been given to the flavour-changing neutral current (FCNC) and charged current (FCCC) semileptonic $B$ and $B_s$ decays. In this paper, we will focus on $\bar{B}\to D(D^*) \ell^-\bar{\nu}_{\ell}$ and $\btopilnu$ decays, which are the FCCC semileptonic decays of $B$ meson. Here, we can identify $\ell$ as the light leptons like $e$ and $\mu$ or the heavy lepton $\tau$. The processes with the light leptons were assumed to be insensitive to any new physics effects, and the data on these modes are used to extract the CKM elements $|V_{cb}|$ and $|V_{ub}|$ from the $\BtoDDstellnu$ and $\btopilnu$ decays, respectively, for the updated analyses, see   \cite{Gambino:2019sif,Jaiswal:2020wer,Leljak:2021vte,Biswas:2021qyq,Bobeth:2021lya, Martinelli:2021myh,Martinelli:2021onb,Bernlochner:2021vlv,Martinelli:2022tte,Bernlochner:2021rel,Gonzalez-Solis:2021pyh,Biswas:2022yvh} and the references therein. Right now, alongside the experimental data on the differential rates \cite{Belle:2015pkj,Abdesselam:2017kjf,Belle:2018ezy,Belle:2023bwv,delAmoSanchez:2010af,Ha:2010rf,BaBar:2012dvs,BaBar:2013pls,Lees:2012vv,Sibidanov:2013rkk}, we have very precise inputs from the lattice on the respective form factors at zero and and non-zero recoils        
\cite{MILC:2015uhg,Na:2015kha,FermilabLattice:2014ysv,FermilabLattice:2015cdh,Flynn:2015mha,FermilabLattice:2021cdg,Colquhoun:2022atw,Aoki:2023qpa}.
With this wealth of data, it is possible to extract $|V_{cb}|$ and $|V_{ub}|$ and the possible new physics effects in $\btoclnu$ and $\btoulnu$ decays along with the respective parameters parametrising the form factors of the decays as mentioned above. In this paper, we have analysed all the available data and the lattice inputs on these modes, extracted the respective CKM elements, and constrained the model-independent new physics (NP) information alongside the shape of the form factors. 

For the same decay modes we can define the ratio of the decays rates $R(D^{(*)}) = \RDRDST$. These observables are potentially sensitive to lepton flavor universality violating (LFUV) new physics effects in these decays. At the moment, measurements are available on $R(D)$ and $R(D^*)$. The measured value of $R(D)$ is given by \cite{hflavold,hflavnew}
\begin{equation}\label{eq:RDexp}
R(D) =  0.356 \pm 0.029 
\end{equation}
which is the average of the measurements in the refs.~\cite{BaBar:2012obs,BaBar:2013mob, Belle:2015qfa,Belle:2019rba,LHCb:2023zxo}.
There is disagreement between the measurement and the SM prediction $R(D)_{SM} = 0.298 \pm 0.004$ \cite{hflavnew} which is an arithmetic average of the estimates obtained in the refs.~ \cite{hflavold,Bigi:2016mdz,Bernlochner:2017jka,Jaiswal:2017rve,Bordone:2019vic,Martinelli:2021onb}. On the other hand, the measured value of $R(D^*)$ is given by 
\begin{equation}\label{eq:RDstexp1}
R(D^*) =  0.291 \pm 0.014, 
\end{equation}
which has been obtained from an average of the estimates in \cite{BaBar:2012obs,BaBar:2013mob,Belle:2015qfa,Belle:2017ilt,Belle:2019rba,LHCb:2023zxo}. Very recently, LHCb has announced their new result on $R(D^*)$ \cite{LHCb:2023cjr}, which, when combined with their earlier measurement \cite{LHCb:2017rln}, the average will be
\begin{equation}\label{eq:RDstlhcbnew}
R(D^*) = 0.257 \pm 0.012 \pm 0.018.
\end{equation}    
On combining this result with the average given in eq.~\ref{eq:RDstexp1}, HFLAV has obtained the new average \cite{hflavnew} 
 \begin{equation}\label{eq:RDstexp2}
 R(D^*) =  0.284 \pm 0.013. 
 \end{equation} 
The SM predictions for the $R(D^*)_{SM} = 0.254 \pm 0.005$, the arithmetic average of the predictions obtained in the refs.~\cite{Bernlochner:2017jka,Jaiswal:2017rve,Gambino:2019sif,BaBar:2019vpl,Bordone:2019vic,
Jaiswal:2020wer,hflavnew}. Apart from \cite{BaBar:2019vpl}, the rest of the analyses are based on the data from Belle \cite{Abdesselam:2017kjf,Belle:2018ezy}, and one input on the form factor from the lattice at zero recoil has been considered \cite{FermilabLattice:2014ysv}. As a recent update, the Fermilab-MILC collaboration estimated the form factors for $\BtoDstlnu$ decays at the non-zero recoils utilising which they have obtained $R(D^*)_{SM} = 0.265 \pm 0.013$. This estimate does not include the input from the experimental measurements.  It is completely based on the lattice inputs and consistent with the measurement presented in eq.~\ref{eq:RDstexp2}. Note that the SM prediction (only lattice) and the measurement have large errors. In the case of $R(D)$, the measurement has a large error compared to the corresponding SM estimate. We have to wait for more precise inputs from the lattice and more precise data from the experimental measurements. There have been many global analyses to accommodate the tension in $R(D^{(*)})$ in the presence of new physics, for eg. ref. \cite{Bhattacharya:2018kig} discusses about some one-operator and two-operator scenarios that can provide plausible explanations to the data with new physics only in the $\tau$ channel. It is important to note that the data on $\BtoDDstellnu$ with light lepton are useful to constrain the shape of the form factors on top of the lattice data. However, doing the fits in the presence of NP contributions in these modes will be more appropriate. Therefore, the present data could be useful to constrain the new physics contributions in $\btoclnu$ and $\btoctaunu$ decays alongside the extraction of $|V_{cb}|$ which is our main focus in this paper. In principle, the NP contributions in $\btoclnu$ and $\btoctaunu$ could be different or they could be of similar nature, we have explored both these possibilities and constrained the new model-independent Wilson coefficients.        

In addition to the $R(D)$ and $R(D^*)$, measurements are available on the $\tau$ polarization asymmetry $P_{\tau}^{D^*}$ \cite{Belle:2017ilt} and the $D^*$ longitudinal polarization fraction $F_{L}(D^*)$ \cite{Belle:2019ewo}. The respective estimates are the following
\begin{equation}\label{eq:angularbelle}
P_{\tau}^{D^*} = - 0.38 \pm 0.51 (stat) ^{ + 0.21}_{-0.16} (sys) , \ \ \ F_L^{\bar{B} \to D^* \tau^- \bar{\nu}} = 0.60 \pm 0.08(stat) \pm 0.04(sys).
\end{equation}
In a recent publication \cite{Belle:2023bwv}, the Belle collaboration has measured the forward-backward asymmetries $A_{FB}^{\bar{B} \to D^* (\mu^-,e^-) \bar{\nu}}$ and $F_L^{\bar{B} \to D^* (\mu^-,e^-) \bar{\nu}}$ .  
Note that the measurement for the $\tau$ polarization has an error of more than 100\%, and the measurement of $F_L^{D^*}$ is not a published article. The SM predictions can be seen from \cite{Jaiswal:2020wer}, which were based on the Belle 2017 and/or 2019 results, and by then, available inputs on the respective form factors in $B\to D^*$ decays at zero recoil. In this paper we will update those numbers. Also, we will update the SM predictions of the relevant observables in $\bar{B}\to D\ell^-\bar{\nu}_{\ell}$ decays, for the older results one could see the ref.~\cite{Bhattacharya:2018kig,Huang:2018nnq}. We haven't considered $P_{\tau}(D^*)$ as input in our analysis. However, we have checked that the predictions are not exceeding the experimental limits. In a few fits, we have included $F_{L}(D^*)$ to check the impact of this measurement.    

The Effective field theory (EFT) is an important framework to deal with phenomenon
that are spread over a multitude of energy or length scales, such as the electroweak scale
determining flavor-changing transitions of quarks and the scale of strong interactions related to the formation of hadrons. Hence, the low energy data could be useful to constrain the scale of the new physics. In the framework of the SM Effective Field Theory (SMEFT), the renormalizable dimension-4 SM Lagrangian is extended by higher dimensional operators suppressed by powers of the new physics scale $\Lambda$ \cite{Henning:2014wua}.
\begin{equation}
   \mathcal{L}_{\text{eff}} = \mathcal{L}_{\text{SM}} + \sum_i \frac{1}{\Lambda^{d_i -4}} C_i \mathcal{O}_i 
\end{equation}
where $\Lambda$ is the cutoff scale of the EFT, $\mathcal{O}_i$ are a set of dimension $d_i$ operators that are invariant under the $\text{SU}(3)_c \times \text{SU}(2)_L \times \text{U}(1)_Y$ gauge group and $C_i$ are their respective Wilson coefficients (WC). New physics at higher energy scales is encoded in the Wilson coefficients of higher dimensional operators. The WCs of the low energy effective operators could be obtained by matching the SMEFT to these operators and following an appropriate renormalization group evolution equations (RGE) \cite{Jenkins:2017jig,Aebischer:2015fzz}. In this analysis, we have obtained the constraints on the ratio $\frac{C_i}{\Lambda^2}$ from the available data on the exclusive $\bar{B} \to (D,D^*,\pi) \ell^- \bar{\nu}_{\ell}$ decays. These decays can be described by writing down the most general dimension-6 operators, which we will discuss in the next section.  

\section{Theory framework and inputs}\label{sec:framework}

Assuming the neutrinos to be left-handed, the most general effective Hamiltonian  with all possible four-fermion operators relevant for $b \to c(u) l^- \bar{\nu_l}$ is given as \cite{Sakaki:2013bfa}
   
\begin{equation} \label{eq:effH}
	\mathcal{H}^{b\to q \ell^-\bar{\nu}_{\ell}}_{\text{eff}}=\frac{4G_F}{\sqrt{2}}V_{qb}[(1 + C_{V_1}^l)\mathcal{O}_{V_1}^l+C_{V_2}^l \mathcal{O}_{V_2}^l+C_{S_1}^l\mathcal{O}_{S_1}^l+C_{S_2}^l\mathcal{O}_{S_2}^l+C_T^l \mathcal{O}_T^l] 
\end{equation}
where the four-Fermi operators are given by,
\begin{eqnarray}\label{eq:effopr}
	\mathcal{O}_{V_1}^l &=& (\bar{q}_L\gamma^\mu b_L)(\bar{l}_L\gamma_\mu\nu_{lL}), \nonumber
	\\
	\mathcal{O}_{V_2}^l &=& (\bar{q}_R\gamma^\mu b_R)(\bar{l}_L\gamma_\mu\nu_{lL}), \nonumber
	\\
	\mathcal{O}_{S_1}^l &=& (\bar{q}_L b_R)(\bar{l}_R\nu_{lL}), \nonumber
	\\
	\mathcal{O}_{S_2}^l &=& (\bar{q}_R b_L)(\bar{l}_R\nu_{lL}), \nonumber
	\\
	\mathcal{O}_{T}^l &=& (\bar{q}_R\sigma^{\mu\nu}b_L)(\bar{l}_R\sigma_{\mu\nu}\nu_{lL}),
\end{eqnarray}
with $q = u$ or $c$. In the Standard Model, the WC $C^{\ell}_i = 0$, hence the WCs defined in eq.~\ref{eq:effH} are associated only with the new operators beyond the SM. We will constrain these WCs from the available data. Using the available experimental data on the $\BtoDDstellnu$, $R(D^{(*)})$, $\btopilnu$ ($\ell = \mu$, e) and the available lattice inputs, we can extract these WCs. In this analysis, we have considered only real WCs.     

In the SMEFT, the operators which will be relevant for the $b \to c (u) \ell^- \bar{\nu}_{\ell}$ transitions are given by \cite{Aebischer:2015fzz}
\begin{eqnarray} \label{eq:opssmeft}
Q^{(3)}_{\ell q} &=& (\bar{\ell}_i \gamma_{\mu} \tau^{I} \ell_j)(\bar{q}_k \gamma^{\mu} \tau^{I} q_l), ~~~~~~ Q_{\phi u d} = i (\tilde{\phi}^{\dagger} D_{\mu} \phi)(\bar{u}_i \gamma^{\mu} d_j) ,\nonumber \\
Q_{\ell edq} &=& (\bar{\ell}_i^a e_j)(\bar{d}_k q_l^a), ~~~~~~~~~~~~~~~~~ Q^{(1)}_{\ell equ} = (\bar{\ell}_i^a e_j)\epsilon_{ab}(\bar{q}_k^b u_l),\nonumber \\
Q^{(3)}_{\ell equ} &=& (\bar{\ell}_i^a \sigma^{\mu \nu} e_j)\epsilon_{ab}(\bar{q}_k^b \sigma_{\mu \nu} u_l), ~~~~Q^{(3)}_{\phi q} = (\phi^{\dagger} i \overset{\leftrightarrow}{D}_{\mu} \phi)(\bar{q}_i \tau^I \gamma^{\mu} q_j)
\end{eqnarray}
In the above equation, $\ell,q$ and $\phi$ represent lepton, quark and Higgs $SU(2)_L$ doublets, while the right-handed isospin singlets are denoted by $e$, $u$ and $d$.

By matching these Standard Model gauge invariant dimension-six operators at the electroweak scale onto the low-energy $B$ physics Hamiltonian by integrating out the top quark, $Z$ and $W$ bosons and the Higgs boson, the Wilson coefficients (WCs) are obtained \cite{Aebischer:2015fzz}. These WCs are then evolved from the electroweak scale $\mu_W$ to the scale $\mu_b$ relevant to $B$ physics measurements by the appropriate RGE equations. As a result, the WCs of the operators defined in eq.~\ref{eq:effH} are obtained as a linear combination of the Wilson coefficients of the corresponding operators of the SMEFT basis. Thus, the low-energy experiments in flavour physics can be used to constrain these coefficients of the SMEFT operator basis.

Following the method discussed above, a tree-level matching of the SMEFT operators (eq.~\ref{eq:opssmeft}) to the effective operator basis defined in eqs.~\ref{eq:effH} and \ref{eq:effopr} will result in the following WCs at the scale $m_b$ for $b\to c\ell^-\bar{\nu}_{\ell}$ decays \cite{Aebischer:2015fzz}
\begin{eqnarray} \label{eq:matchedbtoc}
C_{V_1} &=& -\frac{v^2}{\Lambda^2} \frac{V_{cs}}{V_{cb}} (\tilde{C}^{(3)ll23}_{\ell q}-\tilde{C}^{(3)23}_{\phi q}), ~~~~~~~ C_{V_2} = \frac{v^2}{2 \Lambda^2 {V_{cb}}} \tilde{C}^{23}_{\phi u d}, \nonumber \\
C_{S_1} &=& - \frac{v^2}{2 \Lambda^2} \frac{V_{cs}}{V_{cb}} \tilde{C}^{* ll32}_{\ell edq}, ~~~~~~~~~~~~~~~~~~~~~ C_{S_2} = - \frac{v^2}{2 \Lambda^2} \frac{V_{tb}}{V_{cb}} \tilde{C}^{*(1) ll32}_{\ell equ}, \nonumber \\
C_T &=& - \frac{v^2}{2 \Lambda^2} \frac{V_{tb}}{V_{cb}} \tilde{C}^{*(3) ll32}_{\ell equ}
\end{eqnarray}
and that for $b \to u \ell^- \bar{\nu}$ transitions,
\begin{eqnarray} \label{eq:matchedbtou}
C_{V_1} &=& - \frac{v^2}{\Lambda^2} \frac{V_{ud}}{V_{ub}} (\tilde{C}^{(3)ll13}_{\ell q}-\tilde{C}^{(3)13}_{\phi q}), ~~~~~~~ C_{V_2} =  \frac{v^2}{2 \Lambda^2 {V_{ub}}} \tilde{C}^{13}_{\phi u d}, \nonumber \\
C_{S_1} &=& - \frac{v^2}{2 \Lambda^2} \frac{V_{ud}}{V_{ub}} \tilde{C}^{* ll31}_{\ell edq}, ~~~~~~~~~~~~~~~~~~~~~ C_{S_2} = -\frac{v^2}{2 \Lambda^2} \frac{V_{tb}}{V_{ub}} \tilde{C}^{*(1) ll31}_{\ell equ}, \nonumber \\
C_T &=& - \frac{v^2}{2 \Lambda^2} \frac{V_{tb}}{V_{ub}} \tilde{C}^{*(3) ll31}_{\ell equ}
\end{eqnarray}

From hereon, we will denote $(\tilde{C}^{(3)ll23}_{\ell q}-\tilde{C}^{(3)23}_{\phi q})$ = $\tilde{C}^{(3)}_{l q}$ since we can only constrain the difference from the fits.
Here, $v$ is the Higgs vacuum expectation value, and $\Lambda$ is the new physics scale. In eqs. \ref{eq:matchedbtoc} and \ref{eq:matchedbtou}, we have neglected the sub-dominant CKM elements. Here, the $\tilde{C}$s are the WCs or the couplings of the respective SMEFT operators at the electroweak scale $\mu_{EW}$. We will use the available experimental results on the $\BtoDDstellnu$, $\bar{B}\to \pi\ell^-\bar{\nu}_{\ell}$ and $R(D^{(*)})$ alongside the lattice inputs mentioned in the introduction to fit the ratios $\tilde{C}/\Lambda^2$ given in the above equations. The result will help us to pinpoint the scale of new physics $\Lambda$ for particular choices of the couplings or vice versa. We will discuss this in detail in the following sections.

\subsection{$\bar{B} \to D(\pi) \ell^- \bar{\nu}_{\ell}$}\label{subsec:ObsBtoDpi}

\begin{table}[t]
	\begin{center}
		\begin{tabular}{|c|c|}
			\hline
			Constants & Values\\
			\hline
			$G_F$ & 1.166 $\times$ $10^{-5}$ ~GeV$^{-2}$\\
			$\chi^{T}_{1^-}$ (0)  (for g and $f_+$) & 5.131 $\times$ $10^{-4}$~GeV$^{-2}$ \\
			$\chi^{T}_{1^+}$ (0)  (for f and $F_1$) & 3.894 $\times$ $10^{-4}$~GeV$^{-2}$ \\
			$\chi^{L}_{0^-}$ (0)  (for $F_2$) & 1.9421 $\times$ $10^{-2}$~GeV$^{-2}$ \\
			$\chi^{L}_{0^+}$ (0)  (for $f_0$) & 6.204 $\times$ $10^{-3}$~GeV$^{-2}$ \\
			\hline
		\end{tabular}
		\caption{Various inputs used in this analysis \cite{Bigi:2017jbd}. }
		\label{tab:inputs}
	\end{center}
\end{table} 

\begin{table}[t]
	\begin{center}
		\begin{tabular}{|c|c|}
			\hline
			Form factor involved & $B_c^{(*)}$ pole masses (GeV)\\
			\hline
			$f_+$ and $g$ & 6.32847, 6.91947, 7.030\\
			$f$ and $F_1$ &  6.73847, 6.750, 7.145, 7.150\\
			$F_2$ & 6.27447, 6.8712, 7.250\\
			$f_0$ & 6.70347, 7.122\\
			\hline
		\end{tabular}
		\caption{Pole masses used in the $B \to D^{(*)}$ modes.}
		\label{tab:poleBD}
	\end{center}
\end{table} 

Using the above effective Hamiltonian, the differential decay rate in the presence of new physics for $\bar{B} \to D l^- \bar{\nu}$ transitions can be written as \cite{Sakaki:2013bfa} (both for heavy and light leptons):

\begin{equation}\label{eq:dgamBtoD}
	\begin{split}
		{d\Gamma(\bar{B} \to D l^- \bar{\nu_l}) \over dq^2} =& {G_F^2 \eta_{\text{EW}}^2 |V_{cb}|^2 \over 192\pi^3 m_B^3} q^2 \sqrt{\lambda_D(q^2)} \left( 1 - {m_l^2 \over q^2} \right)^2 \times\biggl\{ \biggr. |1 + C_{V_1}^l + C_{V_2}^l|^2 \left[ \left( 1 + {m_l^2 \over2q^2} \right) H_{V,0}^{s\,2} + {3 \over 2}{m_l^2 \over q^2} \, H_{V,t}^{s\,2} \right] \\ 
		& + {3 \over 2} |C_{S_1}^l + C_{S_2}^l|^2 \, H_S^{s\,2} + 8|C_T^l|^2 \left( 1+ {2m_l^2 \over q^2} \right) \, H_T^{s\,2} + 3 Re[ ( 1 + C_{V_1}^l + C_{V_2}^l ) (C_{S_1}^{l*} + C_{S_2}^{l*} ) ] \\
		& {m_l \over \sqrt{q^2}} \, H_S^s H_{V,t}^s - 12 Re[ ( 1 + C_{V_1}^l + C_{V_2}^l ) C_T^{l*} ] {m_l \over \sqrt{q^2}} \, H_T^s H_{V,0}^s \biggl.\biggr\} \,,
	\end{split}
\end{equation}

where $\lambda_D(q^2)=((m_B-m_D)^2 - q^2)((m_B + m_D)^2 - q^2)$ and 

\begin{align}
			H_{V,0}^s(q^2) =& \sqrt{\lambda_D(q^2) \over q^2} f_+(q^2),~~~~~~~~	H_{V,t}^s(q^2) = {m_B^2-m_D^2 \over \sqrt{q^2}} f_0(q^2) \nn \\
		H_S^s(q^2) =& {m_B^2-m_D^2 \over m_b-m_c} f_0(q^2),~~~~~~~~
			H_T^s(q^2) = -{\sqrt{\lambda_D(q^2)} \over m_B+m_D} f_T(q^2) \,
\end{align}

The form factors are expanded in the BGL method of parametrization as \cite{PhysRevLett.74.4603}

\begin{equation}
\mathcal{F}_i (z) = \frac{1}{P_i (z) \phi_i (z)} \sum_{j=0}^{N} a_{j}^i z^j,
\label{eq:FF-BGL}
\end{equation}
where $z$ is related to the recoil variable $w$ as

\begin{equation}\label{eq:z}
z = \frac{\sqrt{w+1}-\sqrt{2}}{\sqrt{w+1}+\sqrt{2}}.
\end{equation}

$w$ is related to the momentum transferred to the dilepton system ($q^2$) as $q^2 = m_B^2 + m_f^2 - 2 m_B m_f w$.

The functions $P_i (z)$, called the Blaschke factors,  are given by
\begin{equation}
P_i(z) = \prod_p \frac{z-z_p}{1 - z z_p},
\label{eq:Blaschke-fact}
\end{equation}
which are used to eliminate the poles at $z=z_p$ where,
\begin{equation}
z_p = \frac{\sqrt{(m_B + m_f)^2 - m_P^2} - \sqrt{4 m_B m_f}}{\sqrt{(m_B + m_f)^2 - m_P^2} + \sqrt{4 m_B m_f}}.
\label{eq:zp}
\end{equation}

Here $m_P$ denotes the pole masses, details in \cite{Bigi:2017jbd}. The outer functions $\phi_i (z)$ can be any analytic
function of $q^2$ and are chosen to be \cite{Boyd:1997kz}

\begin{eqnarray}
	\phi_{f_+} &=& \frac{8r^2}{m_B} \sqrt{\frac{8 n_I}{3\pi \tilde{\chi}_{1^-}^T (0)}} \frac{(1+z)^2(1-z)^{1/2}}{\left[(1+r)(1-z) + 2\sqrt{r}(1+z)\right]^5}, \nonumber \\
   	\phi_{f_0} &=& r(1-r^2) \sqrt{\frac{8 n_I}{\pi \tilde{\chi}_{1^-}^L (0)}} \frac{(1-z^2)(1-z)^{1/2}}{\left[(1+r)(1-z) + 2\sqrt{r}(1+z)\right]^4},
\end{eqnarray}
where $r = m_D/m_B$.

If we write the double-differential decay distribution as \cite{Becirevic:2019tpx}

\begin{equation} \label{eq:pseudo}
	\frac{d^2\Gamma_l}{dq^2 d \text{cos}\theta } =
	a_l (q^2)+b_l (q^2) \cos \theta + c_l (q^2) \cos^2 \theta,
\end{equation}	
where $\theta$ is the polar angle of the lepton momentum in the rest frame of the $l \bar{\nu}$ pair with respect to the z axis which is defined by the $D$-momentum in the rest frame of $\bar{B}$, the lepton forward-backward asymmetry is defined as

\begin{equation}
	A_{FB} (q^2) = \frac{\int_0^1 \frac{d^2 \Gamma}{d q^2 d \text{cos}\theta_l} d \text{cos}\theta_l-\int_{-1}^0 \frac{d^2 \Gamma}{d q^2 d \text{cos}\theta_l} d \text{cos}\theta_l}{d \Gamma/dq^2} = \frac{b_l (q^2)}{d\Gamma /d q^2}
\end{equation} 

and the lepton polarization asymmetry is defined as

\begin{equation}
	A_{\lambda_l}^{D} (q^2) = - P_{\ell}^D =  \frac{d \Gamma^{\lambda_l = -1/2}/dq^2 - d \Gamma^{\lambda_l = +1/2}/dq^2}{d \Gamma/dq^2}.
\end{equation}
Note that the our definition for the lepton polarization asymmetry has a sign difference with respect to the one used by Belle (eq.~\ref{eq:angularbelle}). 

The rate for the $\btopilnu$ decays are similar to the one given in eq.~\ref{eq:dgamBtoD} with the inputs for the $D$ meson replaced by that for the pions. To get the shape of the decay rate, we need to determine the shape of the form factors for $\bar{B} \to \pi \ell^-\bar{\nu}_{\ell}$. For the form factors, we use the simplified series expansion in $z$ as given in ref.~\cite{Straub:2015ica} by Bharucha-Straub-Zwicky (BSZ): 
\begin{equation}\label{eq:bszexp}
f_i(q^2) = \frac{1}{1 - q^2/m_{R,i}^2} \sum_{k=0}^N a_k^i \, [z(q^2;t_0)-z(0;t_0)]^k, 
\end{equation}
where $z$ is defined as before in eqs.~\ref{eq:z} and \ref{eq:zp}, respectively. In eq \ref{eq:bszexp},  $m_{R,i}$ denotes the mass of the sub-threshold resonances compatible with the quantum numbers of the respective form factors and $a_k^i$s are the coefficients of expansion, the details are given in ref \cite{Straub:2015ica}. Another commonly used approach for the parametrizations of the $B\to \pi$ form factors in the literature is provided by Bourrely-Caprini-Lellouch (BCL)~\cite{BCL}. A comparative study of these two approaches in the extractions of $|V_{ub}|$ from $\bar{B}\to \pi\ell^-\bar{\nu}_{\ell}$ decays has been undertaken in ref.~\cite{Biswas:2021qyq}. The results are extremely consistent with each other in both approaches. For interested readers, a more detailed discussion on this topic could also be seen from our earlier paper \cite{Biswas:2022yvh}.

\subsection{$\bar{B} \rightarrow D^{*} l^- \bar{\nu}$}\label{subsec:ObsBtoDst}

The decay distribution for the four body decay $\bar{B} \to D^*(\to D \pi) l^- \bar{\nu}_l$ can be completely described in terms
of four kinematic variables: the lepton invariant mass squared ($q^2$) and the three angles \cite{Becirevic:2019tpx}

\begin{equation}\label{eq:d4Gamma}
	\frac{d^4\Gamma}{dq^2 dcos\theta_l dcos\theta_V d\phi} =
	\frac{9}{32\pi} I(q^2, \theta_l, \theta_V, \phi)
\end{equation}
where, 
\begin{align} \label{eq:angulardist}
	I(q^2, \theta_l, \theta_V, \phi)& = 
	I_1^s sin^2\theta_V + I_1^c cos^2\theta_V
	+ (I_2^s sin^2\theta_V + I_2^c cos^2\theta_V) cos2\theta_l
	 + I_3 sin^2\theta_V sin^2\theta_l cos2\phi \nonumber \\ 
&	+ I_4 sin2\theta_V sin2\theta_l cos\phi  + I_5 sin2\theta_V sin\theta_l cos\phi
	+ (I_6^s sin^2\theta_V + I_6^c cos^2\theta_V) cos\theta_l \nonumber \\
&	+ I_7 sin2\theta_V sin\theta_l sin\phi 
	+ I_8 sin2\theta_V sin2\theta_l sin\phi
	+ I_9 sin^2\theta_V sin^2\theta_l sin2\phi
\end{align} 

 The angular coefficients ($I$'s) are written in terms of the helicity amplitudes $H_{\lambda}$ as well as their linear combinations \cite{Becirevic:2019tpx} which are in turn expressed as functions of the form factors ($f,g,F_1,F_2$) in the BGL basis and the Wilson coefficients. The outer functions with $r = m_{D^*}/m_B$ are expresssed as \cite{Boyd:1997kz}
\begin{eqnarray}
\phi_f &=& \frac{4r}{m_B^2} \sqrt{\frac{n_I}{6\pi \chi_{1^+}^T (0)}} \frac{(1+z)(1-z)^{3/2}}{\left[(1+r)(1-z) + 2\sqrt{r}(1+z)\right]^4}, \nonumber \\
\phi_g &=& 16r^2 \sqrt{\frac{n_I}{3\pi \tilde{\chi}_{1^-}^T (0)}} \frac{(1+z)^2(1-z)^{-1/2}}{\left[(1+r)(1-z) + 2\sqrt{r}(1+z)\right]^4}, \nonumber\\
\phi_{\mathcal{F}_1} &=& \frac{4r}{m_B^3} \sqrt{\frac{n_I}{6\pi \chi_{1^+}^T (0)}} \frac{(1+z)(1-z)^{5/2}}{\left[(1+r)(1-z) + 2\sqrt{r}(1+z)\right]^5}, \nonumber \\
\phi_{\mathcal{F}_2} &=& 8\sqrt{2}r^2 \sqrt{\frac{n_I}{\pi \tilde{\chi}_{1^+}^L (0)}} \frac{(1+z)^2 (1-z)^{-1/2}}{\left[(1+r)(1-z) + 2\sqrt{r}(1+z)\right]^4} 
\end{eqnarray}

In table \ref{tab:inputs}, we give the values of the susceptibilities $\chi$ relevant to the different form factors. The pole masses used in $B \to D^{(*)}$ channel are given in table \ref{tab:poleBD}.

For this mode, the various observables are defined as follows \cite{Becirevic:2019tpx} :

	\begin{subequations}
	\begin{align}
 	&A_{FB} (q^2) = \frac{3}{8} \frac{(I_6^c+2 I_6^s)}{d\Gamma/dq^2}
 	\,, \\
	&A_{\lambda_l}^{D^*} (q^2) = - P_{\ell}^{D^*} = \frac{d \Gamma^{\lambda_l = -1/2}/dq^2 - d \Gamma^{\lambda_l = +1/2}/dq^2}{d \Gamma/dq^2}
	\,, \\
	& F_L^{D^*} (q^2) = \frac{3 I_1^c-I_2^c}{3I_1^c+6I_1^s-I_2^c-2I_2^s}.
	\end{align}
	\end{subequations}
As defined before, $A_{FB} (q^2)$ and $A_{\lambda_l}^{D^*} (q^2)$ are the forward backward asymmetry and the $\tau$ polarization asymmetry, respectively. In addition, in this channel we have the longitudinal polarization fraction of $D^*$ which is defined by $F_L^{D^*}$.

\subsection{Inputs}\label{subsec:inputs}

In ref.~\cite{Belle:2015pkj}, the Belle collaboration had analyzed the $\bar{B}\to D l^- \bar{\nu}$ decay. They provided measurements for the differential decay width in the recoil variable $w$ in 10 $w$-bins for both the charged and neutral $B$ decays with electrons and muons in the final state. In 2018 and 2023, it presented the results for the differential distributions in $w$, $\text{cos} \theta_l$, $\text{cos} \theta_v$ and $\chi$ in 10 bins for the $\bar{B}\to D^* l^- \bar{\nu}$ decay mode \cite{Belle:2018ezy,Belle:2023bwv}. For $R(D^{(*)})$, we consider the averages as performed by HFLAV \cite{hflavold,hflavnew}, the averages are obtained with and without the input on the very recent measurements on $R(D^*)$ given in eq.~\ref{eq:RDstlhcbnew}. As mentioned in the introduction, in a few fits we have included the experimental data on $F_L^{D^*}$ as input. In addition to these experimental results, we consider the inputs for the hadronic form factors available from various sources. For the $B \to D$ form factors, we take lattice inputs from the Fermilab-MILC collaboration \cite{MILC:2015uhg} and the HPQCD collaboration \cite{Na:2015kha}. In \cite{MILC:2015uhg}, the form factors $f_+$ and $f_0$ are given at three values of the recoil variable $w$ = 1, 1.08 and 1.16, whereas in \cite{Na:2015kha}, the fit results for the form factor parameters following BCL expansion are provided, using which we created synthetic data-points for $f_+$ and $f_0$ at $w$ = 1, 1.06 and 1.12 as their results are directly calculated for $w$ $\lsim$ 1.12. The lattice inputs for the form factors of $B \to D^*$ mode are taken from the Fermilab-MILC and JLQCD collaborations \cite{FermilabLattice:2021cdg,Aoki:2023qpa} where the values of the form factors $g, f, F_1$ and $F_2$ are provided at non-zero values of the recoil parameter, namely $w$ = 1.03, 1.10 and 1.17 and $w$ = 1.025, 1.06 and 1.10 respectively. The lattice predictions are more reliable for the high-$q^2$ or low-$w$ values. In a couple of fits, we have also included the input on the form factors at $q^2=0$ $\text{GeV}^2$ obtained from the Light Cone Sum Rule (LCSR) approach \cite{Gubernari:2018wyi}. 

Data on the differential $\bar{B} \to \pi l^- \bar{\nu}$ decay rates is available from BaBar (2011)~\cite{delAmoSanchez:2010af}, Belle (2011)~\cite{Ha:2010rf}, BaBar (2012)~\cite{Lees:2012vv}, Belle (2013)~\cite{Sibidanov:2013rkk}. The lattice collaborations RBC/UKQCD \cite{Flynn:2015mha} and JLQCD \cite{Colquhoun:2022atw} provide synthetic data points for $f_{+,0}(q^2)$ with full covariance matrices (both systematic and statistical) at three $q^2$ points which we have directly used in our analysis. On the other hand, the Fermilab/MILC collaboration \cite{FermilabLattice:2015cdh,Flynn:2015mha} provides the fit results for the BCL coefficients of the $z$-expansions of the respective form factors. We use their fit results to generate correlated synthetic data points at the same $q^2$ values as RBC/UKQCD, with an extra point for $f_+$ at $q^2 = 20.5$ GeV$^2$. In addition to the lattice inputs, we have used the inputs on the form factors $f_+(q^2=0) = f_0(q^2=0)$ obtained by LCSR approaches \cite{Leljak:2021vte} though the results are insensitive to this input.

 \section{Analysis and results}

In this work, we have minimized the $\chi^2$ statistic defined as
	\begin{equation}
	\label{eq:chisq}
	\chi^2 = \sum_{i,j} (O_i^{\rm exp}- O_i^{\rm theo}).\, Cov^{-1}_{ij}. \,(O_j^{\rm exp}- O_j^{\rm theo}),
	\end{equation} 
	where $O_i^{\rm exp}$ is the vector of data and $Cov$ is the corresponding covariance matrix, $O_j^{\rm theo}$ is the vector of predicted values. The uncertainties of the fit parameters are obtained from the hessian matrix.

We have broadly analysed all the available information following the different fit procedures, which can be understood as enumerated below.
\begin{enumerate}
	\item We have used the available inputs on the form factors from lattice and LCSR to predict the observables in the SM. In these fits, we have not used the experimental data mentioned in section \ref{subsec:inputs}.
	
	\item  Along with the theory inputs, we have used the available data on the $q^2$ and angular distributions in $\BtoDDstmunu$, $\BtoDDstenu$ and $\bar{B}\to \pi (\mu^-, e^-)\bar{\nu}$ decays to extract $|V_{cb}|$, $|V_{ub}|$. Using the fit results, we have predicted the observables related to these decays and $\BtoDDsttaunu$, which we have explicitly mentioned in the last section. We have done these fits with and without any NP contributions in $\BtoDDstmunu$, $\BtoDDstenu$ and $\bar{B}\to \pi (\mu^-, e^-)\bar{\nu}$ decays.  
	
	\item In some additional fits, we have included the measured values of $\rdrdst$ and $F_L^{D^*}$ as inputs alongside the theory inputs and the inputs mentioned in the second enumerated environment. This kind of fit is particularly important for the extraction of NP information in $\BtoDDsttaunu$ decays. 
\end{enumerate}
The motivations for these different type of fits are described in the items below:
\begin{itemize}
	\item In the extractions of $|V_{cb}|$, the experimental data on the rates (see sub-section \ref{subsec:inputs}) of $\BtoDDstmunu$, $\BtoDDstenu$ are needed alongside the inputs from the lattice. Similarly, one uses the data on the rates in $\bar{B}\to \pi (\mu^-, e^-)\bar{\nu}$ decays to extract $|V_{ub}|$. The understanding is that NP effect in these decays is expected to be small, hence, neglected. The more realistic situation will be to consider the NP effect in these decay modes while extracting the CKM elements from experimental data. Here, we have extracted $|V_{cb}|$ or $|V_{ub}|$, allowing NP contributions in the relevant decay modes mentioned and fitting the new physics parameters alongside the CKM elements. The fit results are compared with those fits where NP is not considered. This kind of study will help to check the consistency in the extracted values of the CKM elements in the fits with and without the NP contributions. Also, one will be able to check the size of NP contributions to get an idea whether or not one could neglect them in those fits. 
	
	\item   We need to determine the shape of the form factors to predict (in the SM) the observables related to the decay modes mentioned in item-1 and in $\BtoDDsttaunu$ decays, which we have discussed in the subsections \ref{subsec:ObsBtoDpi} and \ref{subsec:ObsBtoDst}, respectively. These observables are defined in such a way that they are insensitive to the CKM elements. Ideally, the SM predictions should be obtained using only the theory inputs without any biases from the experimental data. It is natural to expect that the experimental data could introduce bias in the SM predictions since the fit usually assumes no NP presence. In principle, one could obtain the shapes of these form factors using only the lattice inputs at the zero and non-zero recoils. Using these shapes, one will obtain the predictions of the observables in the SM. However, due to the unavailability of the lattice data at non-zero recoils, the earlier analyses used the experimental data to fit the form factors without considering any new physics effects in the rates of $\BtoDDstmunu$, $\BtoDDstenu$ decays, following the assumptions mentioned in item-1. Experimental data might introduce biases in the respective predictions, which could deviate from those obtained only from the lattice inputs. To pinpoint the impacts of different inputs and NP, we have extracted the relevant observables using only the theory inputs (lattice or ``lattice + LCSR"), theory inputs alongside the experimental data with and without any NP contributions in the respective rates in $\BtoDDstmunu$, $\BtoDDstenu$ decays and have analysed the results. The similar arguments hold for the observables in $\bar{B}\to \pi (\mu^-, e^-)\bar{\nu}$ decays.   
	
	\item As discussed in the introduction, measurements are available on $\rdrdst$ and $F_L^{D^*}(\BtoDsttaunu)$. These additional data, particularly, are important to estimate the new physics effects in $\BtoDsttaunu$ decays. Hence, in some of the fits, we have included these three data alongside the data on $\BtoDDstmunu$ and $\BtoDDstenu$ decays and the theory inputs mentioned in items 1 and 2. Now, the contribution from NP could be of minimal flavour violating (MFV) type in which the NP effects could be flavour blind. Hence, similar effects will be observed in $\BtoDDstmunu$, $\BtoDDstenu$ and $\BtoDsttaunu$ decays, as for example, see the analysis in \cite{Biswas:2021pic}. Also, it could be a non-MFV scenario (for example, the leptoquark model) where the contributions are flavour dependent; hence, one would expect different contributions in the light and heavy lepton final states. We considered all these possibilities in the model-independent analyses, divided them into two categories and fitted the new physics WCs from the data. In one category, we have considered a similar type of new WC contributing to all the $\BtoDDstmunu$, $\BtoDDstenu$ and $\BtoDsttaunu$ decays. We have assumed different new physics WCs in another category contributing to $\BtoDsttaunu$, $\BtoDDstmunu$, and $\BtoDDstenu$ decays, respectively. This type of comparative study will help us to understand whether or not the new physics contributions allowed in $\BtoDsttaunu$ decays are consistent with the data in $\BtoDDstmunu$, and $\BtoDDstenu$ decays. Also, we will know their order of magnitude if there are differences.

\end{itemize}

\begin{figure}[t]
	\small
	\centering
	\subfloat[]{\includegraphics[width=0.5\textwidth]{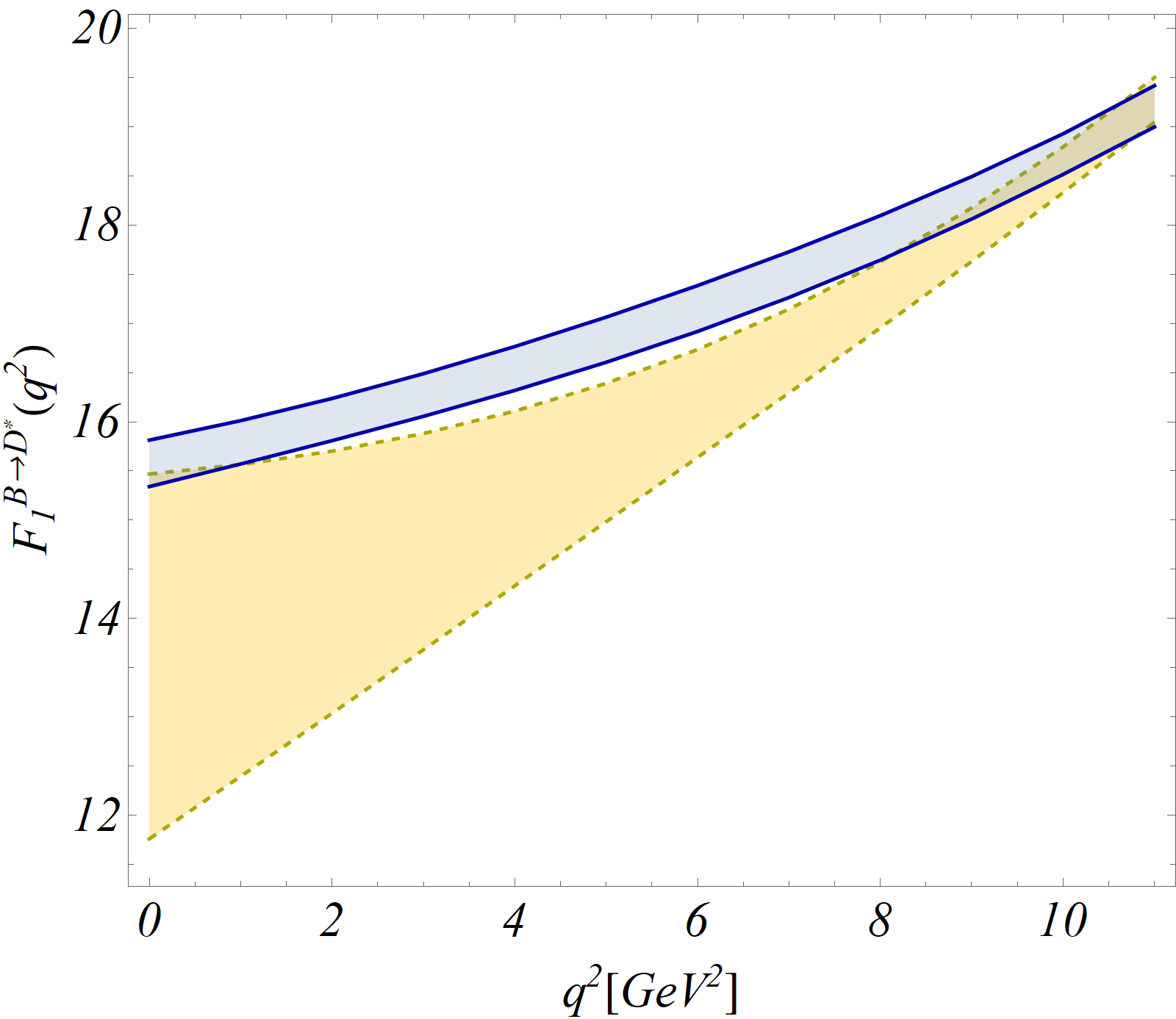}\label{fig:ff1}}~~~~~
	\subfloat[]{\includegraphics[width=0.5\textwidth]{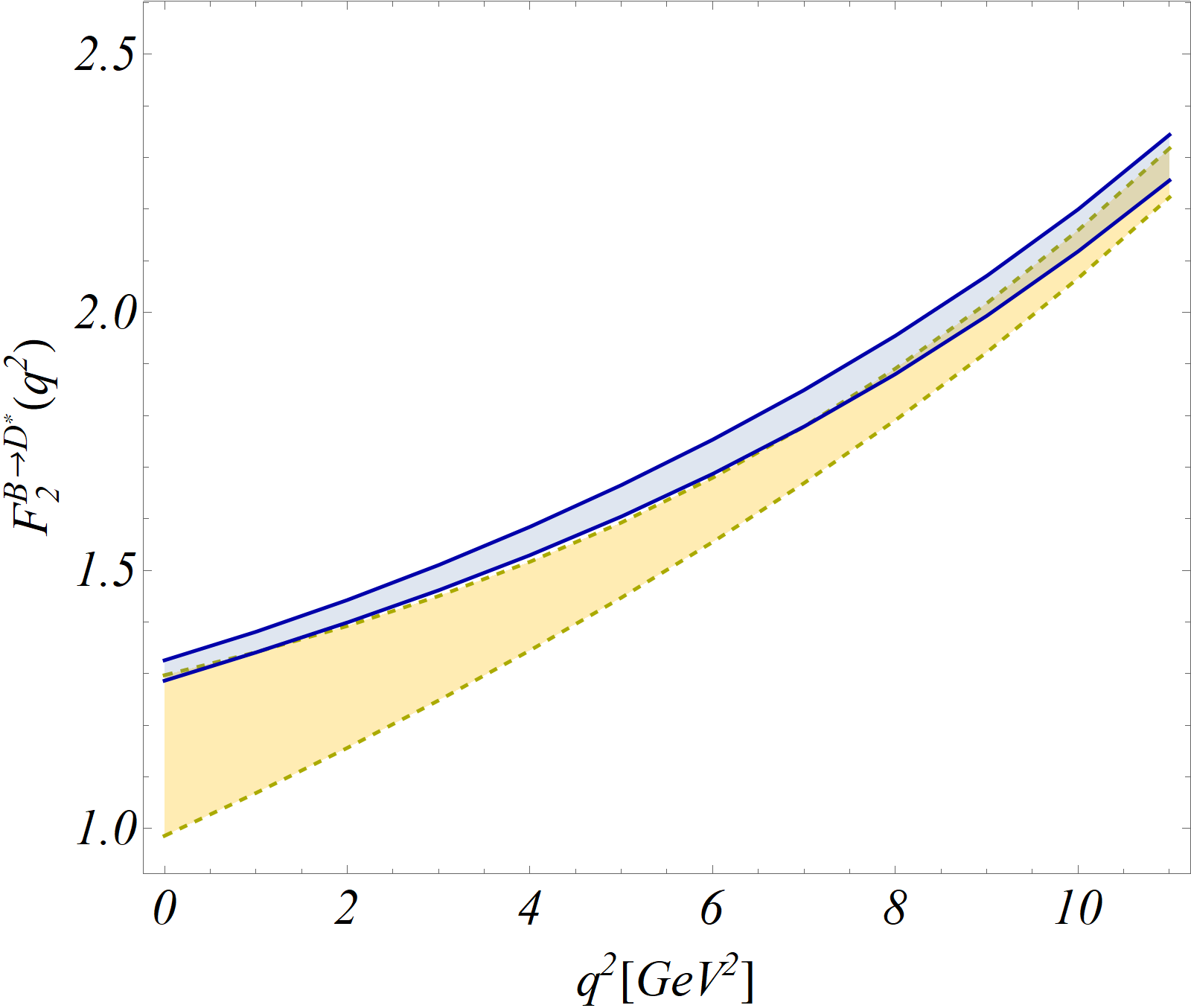}\label{fig:ff2}}\\
	\subfloat[]{\includegraphics[width=0.5\textwidth]{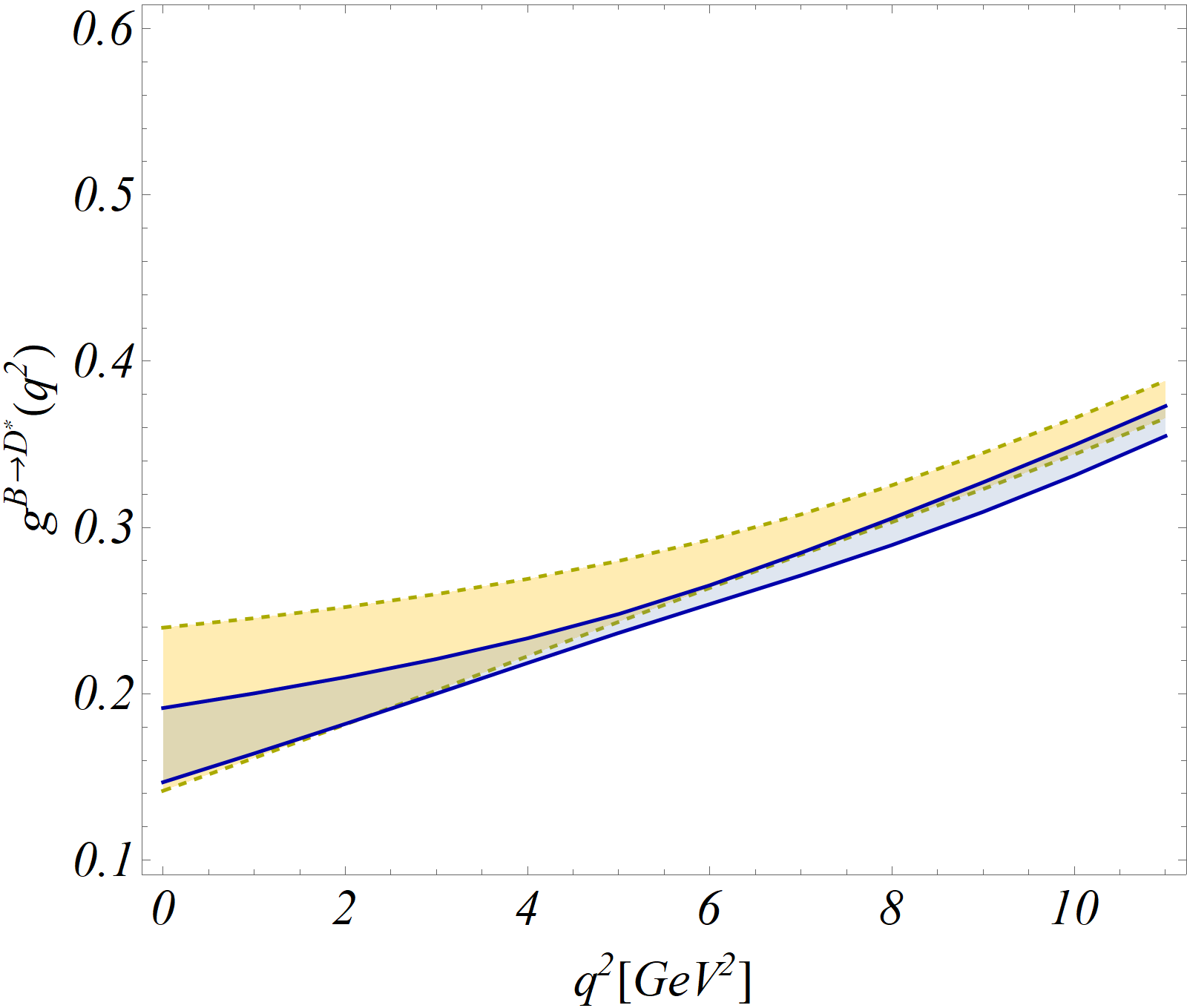}\label{fig:gth}}~~~~~
	\subfloat[]{\includegraphics[width=0.5\textwidth]{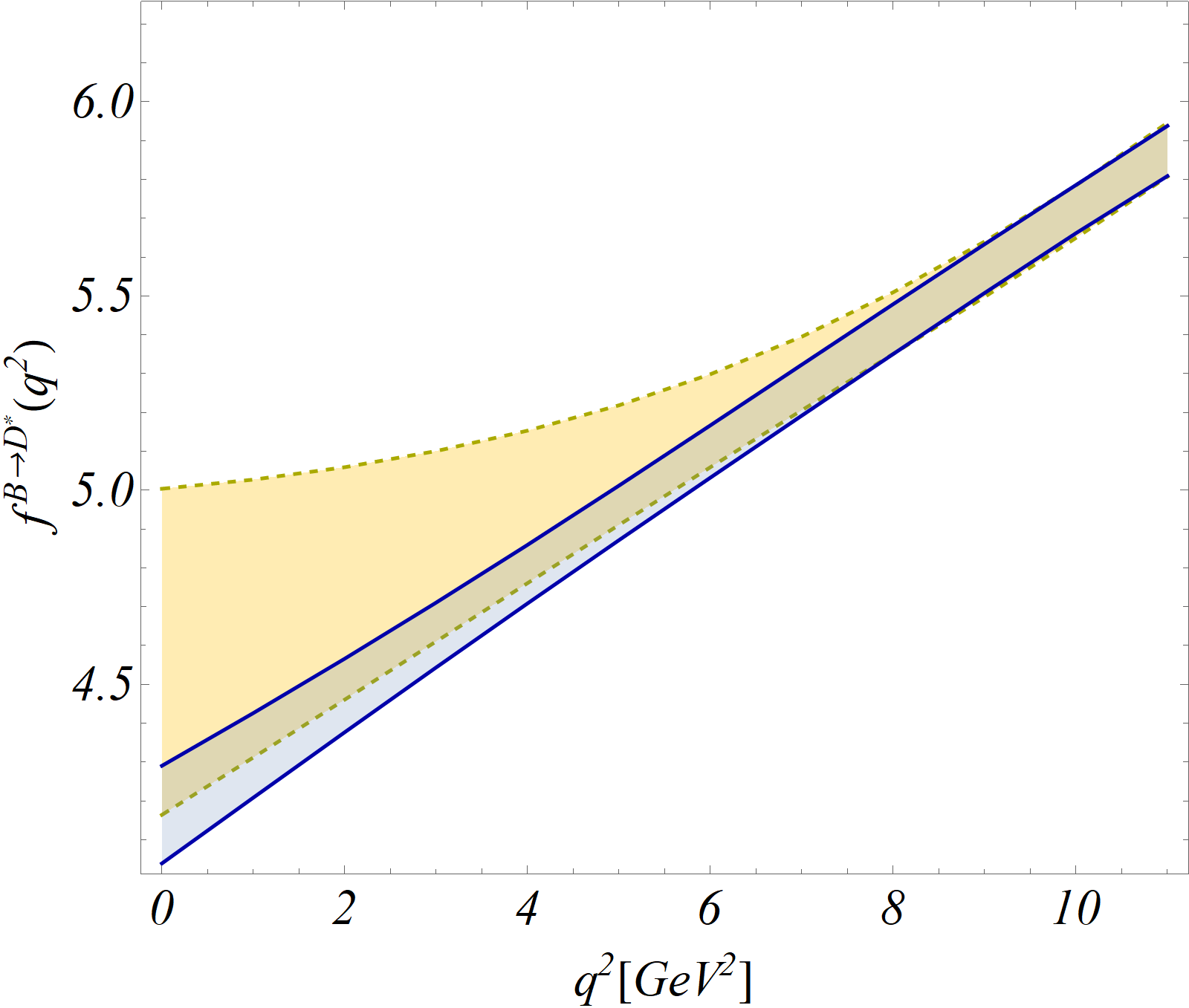}\label{fig:fth}}\\
	{\includegraphics[width=0.5\textwidth]{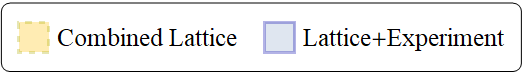}}\\
	\caption{The comparison of the $q^2$ shapes of the $B\to D^*$ form factors obtained in the fit to ``lattice", and ``lattice + experiments".}
	\label{fig:ff}
\end{figure}

\begin{table}[t]
	\begin{center}
		\scriptsize
		\renewcommand*{\arraystretch}{1.6}
		\begin{tabular}{|c|c|c|c|c|c|c|}
			\hline
			Observables & \multicolumn{2}{c|}  {Lattice}  &  Lattice combined  &  All Lattice  & All Lattice   & All Lattice   \\
			\cline{2-4}
			& MILC  & JLQCD  &
			$B\to D$  \cite{MILC:2015uhg,Na:2015kha} & +  & +  &  + LCSR ($q^2$ = 0)  \\
			& \cite{FermilabLattice:2021cdg} & \cite{Aoki:2023qpa} &  $B \to D^*$\cite{FermilabLattice:2021cdg,Aoki:2023qpa} & All Expt.  & LCSR ($q^2$ = 0) & + All Expt  \\
			&  &   & & \cite{Belle:2015pkj,Belle:2018ezy,Belle:2023bwv} & \cite{Gubernari:2018wyi}  &  \\
			\hline
			$R(D)$ & & &  0.304(3) & 0.300(3) & 0.304(3) & 0.300(3) \\
			$R({D^*})$ & 0.271(31) & 0.253(22) & 0.258(12)  & 0.251(1) & 0.253(9) & 0.251(1) \\
			\hline
			$F_L^{\bar{B} \to D^* \tau^- \bar{\nu}}$ & 0.422(11) & 0.446(22) & 0.427(9) & 0.453(3) & 0.439(6) & 0.453(3) \\
			$A_{\lambda_\tau}^{D^*} = - P_{\tau}^{D^*} $ & 0.526(11) & 0.509(15) & 0.519(7)  & 0.507(3) & 0.512(5) & 0.506(3) \\
			$A_{FB}^{\bar{B} \to D^* \tau^- \bar{\nu}}$ & -0.082(12) & -0.053(21) & -0.077(9)  & -0.051(4) & -0.070(7) & -0.051(4)\\
			$A_{\lambda_\tau}^{D}= - P_{\tau}^{D} $ & &  & -0.324(3) & -0.323(3) & -0.324(3) & -0.323(3) \\
			
			$A_{FB}^{\bar{B} \to D \tau^- \bar{\nu}}$ & &  & 0.3596(4) & 0.3600(2) & 0.3596(4) & 0.3600(2) \\
			\hline
			$A_{\lambda_\mu}^{D^*} = - P_{\mu}^{D^*} $ & 0.989(4) & 0.986(4) & 0.987(2) & 0.9852(2) & 0.986(1) & 0.9852(2) \\
			$F_L^{\bar{B} \to D^* \mu^- \bar{\nu}}$ & 0.463(36) & 0.510(60) & 0.480(22) &  0.530(3) & 0.506(15) & 0.530(3) \\
			$A_{FB}^{\bar{B} \to D^* \mu^- \bar{\nu}}$ & -0.247(29) & -0.215(35) & -0.243(17)  & -0.208(3) & -0.233(11) & -0.209(3) \\ $A_{\lambda_\mu}^{D}= - P_{\mu}^{D} $ &  & & 0.9618(2) & 0.9615(2) & 0.9618(2) & 0.9615(2) \\
			$A_{FB}^{\bar{B} \to D \mu^- \bar{\nu}}$ & &  & 0.01369(10) & 0.01380(8) & 0.01369(10) & 0.01379(8) \\
			\hline
			$F_L^{\bar{B} \to D^* e^- \bar{\nu}}$ &  0.463(36)  &  0.510(60) & 0.480(23)  & 0.530(3) & 0.507(15) & 0.530(3)\\
			$A_{FB}^{\bar{B} \to D^* e^- \bar{\nu}}$ & -0.250(29)  & -0.220(34)& -0.247(17) & -0.214(3) & -0.238(10) & -0.214(3) \\
			\hline
		\end{tabular}
		\caption{Predictions of the observables in the $B \to D^{(*)}$ channel in different scenarios. Here, the experimental data is in the $\bar{B}\to D (\mu,e)\bar{\nu}$ and $\bar{B}\to D^* (\mu,e)\bar{\nu}$ decays.}
		\label{tab:rdrdstsm}
	\end{center}
\end{table}

\subsection{Analysis without any NP effects in $b\to c \ell^-\bar{\nu}$}

As pointed above, our first objective is to predict the observables $R(D)$, $R(D^*)$,  $A_{\lambda_{\ell }}^{D^{(*)}}$, $A_{FB}^{D^{(*)}}$ and $F_L^{D^*}$ in the SM for $\BtoDDsttaunu$, $\BtoDDstmunu$ and $\BtoDDstenu$ decays. To understand the gravity of impact of the experimental data, we have done these analyses in scenarios with and without the inputs from the measurements on the decay rates in small $q^2$ or angular bins for the light leptons like muons or electrons. In table \ref{tab:rdrdstsm}, we have predicted the relevant observables in six different analyses depending on the inputs used, which includes a comparative study of the predicted observables in the $\bar{B} \to D^* l \bar{\nu}$ mode using the lattice inputs from Fermilab-MILC \cite{FermilabLattice:2021cdg} and JLQCD \cite{Aoki:2023qpa} collaborations. Also, we have checked the impact of different lattice and LCSR inputs. To predict the observables, we need to constrain the shape of the form factors. Using the results of the fit to ``all lattice" and ``all lattice + all experiments," we have obtained the shapes of all the form factors. For the $B\to D$ form factors, the shapes are very much consistent with each other. The shapes for $B\to D^*$ form factors obtained in these two types of fits are compared in fig.~\ref{fig:ff}; the bands are the errors at 1$\sigma$ CI. As expected, the shapes obtained from the lattice have large errors as compared to those obtained from the fit to ``lattice + experiment", in particular, in the low-$q^2$ regions. The shapes obtained for $F_2$ and $F_1$ in the fits with or without the experimental data are not fully consistent (at 1$\sigma$). There are slight discrepancies in parts of the $q^2$ regions. The same is true for $g$ in the high $q^2$ region, where the two shapes are marginally consistent.

Regarding the predictions of the observables, given the large errors, the respective Fermilab-MILC and JLQCD predictions are consistent at their 1$\sigma$ confidence interval (CI). The predictions of $R(D)$ are consistent in all four scenarios defined in table \ref{tab:rdrdstsm}, so are the $A_{FB}^{\BtoDtaunu}$ and $A_{\lambda_{\tau}}^{D}$, due to the lattice's precise inputs. Whether or not the experimental data are used, the predictions are very much consistent with each other. Also, we note that the input from the LCSR has a negligible impact. As expected, the prediction of $R(D^*)$ using the lattice inputs (combined) has a large error and is marginally consistent with the experimental measurement \cite{hflavnew}. The error is reduced due to the use of the available inputs on the form factors at $q^2=0$ from LCSR. The inputs from the available experiments considerably reduce the prediction error, and the value is inconsistent with the respective measurements given in eq.~\ref{eq:RDstexp1} or eq.~\ref{eq:RDstexp2}. 

We compare the values of $F_L$ and $A_{FB}$ for the ($\mu$,e) channels obtained in the SM which are shown in table \ref{tab:rdrdstsm} with the corresponding measured values by Belle collaboration \cite{Belle:2023bwv}. It is to be noted that the numbers presented in this table  are the most up-to-date predictions obtained in the SM.
\begin{equation}
A_{FB}^{\bar{B}^{(0,-)} \to D^{*(+,0)} \mu^- \bar{\nu}} = 0.252 \pm 0.019 \pm 0.005, \ \ A_{FB}^{\bar{B}^{(0,-)} \to D^{*(+,0)} e^- \bar{\nu}} = 0.230 \pm 0.018 \pm 0.005,
\end{equation}
\begin{equation}
F_L^{\bar{B}^{(0,-)} \to D^{*(+,0)} \mu^- \bar{\nu}} = 0.518 \pm 0.017 \pm 0.005, \ \ F_L^{\bar{B}^{(0,-)} \to D^{*(+,0)} e^- \bar{\nu}} = 0.485 \pm 0.017 \pm 0.005.
\end{equation}
In these two equations, the measured values are presented from the averages of the respective $\bar{B}^0$ and $B^-$ decays. Note that our SM predictions of these observables based on the lattice (only) inputs are consistent with the respective measured values. 

Also, from the results of the table \ref{tab:rdrdstsm}, the predictions are obtained on the following observables
\begin{align}\label{eq:deltaobs}
\Delta A_{FB}^{D^{(*)},\ell_i\ell_j} &= A_{FB}^{\bar{B} \to D^{(*)} \ell_i\nu} -  A_{FB}^{\bar{B} \to D^{(*)} \ell_j\nu} \nonumber \\
\Delta F_L^{D^*,\ell_i \ell_j} &= F_L^{\bar{B} \to D^* \ell_i\nu} -  F_L^{\bar{B} \to D^* \ell_j\nu}.
\end{align}
The predicted values of these observables in the SM using the results of the fit to lattice (only) are given below 
\begin{equation}
\Delta A_{FB}^{D^*,\mu \tau} = -0.166 \pm 0.010, \ \ \Delta A_{FB}^{D^*,\mu e} = 0.004 \pm 0.001,\ \ \Delta A_{FB}^{D^*,e \tau} = -0.170 \pm 0.010, 
\end{equation}
\begin{equation}
\Delta A_{FB}^{D,\mu \tau} = -0.3459 \pm 0.0004, \ \ \Delta A_{FB}^{D,\mu e} = 0.0137 \pm 0.0001,\ \ \Delta A_{FB}^{D,e \tau} = -0.3596 \pm 0.0004, 
\end{equation}
\begin{equation}
\Delta F_L^{D^*,\mu \tau} = 0.053 \pm 0.015, \ \ \Delta F_L^{D^*,\mu e} = (0.07 \pm 0.12)\times 10^{-3}, \ \ \Delta F_L^{D^*,e \tau} = 0.053 \pm 0.015.   
\end{equation}
Note that $\Delta A_{FB}^{D^*,\mu e}$ and $\Delta F_L^{D^*,\mu e} $ are in agreement with the corresponding measured values \cite{Belle:2023bwv}.

In the following we will point out a few other interesting observations from table \ref{tab:rdrdstsm}: 
\begin{itemize}
\item The predicted values of $F_L^{\BtoDsttaunu}$, $A_{\lambda_{\tau }}^{D^{*}}$ and $A_{FB}^{\BtoDsttaunu}$ in the fits with and without the experimental data differ from each other 2.7$\sigma$, 1.6$\sigma$ and 2.6$\sigma$, respectively. This is due to the slight discrepancies in the shape of the form factors shown in fig.~\ref{fig:ff}. In particular, these observables are sensitive to the form factors $F_1(q^2)$ and $g(q^2)$. Experimental data reduces the errors in estimating these observables due to a considerable reduction in error in the extracted form factors.

	\item Similarly for the $\BtoDstmunu$ decays, we can note deviations in the predictions of $F_L^{\BtoDstmunu}$, $A_{\lambda_{\mu }}^{D^{*}}$ and $A_{FB}^{\BtoDstmunu}$ between the fits with or with the experimental data at 2.3$\sigma$, 1.2$\sigma$ and 2.0$\sigma$, respectively. This trend continues for the observables $F_L^{\BtoDstenu}$ and $A_{\lambda_{e }}^{D^{*}}$.
	
	\item We do not see such deviations in the observables in $\BtoDmunu$ and $\BtoDenu$ decays. 
	
	\item  We have estimated $\Delta A_{FB}^{D^*,\mu e}$ in the fits with or without the experimental data and noticed that the predictions of $\Delta A_{FB}^{{D^*,\mu e}}$ between these two types of fits differ at 1.31$\sigma$. In this regard, one could also see the analysis of \cite{Bobeth:2021lya}, where a deviation of $\sim$ 3.9$\sigma $ in $\Delta A_{FB}^{{D^*,\mu e}}$ have been estimated between the data-driven fits and the SM. Also, they have pointed out the discrepancies between data-driven fit and the SM predictions for $A_{FB}^{B\to D^* \mu^-\nu}$ $\sim$ 4$\sigma$ and $A_{FB}^{B\to D^* e^-\nu}$ $\sim$ 2$\sigma$ in \cite{Bobeth:2021lya}. Note that in the analysis of ref. \cite{Bobeth:2021lya}, for the SM predictions, the form factors are obtained using heavy quark flavour and spin symmetries and the inputs from LCSR. The lattice inputs we used in this analysis were not available then. Also, they have used the experimental data available at that point of time. At the moment, more updated experimental data are available, which we are using in this analysis in addition to what had been used in ref.~\cite{Bobeth:2021lya}. From the results of table \ref{tab:rdrdstsm}, we have done similar checks for $\Delta A_{FB}^{D^{(*)},\mu \tau}$ and $\Delta A_{FB}^{D^{(*)},e \tau}$, also for the $\Delta F_L^{D^*,\ell_i \ell_j}$, and we have not observed any significant deviations.          
\end{itemize}

The minor differences mentioned in items one and two are due to the slight differences in the shapes of the form factors relevant to $B\to D^{*}(\mu^-,e^-)\nu$ decays obtained in the fits with and without the experimental data. These discrepancies may become more prominent or disappear with more precise inputs from theory and experiments. Hence, to reach any reasonable conclusions regarding whether or not these are indicative of NP, we need more precise data on the $B\to D^*$ form factors. However, we will take this opportunity to test the NP from the available data. Also, we will see in the following subsections that the new physics WCs in $b\to c\mu^- (e^-)\bar{\nu}$ decays are tightly constrained, and they are negligibly small to give sizeable contributions in the related observables.

\begin{table}[t]
	\begin{center}
		\scriptsize
		\renewcommand*{\arraystretch}{1.6}
		\begin{tabular}{|c|c|c|c|c|}
			\hline
			Scenario & \multicolumn{2}{|c|}{From the simultaneous fitting of all unknowns} & \multicolumn{2}{c|}{BGL coefficients are nuisance parameters} \\
			\cline{2-5}
			& Fit Parameters & Predictions  & Fit Parameters & Predictions \\
			\hline
			$C_{V_2}$ & $\begin{array}{l}
			C_{V_2} = -0.005\pm 0.014\\
			|V_{cb}| = (40.2 \pm 0.6) \times 10^{-3} 
			\end{array}$ & $\begin{array}{l} R(D)= 0.300 \pm 0.003, \\ 
			R(D^*)=	0.251 \pm 0.001 \end{array}$  & $\begin{array}{l}
			C_{V_2} = 0.003\pm 0.015 \\
			|V_{cb}| = (40.5 \pm 0.6) \times 10^{-3} 
			\end{array}$ & $\begin{array}{l} R(D)= 0.298 \pm 0.005, \\ 
			R(D^*)=	0.251 \pm 0.002 \end{array}$ \\
			\hline
			
			$C_T$ & $\begin{array}{l} 
			C_T = (0.2\pm 4.9) \times 10^{-4}\\
			|V_{cb}| = (40.3 \pm 0.5) \times 10^{-3} 
			\end{array}$ & $\begin{array}{l} R(D)= 0.300 \pm 0.003, \\ 
			R(D^*)=	0.251 \pm 0.001 \end{array}$ & $\begin{array}{l} 
			C_T = (0.4\pm 5.0) \times 10^{-4}\\
			|V_{cb}| = (40.5 \pm 0.5) \times 10^{-3} 
			\end{array}$ & $\begin{array}{l} R(D)= 0.298 \pm 0.005, \\ 
			R(D^*)=	0.251 \pm 0.001 \end{array}$ \\
			\hline
			
			$C_{S_1}$ & $\begin{array}{l} 
			C_{S_1} = (-0.003 \pm 0.046)  \\
			|V_{cb}| = (40.3 \pm 0.5) \times 10^{-3} 
			\end{array}$ & $\begin{array}{l} R(D)= 0.299 \pm 0.022, \\ 
			R(D^*)=	0.251 \pm 0.002 \end{array}$ & $\begin{array}{l} 
			C_{S_1} = (-0.005 \pm 0.047)  \\
			|V_{cb}| = (40.5 \pm 0.5) \times 10^{-3} 
			\end{array}$ & $\begin{array}{l} R(D)= 0.295 \pm 0.022, \\ 
			R(D^*)=	0.251 \pm 0.002 \end{array}$ \\
			\hline
			
			$C_{S_2}$ & $\begin{array}{l}
			C_{S_2} = -0.002\pm 0.046\\
			|V_{cb}| = (40.3 \pm 0.5) \times 10^{-3} 
			\end{array}$ & $\begin{array}{l} R(D)= 0.299 \pm 0.021, \\ 
			R(D^*)=	0.251 \pm 0.002 \end{array}$  & $\begin{array}{l}
			C_{S_2} = 0.0001 \pm 0.0464\\
			|V_{cb}| = (40.5\pm 0.5) \times 10^{-3} 
			\end{array}$ & $\begin{array}{l} R(D)= 0.298 \pm 0.022, \\ 
			R(D^*)=	0.251 \pm 0.002 \end{array}$ \\
			\hline
		\end{tabular}
		\caption{The simultaneous fit of $|V_{cb}|$ and the new physics WCs. The inputs are the ``Lattice + LCSR($q^2$ = 0) + data on $\bar{B}\to D (\mu,e)\bar{\nu}$ and $\bar{B}\to D^* (\mu,e)\bar{\nu}$ decays". The fit results for the BGL coefficients are given in a separate file. The predictions of $R(D)$ and $R(D^*)$ using the fit results in different scenarios are also shown. }
		\label{tab:fitNPnoLFU}
	\end{center}
\end{table}

\begin{table}[t]
	\begin{center}
		\scriptsize
		\renewcommand*{\arraystretch}{1.6}
		\begin{tabular}{|c|c|c|}
			\hline
			Scenario & Simultaneous extractions  & Simultaneous fitting of $|V_{cb}|$ and NP,  \\
			& of all the unknowns  &  (BGL coefficients as nuissance)   \\
			\hline
			$C_{V_2}$ & $\begin{array}{l}
			C_{V_2} = -0.005\pm 0.014\\
			|V_{cb}| = (40.3 \pm 0.6) \times 10^{-3} 
			\end{array}$  & $\begin{array}{l}
			C_{V_2} = -0.004\pm 0.014\\
			|V_{cb}| = (40.3 \pm 0.6) \times 10^{-3} 
			\end{array}$    \\
			\hline
			
			$C_T$ & $\begin{array}{l} 
			C_T = (-0.2\pm 4.8) \times 10^{-4}\\
			|V_{cb}| = (40.4 \pm 0.5) \times 10^{-3} 
			\end{array}$ &  $\begin{array}{l} 
			C_T = (-0.2\pm 4.9) \times 10^{-4}\\
			|V_{cb}| = (40.4 \pm 0.5) \times 10^{-3} 
			\end{array}$   \\
			\hline
			
			$C_{S_1}$ & $\begin{array}{l} 
			C_{S_1} = 0.06 \pm 0.03 \\
			|V_{cb}| = (40.3 \pm 0.5) \times 10^{-3} 
			\end{array}$ &  $\begin{array}{l} 
			C_{S_1} = 0.06 \pm 0.03 \\
			|V_{cb}| = (40.3 \pm 0.5) \times 10^{-3} 
			\end{array}$  \\
			\hline
			
			$C_{S_2}$ & $\begin{array}{l}
			C_{S_2} = 0.05\pm 0.03\\
			|V_{cb}| = (40.3 \pm 0.6) \times 10^{-3} 
			\end{array}$ &  $\begin{array}{l}
			C_{S_2} = 0.05\pm 0.03\\
			|V_{cb}| = (40.3 \pm 0.5) \times 10^{-3} 
			\end{array}$ \\
			\hline
		\end{tabular}
		\caption{The simultaneous fit of $|V_{cb}|$ and the new physics WCs. The inputs are ``Lattice + LCSR($q^2$ = 0) + data on $\bar{B}\to D (\mu,e)\bar{\nu}$ and $\bar{B}\to D^* (\mu,e)\bar{\nu}$ decays + $R(D^{(*)})$". These fits assume similar NP in $\tau$, $\mu$ and $e$ final states.  }
		\label{tab:fitNPwLFU}
	\end{center}
\end{table}

\begin{table}[t]
	\begin{center}
		\scriptsize
		\renewcommand*{\arraystretch}{1.6}
		\begin{tabular}{|c|c|c|}
			\hline
			Scenario &	$R(D^{(*)})$ from 2022 HFLAV \cite{hflavold} &  $R(D^{(*)})$ from Winter 2023 HFLAV \cite{hflavnew}   \\
			& Fitted values  & Fitted values  \\
			\hline
			$C_{V_1}$ & $\begin{array}{l} C_{V_1}^{\tau} = 0.08 \pm 0.02\\
			C_{V_1}^{\mu} = unconstrained\\
			|V_{cb}| = (40.3 \pm 0.5) \times 10^{-3} 
			\end{array}$ &  $\begin{array}{l} C_{V_1}^{\tau} = 0.07 \pm 0.02\\
			C_{V_1}^{\mu} = unconstrained\\
			|V_{cb}| = (40.3 \pm 0.5) \times 10^{-3} 
			\end{array}$  \\
			\hline
			
			$C_{V_2}$ & $\begin{array}{l} C_{V_2}^{\tau} = -0.07\pm 0.03\\
			C_{V_2}^{\mu} = -0.002\pm 0.014\\
			|V_{cb}| = (40.3 \pm 0.6) \times 10^{-3} 
			\end{array}$  & $\begin{array}{l} C_{V_2}^{\tau} = -0.06\pm 0.03\\
			C_{V_2}^{\mu} = -0.002 \pm 0.014\\
			|V_{cb}| = (40.3 \pm 0.6) \times 10^{-3} 
			\end{array}$   \\
			\hline
			
			$C_T$ & $\begin{array}{l} C_T^{\tau} = -0.05\pm 0.01\\
			C_T^{\mu} = (0.2\pm 4.9) \times 10^{-4}\\
			|V_{cb}| = (40.3 \pm 0.5) \times 10^{-3} 
			\end{array}$  & $\begin{array}{l} C_T^{\tau} = -0.04\pm 0.01\\
			C_T^{\mu} = (0.2\pm 4.9) \times 10^{-4}\\
			|V_{cb}| = (40.3 \pm 0.5) \times 10^{-3} 
			\end{array}$   \\
			\hline
			
			$C_{S_1}$ & $\begin{array}{l} C_{S_1}^{\tau} = 0.152 \pm 0.040\\
			C_{S_1}^{\mu} = -0.004 \pm 0.047 \\
			|V_{cb}| = (40.3 \pm 0.5) \times 10^{-3} 
			\end{array}$   & $\begin{array}{l} C_{S_1}^{\tau} = 0.150 \pm 0.041\\
			C_{S_1}^{\mu} = -0.004 \pm 0.047 \\
			|V_{cb}| = (40.3 \pm 0.5) \times 10^{-3} 
			\end{array}$  \\
			\hline
			
			$C_{S_2}$ & $\begin{array}{l} C_{S_2}^{\tau} = -1.336\pm 0.044\\
			C_{S_2}^{\mu} = -0.003\pm 0.046\\
			|V_{cb}| = (40.3 \pm 0.5) \times 10^{-3} 
			\end{array}$  & $\begin{array}{l} C_{S_2}^{\tau} = -1.330\pm 0.045\\
			C_{S_2}^{\mu} = -0.004\pm 0.046\\
			|V_{cb}| = (40.3 \pm 0.5) \times 10^{-3} 
			\end{array}$  \\
			\hline			
		\end{tabular}
		\caption{The simultaneous fit of $|V_{cb}|$ and the new physics WCs. The inputs are ``Lattice + LCSR($q^2$ = 0) + data on $\bar{B}\to D (\mu,e)\bar{\nu}$ and $\bar{B}\to D^* (\mu,e)\bar{\nu}$ decays + $R(D^{(*)}) + F_L^{D^*}$". These fit assumes different NP in $\tau$ and $\mu/e$ final states.}
		\label{tab:fitNPnonMFV}
	\end{center}
\end{table}

\begin{table}[t]
	\begin{center}
		\small
		\renewcommand*{\arraystretch}{1.6}
		\begin{tabular}{|*{7}{c|}}
			\hline
			Obs & SM & $C_{V_1}$ & $C_{V_2}$ & $C_T$  & $C_{S_1}$ & $C_{S_2}$  \\
			\hline
			$R(D)$ & 0.304(3) & {\bf 0.344(11)} & 0.269(16) & 0.290(4) & {\bf 0.386(26)} &{\bf 0.330(25)}   \\ 
			$R(D^*)$ & 0.258(12)   & {\bf 0.288(9)} & {\bf 0.277(14)} & {\bf 0.291(13)} & 0.256(2) & 0.306(3)     \\
			$F_L^{D^*}$ & 0.427(9)  & 0.453(3) & 0.457(3) & 0.442(5) & 0.463(4) & {\bf 0.551(5)}  \\
			\hline			
			$A_{\lambda_\tau}^{\bar{B} \to D^* \tau^- \bar{\nu}}$ & 0.519(7)   &  $\text{ 0.506(3)}$  &  $\text{ 0.505(3)}$  & {\bf $\text{0.48(1)}$} & {\bf $\text{0.479(8)}$ } &  {\bf $\text{0.24(1)}^*$} \\
			\hline
			$A_{FB}^{\bar{B} \to D^* \tau^- \bar{\nu}}$ & -0.077(9)  & $\text{ -0.053(4)}$  &  $\text{\bf -0.034(10)} $  &  $\text{\bf 0.009(17)}^*$  \
			&  $\text{\bf -0.039(5)}$  &  $\text{\bf 0.052(4)}$    \\
			\hline
			$A_{\lambda_\tau}^{\bar{B} \to D \tau^- \bar{\nu}}$  & -0.324(3) & $\text{-0.323(3)}$  &  $\text{-0.323(3)}$  &  $\text{\bf -0.348(8)}$  \
			&  $\text{\bf -0.47(4)}$  &  $\text{\bf -0.38(5)}$    \\
			\hline
			$A_{FB}^{\bar{B} \to D \tau^- \bar{\nu}}$ & 0.3596(4)  & $\text{0.3600(2)}$  &  $\text{0.3599(2)}$  &  $\text{\bf 0.343(5)}$  &  $\
			\text{\bf 0.337(7)}$  &  $\text{\bf -0.2724(5)}^*$     \\
			\hline
			$A_{\lambda_\mu}^{\bar{B} \to D^* \mu^- \bar{\nu}}$ & 0.987(2)  & $\text{0.9852(2)}$  &  $\text{0.9852(2)}$  &  $\text{0.9852(2)}$  \
			&  $\text{0.985(2)}$  &  $\text{0.985(2)}$    \\
			\hline
			$F_L^{\bar{B} \to D^* \mu^- \bar{\nu}}$ & 0.480(22)  & $\text{ 0.531(3)}$  &  $\text{ 0.530(3)}$  &  $\text{ 0.531(3)}$  &  $\text{ 0.531(3)}$  &  $\text{ 0.531(3)}$    \\
			\hline
			$A_{FB}^{\bar{B} \to D^* \mu^- \bar{\nu}}$ & -0.243(17)  & $\text{ -0.209(3)}$  &  $\text{ -0.209(3)}$  &  $\text{ -0.209(3)}$  \
			&  $\text{ -0.209(3)}$  &  $\text{ -0.209(3)}$    \\
			\hline
			$A_{\lambda_\mu}^{\bar{B} \to D \mu^- \bar{\nu}}$ & 0.9618(2) & $\text{0.9615(2)}$  &  $\text{0.9615(2)}$  &  $\text{0.9615(2)}$  \
			&  $\text{0.96(2)}$  &  $\text{0.96(1)}$    \\
			\hline
			$A_{FB}^{\bar{B} \to D \mu^- \bar{\nu}}$ & 0.01369(10)  & $\text{0.01380(8)}$  &  $\text{0.01378(8)}$  &  $\text{0.01379(9)}$  \
			&  $\text{0.014(3)}$  &  $\text{0.014(3)}$     \\
			\hline
		\end{tabular}
		
		\caption{Predictions of various observables integrated over the whole $q^2$ bins in the different NP scenarios in the $\bar{B} \to D^{(*)} l \bar{\nu}$ channel. }
		\label{tab:predobsfullq2}
	\end{center}
\end{table}

\begin{table}[t]
	\begin{center}
		\scriptsize		
		\renewcommand*{\arraystretch}{1.6}
		\begin{tabular}{|c|c|c|c|}
			\hline
			Scenarios &  Fit Parameters  & Scenarios & Fit Parameters  \\
			\hline
			$C_{V_1}, C_{V_2}$ &
			$\begin{array}{l} C_{V_1}^{\tau} = 0.08 \pm 0.02\\
			C_{V_2}^{\tau} = 0.008 \pm 0.033\\
			C_{V_2}^{\mu} = -0.005 \pm 0.014 \\
			|V_{cb}| = (40.2 \pm 0.6) \times 10^{-3}
			\end{array}$  & $C_{V_2},C_{S_1} $  & $\begin{array}{l} C_{V_2}^{\tau} = -0.06 \pm 0.03\\
			C_{S_1}^{\tau} = 0.17\pm 0.04\\
			C_{V_2}^{\mu} = -0.005\pm 0.014\\
			C_{S_1}^{\mu} = -0.003 \pm 0.048\\
			|V_{cb}| = (40.2 \pm 0.6) \times 10^{-3}
			\end{array}$  \\
			\hline
			
			$ C_{V_1},C_{S_1}  $ & $\begin{array}{l} C_{V_1}^{\tau} = 0.06\pm 0.03\\
			C_{S_1}^{\tau} = 0.04\pm 0.07\\C_{S_1}^{\mu} = -0.003\pm 0.047\\
			|V_{cb}| = (40.3 \pm 0.5) \times 10^{-3}
			\end{array}$  & $ C_{V_2}, C_{S_2}  $  & $\begin{array}{l} C_{V_2}^{\tau} = 0.05\pm 0.04\\
			C_{S_2}^{\tau} = -1.38\pm 0.06\\
			C_{V_2}^{\mu} = -0.005\pm 0.014\\
			C_{S_2}^{\mu} = -0.0005 \pm 0.0474 \\
			|V_{cb}| = (40.2 \pm 0.6) \times 10^{-3}
			\end{array}$ \\
			\hline
			
			$ C_{V_1}, C_{S_2} $ & $\begin{array}{l} C_{V_1}^{\tau} = -0.05\pm 0.03\\
			C_{S_2}^{\tau} = -1.37 \pm 0.05
			\\C_{S_2}^{\mu} = -0.002 \pm 0.046\\
			|V_{cb}| = (40.3 \pm 0.5) \times 10^{-3}
			\end{array}$  & $C_{V_2}, C_T  $ & $\begin{array}{l} C_{V_2}^{\tau} = 0.11 \pm 0.05\\
			C_T^{\tau} = -0.09\pm 0.02\\
			C_{V_2}^{\mu} = -0.003\pm 0.014\\
			C_T^{\mu} = (0.2\pm 5.0) \times 10^{-4}\\
			|V_{cb}| = (40.3 \pm 0.6) \times 10^{-3}
			\end{array}$  \\
			\hline
			$ C_{V_1}, C_T  $ &
			$\begin{array}{l} C_{V_1}^{\tau} = 0.08 \pm 0.03\\
			C_T^{\tau} = 0.01 \pm 0.03 \\ C_T^{\mu} = (0.2\pm 5.0) \times 10^{-4}\\
			|V_{cb}| = (40.3 \pm 0.5) \times 10^{-3}
			\end{array}$  & $  C_{S_1}, C_T  $ & $\begin{array}{l} C_{S_1}^{\tau} = 0.12 \pm 0.05\\
			C_T^{\tau} = -0.03 \pm 0.01\\ C_{S_1}^{\mu} = -0.003 \pm 0.047\\
			C_T^{\mu} = (0.2\pm 5.0) \times 10^{-4}\\
			|V_{cb}| = (40.3 \pm 0.5) \times 10^{-3}
			\end{array}$   \\
			\hline
			
			$ C_{S_2}, C_{S_1} $  & $\begin{array}{l} C_{S_2}^{\tau} = -1.15 \pm 0.13\\
			C_{S_1}^{\tau} = -0.22 \pm 0.14\\
			C_{S_2}^{\mu} = 0.02 \pm 0.18\\
			C_{S_1}^{\mu} = -0.02\pm 0.18\\
			|V_{cb}| = (40.3 \pm 0.5) \times 10^{-3}
			\end{array}$  & $ C_{S_2}, C_T  $  & $\begin{array}{l} C_{S_2}^{\tau} =-1.36\pm 0.05\\
			C_T^{\tau} = 0.04 \pm 0.02\\
			C_{S_2}^{\mu} = -0.002 \pm 0.046\\
			
			C_T^{\mu} = (0.2\pm 5.0) \times 10^{-4}\\
			|V_{cb}| = (40.3 \pm 0.5) \times 10^{-3}
			\end{array}$   \\
			\hline
		\end{tabular}
		\caption{The simultaneous fit of $|V_{cb}|$, and the new physics WCs. The inputs are ``Lattice + LCSR($q^2$ = 0) + data on $\bar{B}\to D (\mu,e)\bar{\nu}$ and $\bar{B}\to D^* (\mu,e)\bar{\nu}$ decays + $R(D^{(*)}) + F_L^{D^*}$". These fit assumes different NP in $\tau$ and $\mu/e$ final states, and the contributions from the two operators at a time.}
		\label{tab:NPfit2opr}
	\end{center}
\end{table}

\subsection{Analysis considering NP effects in $b\to c\ell^-\bar{\nu}$}

Following the observations made from the results of table \ref{tab:rdrdstsm}, we decide to include new physics contributions in the rates of $\BtoDDstellnu$ decays for light leptons and repeat the fit with the experimental data mentioned in table \ref{tab:rdrdstsm}.  Again, to understand the impact of the different inputs, we divide our analyses into a couple of scenarios. In particular, the analyses are divided based on whether or not $\rdrdst$ are included in the fit. Following are the different fit scenarios: 

\begin{enumerate}
	\item\label{enu1} We have allowed NP effects in $b\to c\mu^-(e^-)\bar{\nu}$ decays and extracted the new physics WCs (one- operator at a time) and $|V_{cb}|$ simultaneously. As inputs, we have used the lattice and the experimental data on the $q^2$ and angular distributions in $\bar{B}\to D(D^*)\mu^-(e^-)\bar{\nu}$ decays. The results are presented in table \ref{tab:fitNPnoLFU} respectively. The results in tables \ref{tab:fitNPnoLFU} and \ref{tab:fitNPwLFU} are obtained following two different methods. We have extracted the BGL coefficients alongside the new physics WCs and $|V_{cb}|$ in one method, here, the $\chi^2_{Total}$ function is defined as 
		\begin{equation}
		\chi^2_{Total} = \chi^2_{rates} + \chi^2_{lattice} + \chi^2_{LCSR (q^2=0)},
		\end{equation}
		where the $\chi^2_{rates}$ is the $\chi^2$ function defined for the $q^2$ and angular distributions of the decay rates with NP and the corresponding experimental data. Here, the rates are expanded as functions of the BGL coefficients. The $\chi^2_{lattice}$ and $\chi^2_{LCSR (q^2=0)}$ are defined for the form factors with the relevant inputs from lattice and LCSR($q^2=0$). Here, also the form factors are expanded in terms of the BGL coefficients. In the other method, the BGL coefficients are obtained from a fit only to the lattice and LCSR ($q^2$ = 0) and are kept as nuisance parameters in the fits to the experimental data, through 		
	\begin{equation}
	\chi^2_{Total} = \chi^2_{rates} + \chi^2_{nuisance},
	\end{equation}
	where $\chi^2_{nuisance} = (F-P)^T V^{-1} (F-P)$ where $F$ is the vector of the BGL coefficients, and the $P$ is the vector of the respective values of the BGL coefficients, and $V$ is the respective covariance matrix.

\item\label{enu2} In addition, we have done a couple of other fits, including the inputs from $\rdrdst$ and $F_L^{D^*}$, alongside the other data mentioned in the above item. Accordingly, we have updated the $\chi^2_{Total}$ defined above. For these analyses, we assume similar or different NP contributions in $\BtoDDstellnu$ and $\BtoDDsttaunu$ decays. The results are shown in table \ref{tab:fitNPwLFU} for similar types of NP. We have not included $F_L^{D^*}$ in this fit. We have checked that including this input will not change the results. In the other scenario, we have assumed different types of NP contributions in the decays to light and heavy leptons, respectively. In this case, we have presented our results, including $F_L^{D^*}$ and $\rdrdst$ in the fits. The corresponding results are shown in table \ref{tab:fitNPnonMFV}.  As seen from table \ref{tab:fitNPnonMFV}, we have done the analyses using both the HFLAV 2022 and 2023 averages on $\rdrdst$ and compared the results. As mentioned in the introduction, the 2023 average on $R(D^*)$ includes the most recent measurement of LHCb, which is yet to be published.  
\end{enumerate}

\subsubsection{Results of similar type of NP in $b\to c \tau^-\bar{\nu}$ and $b\to c (\mu^-,e^-)\bar{\nu}$ decays}
A few observations from the analyses discussed above and the results presented in tables \ref{tab:fitNPnoLFU} and \ref{tab:fitNPwLFU}: 
\begin{itemize}
	\item  The results shown in table \ref{tab:fitNPnoLFU} correspond to the fits described in item- \ref{enu1} above. We have obtained zero consistent values of the new physics WCs. None of the NP scenarios gives a non-zero NP contribution which is as per the expectations since the data on the decay rates in these modes are SM consistent. The allowed value of $C_T$ is very small, however, the allowed values of $C_{V_2}$, $C_{S_1}$ and $C_{S_2}$ have large errors and the allowed values could be $\approx 0.05$.
	
	\item In all four one-operator scenarios, using the fit results, we have estimated the values of $R(D)$ and $R(D^*)$. As we see from the table \ref{tab:fitNPnoLFU}, the estimated values are consistent with those shown in table \ref{tab:rdrdstsm}. However, in the scenarios with  $C_{S_1}$ and $C_{S_2}$, the predicted value of $R(D)$ has significant errors. The allowed value is $\lsim 0.322$, close to the lower limit of the measured value at 1-$\sigma$ confidence interval (CI). Hence, we can conclude that the new physics effects allowed by the data on $\bar{B}\to D(D^*)\mu^-(e^-)\bar{\nu}$ can not change the values of $R(D^{(*)})$ from that of their values obtained without any NP effects. 
	
	\item The results of table \ref{tab:fitNPwLFU} show that the inclusion of $\rdrdst$ does not change the fit values of the new physics WCs and $|V_{cb}|$. They are consistent with those obtained in table \ref{tab:fitNPnoLFU} respectively. Also, the extracted values of $|V_{cb}|$ are consistent with the one obtained from the fit without NP. Following the observation made in the above item, it is not a surprising result since the WCs will be tightly constrained from the data on $\bar{B}\to D(D^*)\mu^-(e^-)\bar{\nu}$ with a negligible impact from $\rdrdst$. Also, in this case, we have checked that the inclusion of $F_L^{D^*}$ marginally changes the fit results, and the allowed ranges of the fit parameters are consistent. Another important point is that in both types of fits mentioned in item-\ref{enu1} the results are very much consistent with each other, which we can observe from the results of tables \ref{tab:fitNPnoLFU} and \ref{tab:fitNPwLFU}, respectively. Hence, for the rest of the fits with the inputs on $\rdrdst$, for eg. tables \ref{tab:fitNPnonMFV}, \ref{tab:NPfit2opr} and \ref{tab:SMEFTfit}, we have considered the BGL coefficients as nuisance parameters which will correctly reproduce the respective SM values without any experimental biases. 
	
	\item In all the fits, we have extracted $|V_{cb}|$ alongside the new physics WCs. Therefore, it will be difficult to extract simultaneously both $|V_{cb}|$ and $C_{V_1}$, which is clear from the expressions of the decay rate distributions.  
 
	\item The results of table \ref{tab:fitNPwLFU} shows that in the case of identical NP WCs for light and heavy lepton final states, only non-zero solutions are allowed for the four-fermi operators with $(S \pm P)$ quark current. The fitted values of $C_{S_1}$ and $C_{S_2}$ have relatively small errors, and the predicted values of $R(D)$ can accommodate the respective measured value. However, we do not see the shift in the respective values of $R(D^*)$ from those presented in table \ref{tab:fitNPnoLFU}. .   
	
	\item  In the case of $C_{V_2}$ and $C_T$, the allowed solutions are zero consistent, though the data allows large values of $C_{V_2}$ and the predicted values of $\rdrdst$ are consistent with the SM, and no deviations are observed. 
	
	\item In the scenarios with scalar and pseudoscalar operators, we have also predicted the angular and the $q^2$-distributions of the rates of $\BtoDDstellnu$ ($\ell$ = light leptons) decays in small bins, which are presented in figure \ref{fig:gamma} in the appendix. The predicted distributions have been compared with the measured values and with the respective predictions obtained from the results of the fits without NP contributions. We do not observe any deviations with respect to SM predictions for the light leptons, which is not surprising since the scalar and pseudoscalar contributions are proportional to the mass of the respective leptons.
\end{itemize}

\begin{table}[t]
	\begin{center}
		\scriptsize
		\renewcommand*{\arraystretch}{2.0}
		\begin{tabular}{|c|c|c|c|}
			\hline
			Scenarios & \multicolumn{3}{c|}{Estimates from the fit results of Tab. \ref{tab:NPfit2opr} }  \\
			\cline{2-4}
			& $R(D)$ & $R(D^*)$ &   $F_L^{D^*}$ \\ 
			\hline
			$C_{V_1}, C_{V_2}$  & $0.355 \pm 0.029$ & $0.284 \pm 0.013$ & $0.452 \pm 0.004$ \\
			\hline 
			$ C_{V_2}, C_{S_1} $  &  $ 0.357 \pm 0.029$ & $ 0.284 \pm 0.013$ & $ 0.468 \pm 0.004$ \\
			\hline
			$ C_{V_1}, C_{S_1}  $ &  $ 0.357 \pm 0.029$ & $ 0.283 \pm 0.013 $ & $ 0.455 \pm 0.005$ \\
			\hline
			$\bf C_{V_2},C_{S_2}  $  & $\bf 0.357 \pm 0.029$ &  $\bf 0.284 \pm 0.013$ & $\bf 0.557 \pm 0.007$\\
			\hline	
			$\bf C_{V_1}, C_{S_2}  $ &  $\bf 0.357 \pm 0.029$ &$\bf 0.284 \pm 0.013$ & $\bf 0.561 \pm 0.007$  \\
			\hline 
			$C_{V_2}, C_T $ &  $0.348 \pm 0.029$ & $ 0.284 \pm 0.013 $& $ 0.416 \pm 0.015$ \\
			\hline
			$ C_{V_1}, C_T  $ &  $ 0.357 \pm 0.028 $& $0.284 \pm 0.013 $ & $ 0.455 \pm 0.003$ \\
			\hline 
			$ C_{S_1},C_T  $ & $ 0.359 \pm 0.029 $ & $0.283 \pm 0.013 $ & $ 0.453 \pm 0.007$  \\
			\hline
			$\bf C_{S_2}, C_{S_1} $  & $\bf 0.354 \pm 0.029$& $\bf 0.287 \pm 0.013 $& $\bf 0.521 \pm 0.021$\\
			\hline
			$\bf C_{S_2},C_T  $  &  $\bf 0.357 \pm 0.029 $& $\bf 0.284 \pm 0.013 $& $\bf 0.564 \pm 0.008$  \\
			\hline
		\end{tabular}
		\caption{The estimated values of $R(D^{(*)})$ and $F_L^{D^*}$ obtained using the respective fit results of table~ \ref{tab:NPfit2opr}.}
		\label{tab:predmain2opr}
	\end{center}
\end{table}

\begin{table}[t]
	\begin{center}
		\small
		\renewcommand*{\arraystretch}{1.6}
		\begin{tabular}{|*{6}{c|}}
			\hline
			Obs & SM & $[C_{V_1},C_{S_2}] $ & $[C_{V_2},C_{S_2}]$ & $[C_T,C_{S_2}]$  & $[C_{S_1},C_{S_2}]$  \\
			\hline			
			$A_{\lambda_\tau}^{\bar{B} \to D^* \tau^- \bar{\nu}}$ & 0.519(7)   &  $\text{0.21(2)}$  &  $\text{\ 0.21(2)}$  & $\text{0.21(2)}$ &  $\text{0.32(6)}$   \\
			\hline
			$A_{FB}^{\bar{B} \to D^* \tau^- \bar{\nu}}$ & -0.077(9)  & $\text{ 0.059(6)}$  &  $\text{ 0.044(7)} $  &  $\text{ 0.01(3)}$  &  $\text{ 0.03(2)}$   \\
			\hline
			$A_{\lambda_\tau}^{\bar{B} \to D \tau^- \bar{\nu}}$  & -0.324(3) & $\text{-0.48(6)}$  &  $\text{-0.37(5)}$  &  $\text{ -0.40(4)}$  &  $\text{ -0.43(5)}$      \\
			\hline
			$A_{FB}^{\bar{B} \to D \tau^- \bar{\nu}}$ & 0.3596(4)  & $\text{-0.268(5)}$  &  $\text{-0.2725(2)}$  &  $\text{ -0.285(5)}$  &  $\
			\text{ -0.271(2)}$     \\
			\hline
		\end{tabular}		
		\caption{Predictions of $q^2$ integrated $A_{FB}$ and $A_{\lambda_{\tau}}$  in $B\to D^{(*)}\tau^-\bar{\nu}$ decays in a few two operator scenarios allowed by the data. The predictions are obtained using the fit results of table~ \ref{tab:NPfit2opr}.  }
		\label{tab:pred2oprobsfullq2}
	\end{center}
\end{table} 

\subsubsection{Results of analyses with different type of NP in $b\to c \tau^-\bar{\nu}$ and $b\to c (\mu^-,e^-)\bar{\nu}$ decays}

In this subsection, we will discuss the results of the analyses mentioned in enumerate-\ref{enu2}. Table \ref{tab:fitNPnonMFV} presents the fit results considering different new physics WCs in the heavy and light lepton channels, respectively. In this part of the analysis, we have included the measured value of $F_L^{D^*}$ alongside the abovementioned inputs. Here, $C_i^{\tau}$ represents the WC in the decays with a $\tau$ in the final state, while $C_i^{\mu}$ represents that in the decays with a muon in the final states. The new data on $R(D^*)$ from winter 2023 does not change the fit results of the WCs from what we have obtained using the old results. Note that in all the cases, non-zero values of the WCs are allowed in $\BtoDDsttaunu$ decays. In particular, we have obtained a relatively large negative value for $C_{S_2}^{\tau}$, which is a requirement to explain the data on $F_{L}^{D^*}$. However, the values obtained for $C_{S_2}^{\tau}$ will lead to a large branching fraction $\mathcal{B}(B_c\to \tau^-\bar{\nu}) \approx 80$\%. Though, at the moment, we do not have a measurement of this branching fraction\footnote{Regarding the allowed ranges of $\mathcal{B}(B_c\to \tau \nu_{\tau})$ the reader could see the refs.~\cite{Alonso:2016oyd,Blanke:2018yud}.}. In all the other scenarios, the predictions for $\mathcal{B}(B_c\to \tau^-\bar{\nu})$ will be less than 10\%. As expected, for the decays with the light leptons, the allowed values of the WCs are zero consistent. Another important point to note is that the extracted values of $|V_{cb}|$ in all the NP scenarios are consistent with each other, and they are also consistent with the one extracted from the fits without taking into account the contributions from NP.

Using the results of table \ref{tab:fitNPnonMFV}, we have estimated the best-fit values of observables discussed earlier along with their respective errors at 1-$\sigma$ CI, which we have presented in table \ref{tab:predobsfullq2}. The estimated values with a pull $\gsim$ 1.5 from the respective SM predictions have been indicated in bold font, and estimated values largely deviated from their SM predictions are marked with a star. Note that we have pointed out only those discrepancies which are explicitly due to the contributions from the NP. There are also estimates which have deviations from the respective SM predictions. However, those are not due to NP but due to a discrepancy between the form factors extracted with and without the inputs from the experimental data which we have mentioned earlier.

From the results of table \ref{tab:predobsfullq2}, we have the following essential observations:

\begin{itemize}	
	
	\item It is to be noted that only the one operator scenario $\mathcal{O}_{V_1}$ can explain both the measured values of $R(D)$ and $R(D^*)$ simultaneously, but not $F_L^{D^*}$. The rest of the one-operator scenarios can only partially explain one observation. For example the one operator scenarios $\mathcal{O}_{V_2}$ and $\mathcal{O}_T$ could explain the observation on $R(D^*)$ but not the $R(D)$. None of these scenarios can accommodate the measured value of $F_L^{D^*}$. Similarly, the scenario with $\mathcal{O}_{S_1}$ or $\mathcal{O}_{S_2}$ can accommodate the observation in $R(D)$ but not the $R(D^*)$. Also, amongst all the one operator scenarios, only $\mathcal{O}_{S_2}$ can accommodate the measured value of $F_L^{D^*}$. In addition, note that the required value of $C_{S_2}^{\tau}$ has a magnitude of order one and negative. 
	
	\item From table \ref{tab:predobsfullq2}, we see that in the NP scenarios with scalar or tensor operators, there are deviations in the $\tau$-polarization and forward-backward asymmetries in $\bar{B}\to D^{(*)} \tau^-\bar{\nu}$ decays from their respective SM predictions. Like in $\bar{B}\to D^*\tau^-\bar{\nu}$ decays, for the scenario $\mathcal{O}_{S_1}$ or $\mathcal{O}_{T}$ the discrepancies are more than 3.5 $\sigma$. It is much higher in the case of $\mathcal{O}_{S_2}$. However, a measurement on $\mathcal{B}(B_c\to \tau^-\bar{\nu})$ might put restriction on such a large enhancement in this scenario. Also, in the scenarios with $\mathcal{O}_{V_2}$, we note deviations in $A_{FB}^{B\to D^* \tau\nu}$ by more than 3$\sigma$. For the $A_{\lambda_\tau}^{B\to D^*\tau\nu}$, we note a mismatch between the SM predictions and the estimates from the fit with $C_{V_1}$ and $C_{V_2}$. However, these estimated values agree with those obtained from the fit without NP (table \ref{tab:rdrdstsm}). Therefore, such disagreements are not due to NP but due to the reason we have discussed earlier. Similar argument holds for $A_{FB}^{B\to D^*\tau\nu}$ in the scenario with $\mathcal{O}_{V_1}$.    
	
	Similarly, for $\bar{B}\to D \tau^-\bar{\nu} $ decays, we observe deviations of more than 3$\sigma$ in $A_{FB}$ and $A_{\lambda_\tau}$ in the scenarios with $\mathcal{O}_{S_1}$ or $\mathcal{O}_T$, respectively. However, the scenarios with a vector operator are consistent with the respective SM predictions. Note that in the scenario with $\mathcal{O}_{S_2}$, the deviation is huge in the case of $A_{FB}$. However, it is slightly greater than 1$\sigma$ in $A_{\lambda_\tau}$, though, the shift in its best-fit value from respective SM prediction is roughly about 18\%. Therefore, the deviation is small due to a large error in the estimate of $A_{\lambda_\tau}$.  
	
	\item In all the NP scenarios, the estimated values of the angular observables in $B\to D^*\mu^-\bar{\nu}$ decays are the same. Also, these estimates are the same as those obtained in table \ref{tab:rdrdstsm} from the fit to the ``experimental data + lattice". Like before, we note discrepancies in $A_{FB}^{D^*}$ and $F_L^{D^*}$ which are around 2$\sigma$ in all the NP scenarios. These discrepancies are not due to a NP contributions since in the allowed regions of $C_i^{\mu}$ the contributions from NP are negligibly small; we have explicitly checked this. As we have mentioned, this is due to the slightly different shapes of the form factors obtained in the fit with the experimental data than the lattice.
		
\end{itemize}

As we have noted, none of the one operator scenarios can accommodate all the three measurements in $B\to D^{(*)}$ decays which was the case with the old data as well \cite{Bhattacharya:2018kig,Blanke:2018yud,Becirevic:2019tpx,Bardhan:2019ljo,Fedele:2022iib}.
Note that the test of the NP sensitivities of $A_{FB}$ and $A_{\lambda_{\tau}}$ in $B\to D^{(*)} \tau^-\bar{\nu}$ had been pointed out earlier. In ref.~\cite{Bhattacharya:2018kig}, the correlations of these observables with $R(D^{(*)})$ in the different NP scenarios have been worked out. The sensitivity of $A_{FB}^{B\to D\tau\nu}$ and $A_{\lambda_{\tau}}^{B\to D}$ towards the scalar or tensor current operators had been shown in ref.~\cite{Becirevic:2019tpx,Datta:2023mmb}. Also, it has been shown that $A_{FB}^{B\to D^*\tau\nu}$ can deviate from its SM prediction for the allowed solutions of $C_{V_2}$ obtained from a fit to the data on $R(D^{(*)})$ available at that time. These analyses were based on the old data set, and the lattice data on $B\to D^*$ form factors in the full $q^2$ regions were unavailable. Hence, the SM predictions had biases from the experimental data on the rates of $B\to D^*\mu^-(e^-)\bar{\nu}$ decays or the predictions were based on the assumption of heavy quark symmetries. The SM predictions of the different angular observables for all the modes with $\tau$ or $\mu$ or $e$ are the completely new results of this analysis. Hence, we are able to test the NP sensitivities of these observables in a more meaningful or statistically consistent way, comparing the results of various available inputs. The predictions obtained on the observables in different NP scenarios are part of the new results of this analysis.

We have also done a fit where we have considered $|V_{cb}| = (41.1 \pm 0.7) \times 10^{-3}$ as input which has been taken from the CKMfitter \cite{ckmfitter1}. This $|V_{cb}|$ value has been obtained from fitting the Wolfenstein parameters from measurements of the modes other than $\btoclnu$ semileptonic decays. This fit, in particular, is required to extract the new physics contribution $C_{V_1}$, which is otherwise challenging to fit. The constraints obtained on $C_{V_1}^{\tau}$ and  $C_{V_1}^{\mu}$ are given by 
\begin{equation}
C_{V_1}^{\tau} = 0.10 \pm 0.02, \ \ \ \ C_{V_1}^{\mu} = 0.03 \pm 0.01.
\end{equation}
Note that in this scenario, we need non-zero solutions for both the WCs allowed by the available data. 

\subsubsection{Two operator scenarios}

Following the discussion above, an obvious extension is to look for the two operator scenarios which accommodate all the three measured values in $B\to D^*\tau^-\bar{\nu}$ decays. Also, in these analyses, we have considered different NP in decay modes with $\tau$, $\mu$ or $e$ in the final states, respectively. The results of the fits with two operator scenarios are shown in table \ref{tab:NPfit2opr}. Using these fit results, we have estimated the values of $R(D)$, $R(D^*)$ and $F_L^{D^*}$ in table \ref{tab:predmain2opr}. Note that all the two operator scenarios can explain the measured values of $R(D)$ and $R(D^*)$. However, only the scenarios with $\mathcal{O}_{S_2}$ as one of the operators could explain $\rdrdst$ and $F_L^{D^*}$\footnote{A slight enhancement in the value of $F_L^{D^*}$ has been pointed out in ref. \cite{Blanke:2018yud,Becirevic:2019tpx} in the two operator scenarios $[\mathcal{O}_{S_1},\mathcal{O}_{S_2}]$.}. Note that apart from the scenario $[\mathcal{O}_{S_1},\mathcal{O}_{S_2}]$, in the rest of the two operator scenarios with $\mathcal{O}_{S_2}$, we have obtained a branching fraction $\mathcal{B}(B_c\to \tau \nu_{\tau} ) \approx$ 80\% which is a relatively high value, also observed in ref.\cite{Jaiswal:2020wer}. In the scenario $[\mathcal{O}_{S_1},\mathcal{O}_{S_2}]$, the $\mathcal{B}(B_c\to \tau \nu_{\tau}) = 0.50 \pm 0.23$  and the scenarios without $\mathcal{O}_{S_2}$ have $\mathcal{B}(B_c\to \tau \nu_{\tau} ) \lsim 10$\%. Hence, if a future measurement does not allow a very large value of this branching fraction, then the only two operator scenario $[\mathcal{O}_{S_1},\mathcal{O}_{S_2}]$ will be allowed by all the data. For a better understanding, we have to wait for more precise data on $D^*$ polarization as well as for the measurement of $\mathcal{B}(B_c\to \tau^-\bar{\nu})$. 

In all the two operator scenarios, we have predicted the $q^2$ integreted values of all the angular observables listed in table \ref{tab:rdrdstsm} which are shown in tables \ref{tab:predB2DOP2a}, \ref{tab:predB2DOP2b}, \ref{tab:predB2Dst2OPa} and \ref{tab:predB2DstOP2b}, respectively. However, for the four two operator scenarios which can accommodate the measured values of $R(D^{(*)})$ and $F_L^{D^*}$, we have shown the predicted values of $A_{FB}$ and $A_{\lambda_{\tau}}$ in table \ref{tab:pred2oprobsfullq2}. We have also compared the estimated values with the respective SM predictions. Note that in all these four scenarios we note deviations in the estimated values of $A_{FB}^{B\to D^{(*)}\tau\nu}$ and $A_{\lambda_{\tau}}^{B\to D^{(*)}\tau\nu}$. However, as mentioned above the restrictions might come once the measurement of $\mathcal{B}(B_c\to \tau^-\bar{\nu})$ will be available. In a couple of other two operator scenarios we have seen deviations which can be seen from the tables in the appendix.           

 Note that the fitted values of the corresponding WCs do not allow significant deviations of these observables in $\bar{B}\to D^{(*)}\mu^-\bar{\nu}$ decays. In the appendix, we have presented the expressions for $\rdrdst$, $F_L^{D^*}$ and the other related observables in the presence of the NP which could be useful for different phenomenological analysis. In those expressions, we have considered $C_i^{\tau} \ne 0$, however, $C_i^{\mu, e} = 0$.

\begin{table}[t]
	\begin{center}
		\small
		\renewcommand*{\arraystretch}{1.6}
		\begin{tabular}{|c|c|c|}
			\hline
			Scenario  & Simultaneous fit with & Fit results with\\
			& $|V_{ub}|$ 	& $|V_{ub}| = (3.64 \pm 0.07) \times 10^{-3}$ \cite{ckmfitter1} \\
			\hline
			$C_{V_1}$ & N.A. & $C_{V_1}^{\mu}$ = $-0.01\pm 0.03$  \\
			\hline
			$C_{V_2}$ & N.A.& $C_{V_2}^{\mu}$ = $-0.01\pm 0.03$   \\
			\hline
			$C_T$  & $\begin{array}{l} 
			C_T^{\mu} = 0.13\pm 0.13 \\
			|V_{ub}| = (3.42 \pm 0.30) \times 10^{-3} 
			\end{array}$ & $C_T^{\mu}$ = $-0.06\pm 0.13$   \\
			\hline			
			$C_{S_1}$ & $\begin{array}{l} 
			C_{S_1}^{\mu} = -0.02 \pm 0.23  \\
			|V_{ub}| = (3.60 \pm 0.11) \times 10^{-3} 
			\end{array}$ & $C_{S_1}^{\mu}$ = $-0.02 \pm 0.15$   \\
			\hline
			
			$C_{S_2}$ & $\begin{array}{l} 
			C_{S_2}^{\mu} = -0.02\pm 0.23\\
			|V_{cb}| = (3.60 \pm 0.11) \times 10^{-3} 
			\end{array}$ & $C_{S_2}^{\mu}$ = $-0.02\pm 0.15$  \\
			\hline
		\end{tabular}
		\caption{The extractions of new physics WCs from a fit to the data on rates in $\btopilnu$ decays and the relevant inputs on the respective form factors. In one fit, $|V_{ub}|$ has been extracted simultaneously with the WCs. In another fit, the value $|V_{ub}| = (3.64 \pm 0.07) \times 10^{-3}$ from CKMfitter \cite{ckmfitter1} has been used as input.}
		\label{tab:fitNPBtopi}
	\end{center}
\end{table}

\begin{table} [t]
	\begin{center}
		\small
		\renewcommand*{\arraystretch}{1.6}
		\begin{tabular}{|c|c|c|c|}
			\hline
			Scenario & 	\multicolumn{1}{|c|}{Simultaneous fit with the CKM elements} & \multicolumn{1}{c|}{Fits with $\begin{array}{l} |V_{ub}| = (3.64 \pm 0.07) \times 10^{-3} \\
				|V_{cb}| = (41.1 \pm 0.7) \times 10^{-3}  \end{array}$} \\
			\hline
			&  Fit Parameters ($\frac{\tilde{c}}{\Lambda^2}$) & Fit Parameters ($\frac{\tilde{c}}{\Lambda^2}$) \\
			\hline
			$C_{V_1}$ & $\begin{array}{l} |V_{ub}| = (3.60 \pm 0.11) \times 10^{-3} \\
			|V_{cb}| = (40.3 \pm 0.5) \times 10^{-3} \\ \tilde{C}^{(3) b \to c}_{\tau q}/\Lambda^2 = (-0.48 \pm 0.12) \times 10^{-7} 
			\end{array}$ & $\begin{array}{l} \tilde{C}^{(3)b \to c}_{\tau q}/\Lambda^2 = (-0.35 \pm 0.15) \times 10^{-7} \\
			\tilde{C}^{(3)b \to c}_{\mu q}/\Lambda^2 = (0.14 \pm 0.09) \times 10^{-7}\\
			\tilde{C}^{(3)b \to u}_{\mu q}/\Lambda^2 = (0.06 \pm 0.20) \times 10^{-8}	
			\end{array}$  \\
			\hline
			
			$C_{V_2}$ & $\begin{array}{l} 
			|V_{ub}| = (3.60 \pm 0.11) \times 10^{-3} \\
			|V_{cb}| = (40.3 \pm 0.6) \times 10^{-3} \\ 
			\tilde{C}^{b \to c}_{\phi u d}/\Lambda^2 = (-0.06 \pm 0.18) \times 10^{-7} 
			\end{array}$  & $\begin{array}{l} \tilde{C}^{b \to c}_{\phi u d}/\Lambda^2 = (0.05 \pm 0.17) \times 10^{-7} \\
			\tilde{C}^{b \to u}_{\phi u d}/\Lambda^2 = (-0.01 \pm 0.04) \times 10^{-7} \end{array}$ \\
			\hline
			
			$C_{S_1}$ & $\begin{array}{l} 
			|V_{ub}| = (3.60 \pm 0.11) \times 10^{-3} \\
			|V_{cb}| = (40.3 \pm 0.5) \times 10^{-3} \\ 
			\tilde{C}^{* b \to c}_{\tau edq}/\Lambda^2 = (-2.02 \pm 0.56) \times 10^{-7} \\
			\tilde{C}^{* b \to c}_{\mu edq}/\Lambda^2 = (0.05 \pm 0.64) \times 10^{-7} \\
			\tilde{C}^{* b \to u}_{\mu edq}/\Lambda^2 = (0.03 \pm 0.29) \times 10^{-7} 
			
			\end{array}$ & $\begin{array}{l}  
			\tilde{C}^{* b \to c}_{\tau edq}/\Lambda^2 = (-2.03 \pm 0.57) \times 10^{-7} \\
			\tilde{C}^{* b \to c}_{\mu edq}/\Lambda^2 = (0.08 \pm 0.60) \times 10^{-7} \\
			\tilde{C}^{* b \to u}_{\mu edq}/\Lambda^2 = (0.03 \pm 0.18) \times 10^{-7} 
			\end{array}$\\
			\hline
			
			$C_{S_2}$ & $\begin{array}{l} 
			|V_{ub}| = (3.61 \pm 0.11) \times 10^{-3} \\
			|V_{cb}| = (40.3 \pm 0.5) \times 10^{-3}\\
			\tilde{C}^{*(1) b \to c}_{\tau equ}/\Lambda^2 = (1.77\pm 0.06) \times 10^{-6}\\
			\tilde{C}^{*(1) b \to c}_{\mu equ}/\Lambda^2 = (0.06\pm 0.61) \times 10^{-7} \\
			\tilde{C}^{*(1) b \to u}_{\mu equ}/\Lambda^2 = (0.03\pm 0.27) \times 10^{-7}
			\end{array}$  & $\begin{array}{l}
			\tilde{C}^{*(1) b \to c}_{\tau equ}/\Lambda^2 = (1.81\pm 0.06) \times 10^{-6} \\
			\tilde{C}^{*(1) b \to c}_{\mu equ}/\Lambda^2 = (0.09\pm 0.58) \times 10^{-7} \\
			\tilde{C}^{*(1) b \to u}_{\mu equ}/\Lambda^2 = (0.03\pm 0.18) \times 10^{-7}
			\end{array}$\\
			\hline
			
			$C_T$ & $\begin{array}{l} 
			|V_{ub}| = (3.44 \pm 0.26) \times 10^{-3} \\		
			|V_{cb}| = (40.3 \pm 0.5) \times 10^{-3}\\
			\tilde{C}^{*(3) b \to c}_{\tau equ}/\Lambda^2 = (0.61\pm 0.17) \times 10^{-7}\\
			\tilde{C}^{*(3) b \to c}_{\mu equ}/\Lambda^2 = (- 0.28\pm 6.57) \times 10^{-10} \\
			\tilde{C}^{*(3) b \to u}_{\mu equ}/\Lambda^2 = (- 0.14 \pm 0.12) \times 10^{-7}
			\end{array}$  & $\begin{array}{l} 
			\tilde{C}^{*(3) b \to c}_{\tau equ}/\Lambda^2 = (0.62\pm 0.17) \times 10^{-7}\\
			\tilde{C}^{*(3) b \to c}_{\mu equ}/\Lambda^2 = (-0.33\pm 6.61) \times 10^{-10} \\
			\tilde{C}^{*(3) b \to u}_{\mu equ}/\Lambda^2 = (0.07 \pm 0.16) \times 10^{-7}
			\end{array}$\\
			\hline
		\end{tabular}
		\caption{Values of SMEFT coefficients in various scenarios. The fixed values of $|V_{ub}|$ and $|V_{cb}|$ are taken from \cite{ckmfitter1}.}
		\label{tab:SMEFTfit}
	\end{center}
\end{table}

The predictions in small $q^2$ bins are also obtained which are presented in tables \ref{tab:predB2D1OP}, \ref{tab:predB2Dst1OP} for the one operator scenarios. The respective predictions in the two operator scenarios are given in tables \ref{tab:predB2DOP2a}, \ref{tab:predB2DOP2b} for $B\to D$, and in tables \ref{tab:predB2Dst2OPa} and \ref{tab:predB2DstOP2b} for $B\to D^*$ decays, respectively. In all the tables we have presented the corresponding SM predictions which are based on the form factors obtained from the lattice inputs. From the table one can read the $q^2$-bin wise NP sensitivities of different observables in $\BtoDDsttaunu$ decays.       

\subsubsection{Fit to SMEFT Wilson coefficients}
Before we explain the fit of the SMEFT WCs, and the scale $\Lambda$, we would like to focus on $\btopilnu$ decays. Like $\BtoDDstellnu$ decays, we have extracted $|V_{ub}|$ alongside the new physics WCs from the available inputs on $\btopilnu$ decays. Also, we have simultaneously extracted the BSZ coefficients, which we have presented in a separate file. The fit results are shown in table \ref{tab:fitNPBtopi}. Here, we have followed two different approaches. We have extracted $|V_{ub}|$ alongside the new physics WCs and BSZ coefficients in one method. In the other method, we have considered $|V_{ub}| = (3.64 \pm 0.07)\times 10^{-3}$ \cite{ckmfitter1} as input which has been obtained from the Wolfenstein parameters obtained from a fitting to the available inputs other than the measured value of $|V_{ub}|$. In the simultaneous fit, we can not extract $C_{V_1}^{\mu}$ and $C_{V_2}^{\mu}$. However, the rest of the WCs can be extracted. Note that in such fits, the extracted values of $C_{S_1}^{\mu}$ and $C_{S_2}^{\mu}$ are consistent with zero. While the fitted value of $C_T^{\mu}$ is marginally consistent with zero, in this case, the extracted value of $|V_{ub}|$ is lower than the one obtained in the fit without any NP. In all these three cases, large values of WCs are allowed. In the other method, we can extract $C_{V_1}^{\mu}$ and $C_{V_2}^{\mu}$ on top of the other one operator scenarios. Also, in this method with a fixed $|V_{ub}|$, in all the one-operator scenarios, the extracted values of the new physics WCs are consistent with zero and could be large. The extracted values of $C_{V_1}^{\mu}$ and $C_{V_2}^{\mu}$ could be smaller than those in $C_{T,S_1,S_2}^{\mu}$. The SM predictions of the associated observables can be seen from the ref.~\cite{Biswas:2021cyd}. In this reference, the fit did not include any new physics contribution in $b\to u\ell^-\bar{\nu}$ decays. Here, we have predicted the relevant observables in small $q^2$ bins, which are presented in table \ref{tab:predB2Pi} in the appendix.    

Finally, we have extracted the ratio $\tilde{C}/\Lambda^2$, defined in the SMEFT, from the data on $\BtoDDstellnu$ and $\btopilnu$ both from theory and experiments. Here, $\tilde{C}$s are the couplings of the different dimension-6 operators defined in eq.~\ref{eq:opssmeft}, and $\Lambda$ is the scale of the new physics. The corresponding WCs at the scale $m_b$ are defined in eqs.~\ref{eq:matchedbtoc} and \ref{eq:matchedbtou}, respectively, for $b\to c$ and $b\to u$ transitions. Here also, we have kept the form factor coefficients obtained from a fit only to the lattice and LCSR ($q^2$ = 0) as nuisance parameters in the fits with the experimental data. We divide this part of the analysis into two broad categories. In one of them, we have fitted the ratio $\tilde{C}/\Lambda^2$ along with $|V_{ub}|$ and $|V_{cb}|$ and in another, for the fixed values of these CKM elements directly. The results are shown in table \ref{tab:SMEFTfit}. As mentioned earlier, we can not simultaneously extract the CKM elements and $\tilde{C}^{3}_{\mu q}$. However, in the fit for fixed values of the CKM elements, we could simultaneously extract $\tilde{C}^{3}_{\tau q}$ and $\tilde{C}^{3}_{\mu q}$, respectively. The general trend shows that the couplings associated with the $b\to u$ decays with light leptons are more tightly constrained than similar couplings in $b\to c $ decays. Also, within their allowed ranges, apart from $\tilde{C}^{*(3) b \to u}_{\mu equ}$ in the rest of the couplings $\tilde{C}^{b\to u} < \tilde{C}^{b\to c}$, the coupling $\tilde{C}^{*(3) b \to u}_{\mu equ} >> \tilde{C}^{*(3) b \to c}_{\mu equ}$. Also, we notice that the couplings associated with the $\tau$ are one order of magnitude larger than the similar couplings with the $\mu$s. However, $\tilde{C}^{*(3) b \to c}_{\mu equ}  \lsim 10^{-3} \times \tilde{C}^{*(3) b \to c}_{\tau equ}$.     

\section{Renormalization-group running of the Wilson Coefficients}

Note that the bounds obtained on the new physics WCs from data will be relevant at the scale $m_b$. 
Having extracted the Wilson coefficients at the low energy scale ($\mu$ $\sim$ $m_b$), it is very interesting to check their behaviour at high energy scales ($\mu$ $\gsim$ 1 TeV) which are accessible at the colliders. For this, it is necessary to solve the renormalization-group (RG) evolution equations of the WCs. The anomalous dimension matrices corresponding to $\mathcal{O}_{V_1}$ and $\mathcal{O}_{V_2}$ vanish, thus the corresponding WCs are scale independent and don't mix with the other operators ($\mathcal{O}_{S_1}$,$\mathcal{O}_{S_2}$ and $\mathcal{O}_T$) \cite{Sakaki:2013bfa,Alonso:2013hga}. The RG evolution for the coefficients $C_i$ = ($C_{S_1}$, $C_{S_2}$ and $C_T$) at the one-loop level from $\mu_{NP}$ to $\mu_{M_Z}$ is given as \cite{Gonzalez-Alonso:2017iyc,Feruglio:2018fxo,Alonso:2013hga,Jenkins:2013wua}

\begin{align} 
\frac{d C_{i} (\mu)}{ d \ln \mu} = \frac{1}{16 \pi^2} \left[ g_{s}(\mu)^2 \gamma_{s}^{T} + \gamma_{w}^T(\mu)   + y_t(\mu)^2 \gamma^T_t \right]_{ij}  C_j (\mu),
\label{eq:RGE}
\end{align}
with
\begin{align}
\gamma_{s}^T &= \left\{ \gamma_{S}, \gamma_{S}, \gamma_{T}\right\}_{\rm diag},\\
\gamma_{w}^T(\mu) & = 
\begin{pmatrix} 
 - \frac{8}{3} g^{\prime 2} (\mu)  & 0 &0 \\
 0 & - \frac{11}{3} g^{\prime 2}(\mu) & 18 g^2 (\mu) + 30 g^{\prime 2}(\mu)\\
 0 & \frac{3}{8} g^2(\mu) + \frac{5}{8} g^{\prime 2}(\mu) & - 3 g^2(\mu) + \frac{2}{9} g^{\prime 2} (\mu)
\end{pmatrix} \label{eq:gamw},\\
\gamma_{t}^T &= \left\{ 0, 1/2, 1/2\right\}_{\rm diag},
\end{align}
where 
$\gamma_S = - 6 C_F = -8$ and $\gamma_T = 2 C_F = 8/3$. Here, $y_t(\mu)$ refers to the energy scale dependence of the top Yukawa coupling. Its RG running is dominated by QCD corrections, and a subdominant contribution comes from the top Yukawa itself \cite{Goncalves:2018pkt}

\begin{equation}
\frac{d y_t}{dt} = \frac{y_t} {16 \pi^2} \biggl{(}\frac{9}{2} y_t^2 -8 g_s^2 - \frac{9}{4} g^2 - \frac{17}{20} g^{\prime 2} \biggl{)}
\end{equation}

with t = $\text{ln}[\mu]$.
 From eq \ref{eq:gamw}, we see that there is operator mixing between $\mathcal{O}_{S_2}$ and $\mathcal{O}_T$ which arises from the electroweak anomalous dimension above the electroweak symmetry breaking scale. On the other hand, the main contributions to RG evolution below the electroweak scale come from QCD. In ref. \cite{Gonzalez-Alonso:2017iyc} (also see \cite{Iguro:2018vqb}), a numerical solution for the RG evolution is provided at the
three-loop in QCD and the one-loop in QED 

\begin{align} \label{eq:WCrelation}
\begin{pmatrix}
C_{S_1} (\mu_b) \\
C_{S_2} (\mu_b) \\
C_T (\mu_b) 
\end{pmatrix}
\simeq 
\begin{pmatrix}
1.46 & 0 &  0\\
0 & 1.46 & -0.0177 \\
 0 & -0.0003 & 0.878
\end{pmatrix}
\begin{pmatrix}
C_{S_1} (m_Z) \\
C_{S_2} (m_Z) \\
C_T (m_Z)
\end{pmatrix} .
\end{align}

\begin{figure}
	\small
	\centering
	\subfloat[]{\includegraphics[width=0.31\textwidth]{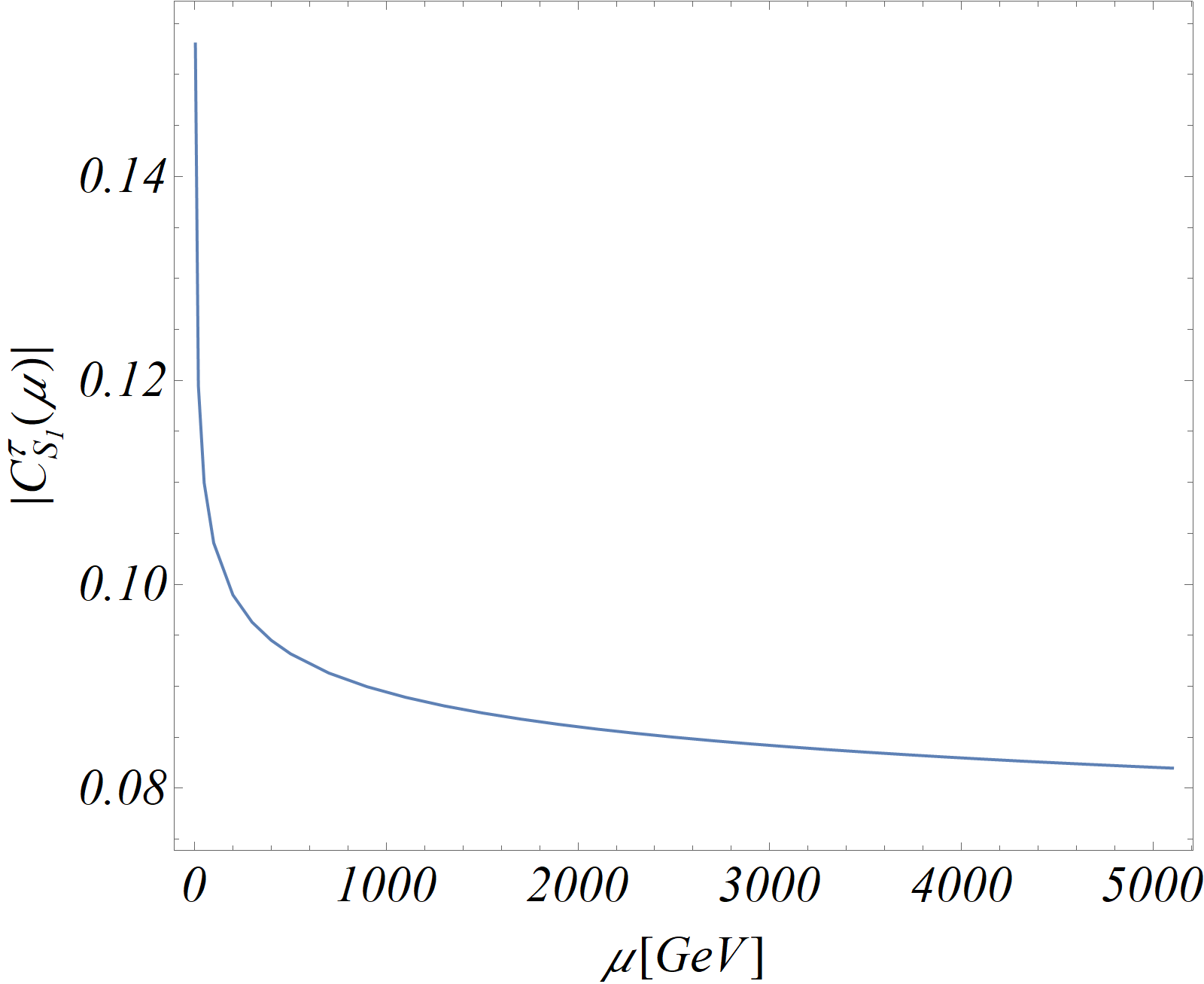}\label{fig:cs1}}~~~
	\subfloat[]{\includegraphics[width=0.31\textwidth]{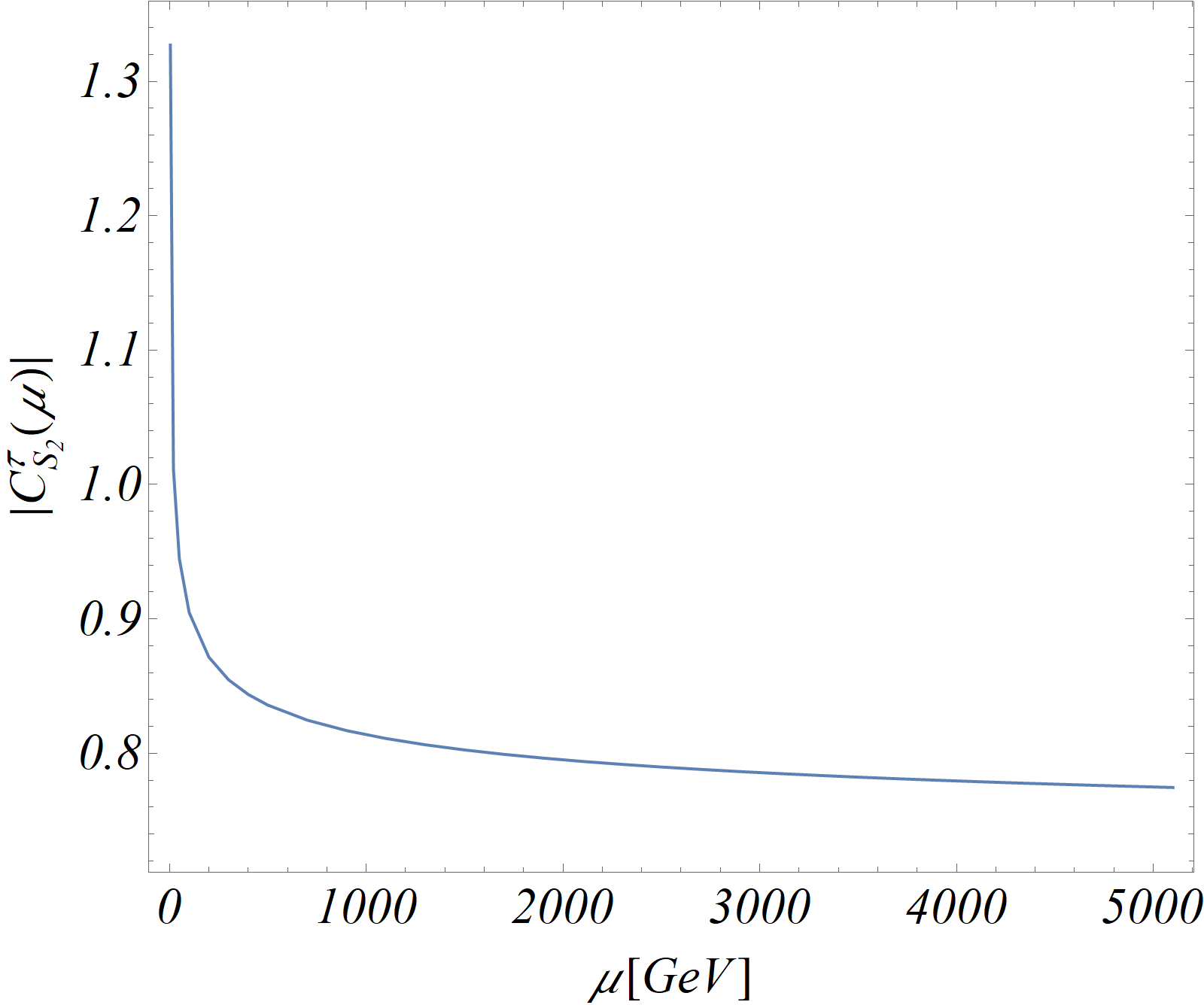}\label{fig:cs2}}~~~
	\subfloat[]{\includegraphics[width=0.31\textwidth]{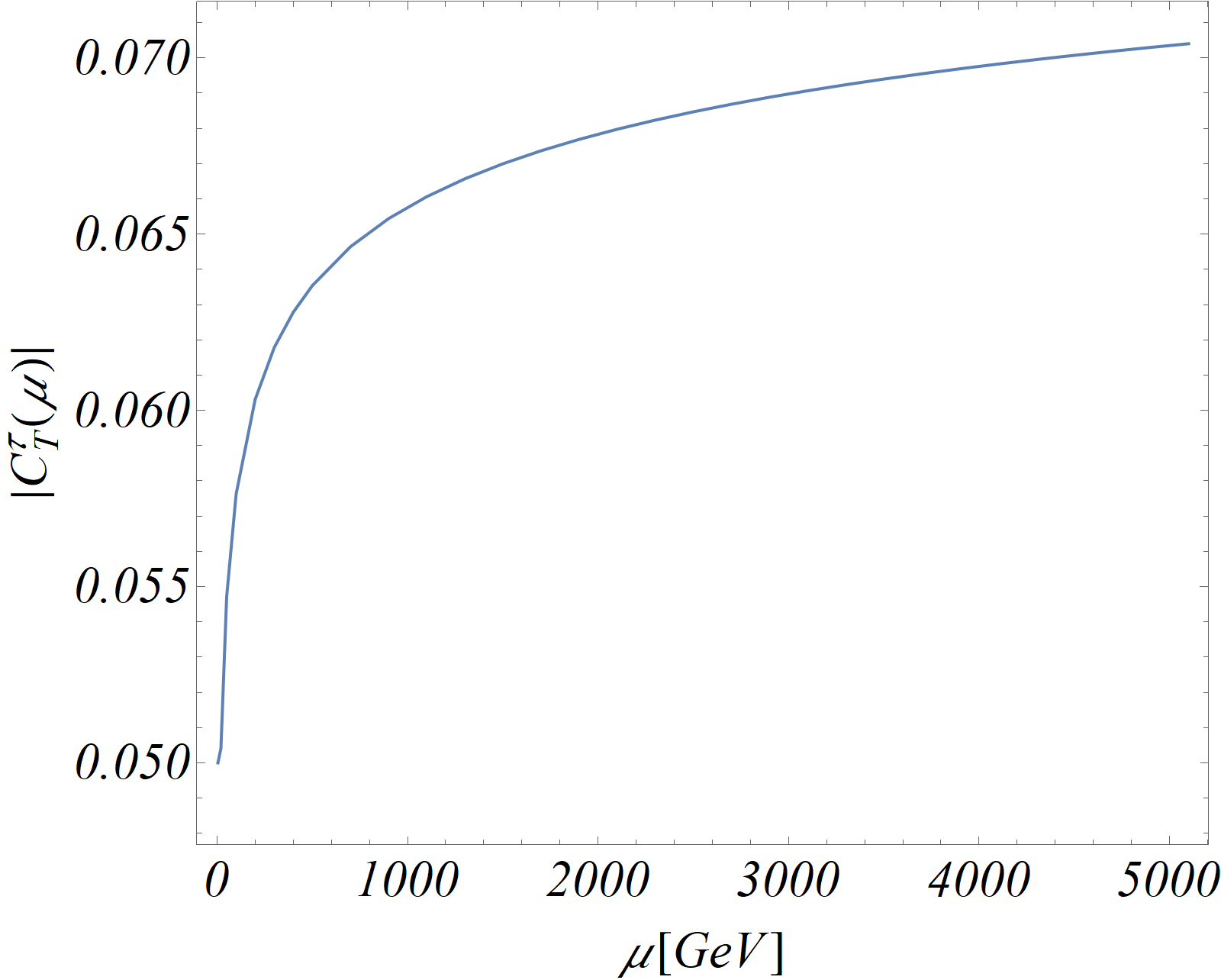}\label{fig:ct}}
 \caption{The variation of the WCs $C_{S_1}^{\tau}$, $C_{S_2}^{\tau}$ and $C_T^{\tau}$ with the energy scale $\mu$.}
	\label{fig:WC}
\end{figure}

Thus, using the equations \ref{eq:RGE} and \ref{eq:WCrelation}, we will get the relation between the WCs at $\mu_{NP}$ and $\mu_b$. In figs. \ref{fig:cs1}, \ref{fig:cs2} and \ref{fig:ct}, we show the variation of the WCs $C_{S_1}^{\tau}$, $C_{S_2}^{\tau}$ and $C_T^{\tau}$ with the energy scale, taking the values of the WCs at $\mu_b$ from table \ref{tab:fitNPnonMFV}. These projected bounds will be useful in collider searches. 
 
	\section{Summary} 
	
	This paper considers NP effects in the $\BtoDDstellnu$ decays with the heavy ($\tau$) and light ($\mu, e$) leptons. The analyses include the most up-to-date results from the experiments and the lattice. For example, inputs from the lattice include the $B\to D^*$ form factors at non-zero recoils. Also, amongst the data, the most updated results on $\rdrdst$ and $d\Gamma(\bar{B}\to D^* (\mu^-,e^-)\bar{\nu})/dq^2$ are included in the fits. We review the standard model (SM) predictions of the different observables, like $\rdrdst$, $A_{FB}^{D^{(*)}}$, $P_{\ell}^{D^{(*)}}$, and $F_L^{D^*}$ related to these decay channels with $\tau$, $\mu$ and $e$. Our predictions of the $\rdrdst$ based on the lattice and LCSR inputs are given as 
	$$ R(D)_{\text{Lat+LCSR}}=0.304 \pm 0.003, \ \ \ \ R(D^*)_{\text{Lat+LCSR}} = 0.253 \pm 0.009 .$$
	Whereas the results obtained using only the lattice inputs are given by 
		$$ R(D)_{\text{Lat}}=0.304 \pm 0.003, \ \ \ \ R(D^*)_{\text{Lat}} = 0.258 \pm 0.012 .$$
		
		The experimental values according to the HFLAV average \cite{hflavnew}

$$ R(D) = 0.356 \pm 0.029, \ \ \ R(D^*) = 0.284 \pm 0.013 .$$

	 For all the different lepton final states, we have noticed that the predicted values of $F_L^{\bar{B} \to D^* \ell^- \bar{\nu}}$ and $A_{FB}^{\bar{B} \to D^* \ell^- \bar{\nu}}$ between the fits with or without the experimental data deviate from each other. For the decays to $\tau$, it is $\gsim 2 \sigma$ while for the $\mu$ and $e$ it is $\sim 2 \sigma$. Also, we have predicted in the SM the differences $\Delta A_{FB}^{D^{(*)}}$ and $\Delta F_L^{D^*}$ between the tauons, muons and electrons in different pairs.  
	
	From the available data on $\bar{B}\to D^{(*)}(\mu^-, e^-)\bar{\nu}$ decays and the available lattice inputs we have simultaneously extracted $|V_{cb}|$, the new physics WCs and the form factors. In all the different NP scenarios, the extracted values of $|V_{cb}|$ is given by 
	$$|V_{cb}| = (40.3 \pm 0.5)\times 10^{-3},$$ which is consistent with the fit without NP. In this case, the fitted values of the new physics WCs are consistent with zero but could be considerable. Only in the scenarios with the WC $C^{\mu}_{S_1}$ or $C^{\mu}_{S_2}$, we have noted an enhancement in the predictions of $R(D)$ (but not in $R(D^*)$) from that of SM but not large enough to explain the 1-$\sigma$ lower limit of the corresponding data. 
	
	Also, we have analysed the most up-to-date data on $\rdrdst$ with or without $F_L^{D^*}$ alongside the inputs on $\Gamma(\bar{B}\to D^{(*)}(\mu^-, e^-)\bar{\nu})$ decays. In this study, we divide our analyses based on whether there are similar types or different types of NP in the decays with $\tau$ and $\mu$ or $e$ in the final state. In both types of fits, the extracted values of $|V_{cb}|$ do not shift from the one mentioned above. For similar types of NP, only the one operator scenarios $\mathcal{O}_{S_1}$ or $\mathcal{O}_{S_2}$ could explain the observed data in $R(D)$ but not $R(D^*)$. Also, amongst these two scenarios, only the contribution from $\mathcal{O}_{S_2}$ shows an enhancement in the value of $F_L^{D^*}$ but not sufficient to explain the respective measured value.

	In the case of different types of NP in the heavy and light leptons, the one operator scenario $\mathcal{O}_{V_1}^{\tau}$ can accommodate the observations in $\rdrdst$. However, it can not explain $F_L^{D^*}$. On the other hand, the scenario $\mathcal{O}^{\tau}_{S_2}$ can explain $R(D)$ and $F_{L}^{D^*}$ but not $R(D^*)$. The scenarios $\mathcal{O}^{\tau}_{V_2,T}$ can explain only $R(D^*)$ and the $\mathcal{O}^{\tau}_{S_1}$ can explain only $R(D)$, not the other two observations. However, all the two operator scenarios with $\mathcal{O}^{\tau}_{S_2}$ as one of the operators could explain all the measured values. Note that apart from the scenario $[\mathcal{O}^{\tau}_{S_1},\mathcal{O}^{\tau}_{S_2}]$, in all the other two operator scenarios with one of the operator $\mathcal{O}^{\tau}_{S_2}$, the branching fractions $\mathcal{B}(B_c\to \tau\nu)\approx 80$\%. However, in the scenario $[\mathcal{O}^{\tau}_{S_1},\mathcal{O}^{\tau}_{S_2}]$ the fit result predicts $\mathcal{B}(B_c\to \tau\nu) = 0.50 \pm 0.23$. The bounds obtained on the new physics WCs will be relevant at the scale $m_b$; following an RGE, we have projected those bounds to high-scale NP, which could be useful for collider searches. 
	
	In addition to $\BtoDDstellnu$ decays, we have also studied the available data on $\btopilnu$ decays in the presence of NP. The CKM element $|V_{ub}|$ has been extracted alongside the new physics WCs. We have noticed that the extracted new physics WCs have large errors and are consistent with zero, but large values are allowed by the data. Finally, in the SM and the NP scenarios, we have predicted various observables associated with the above decay modes for the light leptons.                   
	
	Using all these available data for the light and heavy leptons in $\BtoDDstellnu$ and $\btopilnu$ decays, we have given bounds on the couplings of the relevant SMEFT operators and the probable NP scale $\Lambda$. We have noticed that in the $\tau$ channel for the order one magnitude of the couplings ($\tilde{C}^{\tau}$), the NP scale $\Lambda \approx 1$ TeV, while for muon channel for a coupling strength $\tilde{C}^{\mu} \approx 1$, we need $\Lambda \approx 5$ TeV. Therefore, based on the current data, for a fixed value of $\Lambda$, $\tilde{C}^{\tau} \approx 10\times \tilde{C}^{\mu}$ which is as per the expectation.     

\acknowledgments
We would like to thank Markus Prim, Florian Bernlochner and Syuhei Iguro for some useful communications.

\section{Appendix}

The expressions of $R(D)$, $R(D^*)$,  $F_L^{D^*}$, $A_{FB}^{D^*}$, $A_{\lambda_{\tau}}^{D^*}$, $A_{FB}^D$ and $A_{\lambda_{\tau}}^D$ in the presence of NP are given in the equations \ref{eq:RD}, \ref{eq:RDst}, \ref{eq:FL}, \ref{eq:AFBDst}, \ref{eq:leptonDst}, \ref{eq:AFBD} and \ref{eq:leptonD} respectively.

\begin{equation} \label{eq:RD}
	\begin{split}
		R(D) = &  (0.304 \pm 0.003) \times (1+ 1.35 C_{S_1}^{\tau 2}+C_{S_1} (2.70 C_{S_2}^{\tau}+1.72 C_{V_1}^{\tau}+1.72C_{V_2}^{\tau}+1.72)+1.35
		C_{S_2}^{\tau 2} + \\ & 0.83 C_T^{\tau} C_{V_1}^{\tau} +C_{S_2}^{\tau} (1.72C_{V_1}^{\tau}+1.72 C_{V_2}^{\tau} + 1.72)+0.83
		C_T^{\tau} C_{V_2}^{\tau}+(0.49 C_T^{\tau} +0.83) C_T^{\tau} + C_{V_1}^{\tau 2}+ \\ &  2.00 C_{V_1}^{\tau}
		C_{V_2}^{\tau}+ 2.00 C_{V_1}^{\tau} + C_{V_2}^{\tau 2} + 2.00 C_{V_2}^{\tau}).
	\end{split}
\end{equation}

\begin{equation} \label{eq:RDst}
	\begin{split}
		R(D^*) = &  (0.258 \pm 0.012) \times (1+ 0.04 C_{S_1}^{\tau 2}+C_{S_1}^{\tau} (-0.07C_{S_2}^{\tau}+0.10 C_{V_1}^{\tau}-0.10 C_{V_2}^{\tau}+0.10)+0.04
		C_{S_2}^{\tau 2}+ \\ & C_{S_2}^{\tau} (-0.10 C_{V_1}^{\tau}+0.10C_{V_2}^{\tau}-0.10)-2.94C_T^{\tau} C_{V_1}^{\tau}+4.79
		C_T^{\tau} C_{V_2}^{\tau}+C_T^{\tau} (10.65 C_T^{\tau}-2.94)+ C_{V_1}^{\tau 2}- \\ & 1.79 C_{V_1}^{\tau} C_{V_2}^{\tau}+2.
		C_{V_1}^{\tau}+ C_{V_2}^{\tau 2}-1.79 C_{V_2}^{\tau}).
	\end{split}
\end{equation}

\begin{equation} \label{eq:FL}
	\begin{split}
		F_L^{D^*} =& (0.427 \pm 0.009) \times (27.26 + 2.34 C_{S_1}^{\tau 2}+C_{S_1}^{\tau} (-4.68 C_{S_2}^{\tau}+6.65 C_{V_1}^{\tau}-6.65C_{V_2}^{\tau}+6.65)+ 2.34 C_{S_2}^{\tau 2} - \\ & 66.82C_T^{\tau} C_{V_1}^{\tau} + C_{S_2}^{\tau} (-6.65 C_{V_1}^{\tau}+6.65 C_{V_2}^{\tau}-6.65) +66.82
		C_T^{\tau} C_{V_2}^{\tau}+C_T^{\tau} (69.62 C_T^{\tau}-66.82)+ \\ & 27.26C_{V_1}^{\tau 2}  -54.52 C_{V_1}^{\tau}
		C_{V_2}^{\tau}+  54.52 C_{V_1}^{\tau}+ 27.26 C_{V_2}^{\tau 2}-54.52 C_{V_2}^{\tau} ) / (27.26 + C_{S_1}^{\tau 2}+ -80.15 C_T^{\tau} C_{V_1}^{\tau}+ \\ & 130.72 C_T^{\tau} C_{V_2}^{\tau} +  C_{S_1}^{\tau} (-2.
		C_{S_2}^{\tau}+2.84C_{V_1}^{\tau}-2.84 C_{V_2}^{\tau}+2.84)+C_{S_2}^{\tau 2}+  C_{S_2}^{\tau} (-2.84C_{V_1}^{\tau}+2.84
		C_{V_2}^{\tau}  \\ & -2.84) +  C_T^{\tau} (290.34
		C_T^{\tau}-80.15)+27.26 C_{V_1}^{\tau 2}- 48.81 C_{V_1}^{\tau} C_{V_2}^{\tau}+54.52 C_{V_1}^{\tau}+27.26
		C_{V_2}^{\tau 2} -48.81 C_{V_2}^{\tau}).
	\end{split}
\end{equation}

\begin{equation} \label{eq:AFBDst}
\begin{split}
A_{FB}^{D^*} = & (-0.077 \pm 0.009 ) \times (0.09 + C_{S_1}^{\tau} (0.27 C_T^{\tau}-0.10 C_{V_1}^{\tau}+0.10 C_{V_2}^{\tau}-0.10)+ 0.29 C_{V_1}^{\tau}
C_{V_2}^{\tau}+0.19 C_{V_1}^{\tau} + \\ & 0.29 C_{V_2}^{\tau} +  C_{S_2}^{\tau} (-0.27
C_T^{\tau}+0.10 C_{V_1}^{\tau}-0.10 C_{V_2}^{\tau}+0.10)-4.19 C_T^{\tau 2}+ 1.50 C_T^{\tau}
C_{V_1}^{\tau}-2.63 C_T^{\tau} C_{V_2}^{\tau}+ \\ & 1.50 C_T^{\tau}+0.09 C_{V_1}^{\tau 2} -0.38 C_{V_2}^{\tau 2} )/ (0.09 + 0.003
C_{S_1}^{\tau 2}+C_{S_1}^{\tau} (-0.01 C_{S_2}^{\tau}+0.01 C_{V_1}^{\tau}-0.01 C_{V_2}^{\tau}+0.01)+ \\ & 0.003
C_{S_2}^{\tau 2}+ 0.19 C_{V_1}^{\tau}+  C_{S_2}^{\tau} (-0.01 C_{V_1}^{\tau}+0.01 C_{V_2}^{\tau}-0.01)-0.28 C_T^{\tau}
C_{V_1}^{\tau}+0.45 C_T^{\tau} C_{V_2}^{\tau}+ \\ & C_T^{\tau} ( C_T^{\tau}-0.28)+0.09 C_{V_1}^{\tau 2}-0.17
C_{V_1}^{\tau} C_{V_2}^{\tau}+  0.09 C_{V_2}^{\tau 2}-0.17 C_{V_2}^{\tau} )
\end{split}
\end{equation}

\begin{equation} \label{eq:leptonDst}
	\begin{split}
		A_{\lambda_{\tau}}^{D^*} = & (0.519 \pm 0.007) \times (3.29 -0.23 C_{S_1}^{\tau 2}+C_{S_1}^{\tau} (0.46 C_{S_2}^{\tau}-0.66 C_{V_1}^{\tau}+0.66 C_{V_2}^{\tau}-0.66)-0.23 C_{S_2}^{\tau 2}+ \\ & 3.29 C_{V_2}^{\tau 2} +  C_{S_2}^{\tau} (0.66 C_{V_1}^{\tau}-0.66 C_{V_2}^{\tau}+0.66)-6.21 C_T^{\tau} C_{V_1}^{\tau}+10.13
		C_T^{\tau} C_{V_2}^{\tau}+(-3.52 C_T^{\tau}-6.21) C_T^{\tau}+ \\ & 3.29 C_{V_1}^{\tau 2}-5.75 C_{V_1}^{\tau}
		C_{V_2}^{\tau}+  6.58 C_{V_1}^{\tau}-5.75 C_{V_2}^{\tau}) / (3.29 +0.12  C_{S_1}^{\tau 2} -9.67 C_T^{\tau} C_{V_1}^{\tau}+15.77 C_T^{\tau} C_{V_2}^{\tau} \\ & +C_{S_1}^{\tau}
		(-0.24 C_{S_2}^{\tau}+0.34 C_{V_1}^{\tau}-0.34 C_{V_2}^{\tau}+0.34)+0.12 C_{S_2}^{\tau 2}+ 6.58 C_{V_1}^{\tau} -5.89 C_{V_1}^{\tau} C_{V_2}^{\tau} -5.89 C_{V_2}^{\tau} \\ & +  C_{S_2}^{\tau} (-0.34
		C_{V_1}^{\tau}+0.34 C_{V_2}^{\tau}-0.34)  +C_T^{\tau}
		(35.02 C_T^{\tau}-9.67)+3.29 C_{V_1}^{\tau 2} +  3.29
		C_{V_2}^{\tau 2})
	\end{split}
\end{equation}

\begin{equation} \label{eq:AFBD}
\begin{split}
A_{FB}^D = & (0.3596 \pm 0.0004) \times (2.06 + C_{S_1}^{\tau} (5.61 C_T^{\tau}+2.84 C_{V_1}^{\tau}+2.84 C_{V_2}^{\tau}+2.84)+ 3.82 C_T^{\tau} C_{V_1}^{\tau} + \\ & 3.82 C_T^{\tau} C_{V_2}^{\tau} 
+ 3.82 C_T^{\tau}+  C_{S_2}^{\tau} (5.61
C_T^{\tau}+2.84 C_{V_1}^{\tau}+2.84 C_{V_2}^{\tau}+2.84) + 2.06 C_{V_1}^{\tau 2}  +4.12 C_{V_1}^{\tau} C_{V_2}^{\tau} \\ & +4.12 C_{V_1}^{\tau}+2.06
C_{V_2}^{\tau 2}  +4.12 C_{V_2}^{\tau} )/ (2.06 +  2.78 C_{S_1}^{\tau 2}+C_{S_1}^{\tau} (5.57 C_{S_2}^{\tau}+3.54
C_{V_1}^{\tau}  + 3.54 C_{V_2}^{\tau} \\ & +3.54)+  2.78 C_{S_2}^{\tau 2}   +C_{S_2}^{\tau} (3.54 C_{V_1}^{\tau}  + 3.54
C_{V_2}^{\tau}+3.54) +  1.71 C_T^{\tau} C_{V_2}^{\tau}  +1.71 C_T^{\tau} C_{V_1}^{\tau} \\ & +C_T^{\tau} (C_T^{\tau}+1.71)+  2.06 C_{V_1}^{\tau 2}+  4.12 C_{V_1}^{\tau} C_{V_2}^{\tau}+  4.12 C_{V_1}^{\tau}+2.06
C_{V_2}^{\tau 2}+4.12 C_{V_2}^{\tau} )
\end{split}
\end{equation}

\begin{equation} \label{eq:leptonD}
	\begin{split}
		A_{\lambda_{\tau}}^D = & (-0.324 \pm 0.003) \times (0.74 +3.09 C_{S_1}^{\tau 2}+C_{S_1}^{\tau} (6.17 C_{S_2}^{\tau}+3.93 C_{V_1}^{\tau}+3.93 C_{V_2}^{\tau}+3.93)+3.09
		C_{S_2}^{\tau 2} \\ & -0.63 C_T^{\tau}   C_{V_1}^{\tau} +  C_{S_2}^{\tau}(3.93 C_{V_1}^{\tau}+3.93 C_{V_2}^{\tau}+3.93)-0.63
		C_T^{\tau} C_{V_2}^{\tau}+(0.07 C_T^{\tau}-0.63) C_T^{\tau}+ 0.74 C_{V_1}^{\tau 2} \\ & +1.48 C_{V_1}^{\tau}
		C_{V_2}^{\tau}  +1.48C_{V_1}^{\tau}+  0.74 C_{V_2}^{\tau 2}+1.48 C_{V_2}^{\tau} )/ (0.74 + C_{S_1}^{\tau 2} +0.61 C_T^{\tau} C_{V_1}^{\tau}+0.61 C_T^{\tau} C_{V_2}^{\tau} \\ &  + C_{S_1}^{\tau} (2.
		C_{S_2}^{\tau}+1.27 C_{V_1}^{\tau}+1.27 C_{V_2}^{\tau}+1.27)+C_{S_2}^{\tau2} +(0.36 C_T^{\tau}+0.61)
		C_T^{\tau}+0.74 C_{V_1}^{\tau 2} +1.48
		C_{V_2}^{\tau} \\ & C_{S_2}^{\tau} (1.27C_{V_1}^{\tau} +1.27
		C_{V_2}^{\tau}+1.27) +1.48 C_{V_1}^{\tau} C_{V_2}^{\tau}+ 1.48 C_{V_1}^{\tau}+0.74 C_{V_2}^{\tau 2} )
	\end{split}
\end{equation}

One can obtain the SM predictions by setting the new physics WCs $C_i = 0$. We have obtained the results for the SM following the fit results of the form factors obtained from only the use of lattice. We have also provided the corresponding errors. We have not quoted the errors associated with the terms of the new physics WCs. Depending on different models, these expressions are useful in phenomenological analyses to constrain the new physics contributions in the $C_i$s. The numerical factors in front of each of the coefficients also depend on the fitted values of the BGL coefficients.

\begin{figure*}[t]
	\small
	\centering
	\subfloat[]{\includegraphics[width=0.32\textwidth]{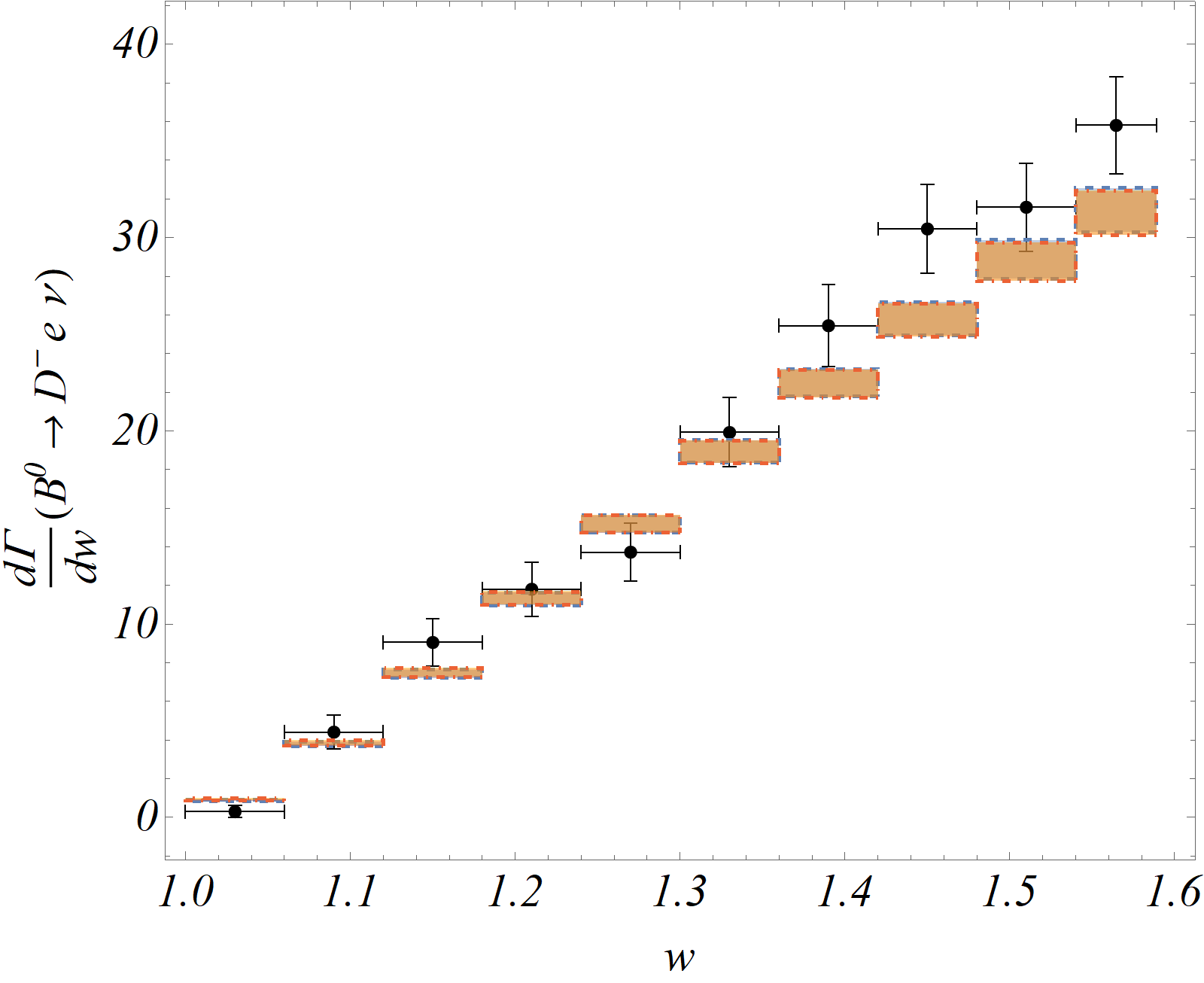}\label{fig:gaepl}}~~~
	\subfloat[]{\includegraphics[width=0.32\textwidth]{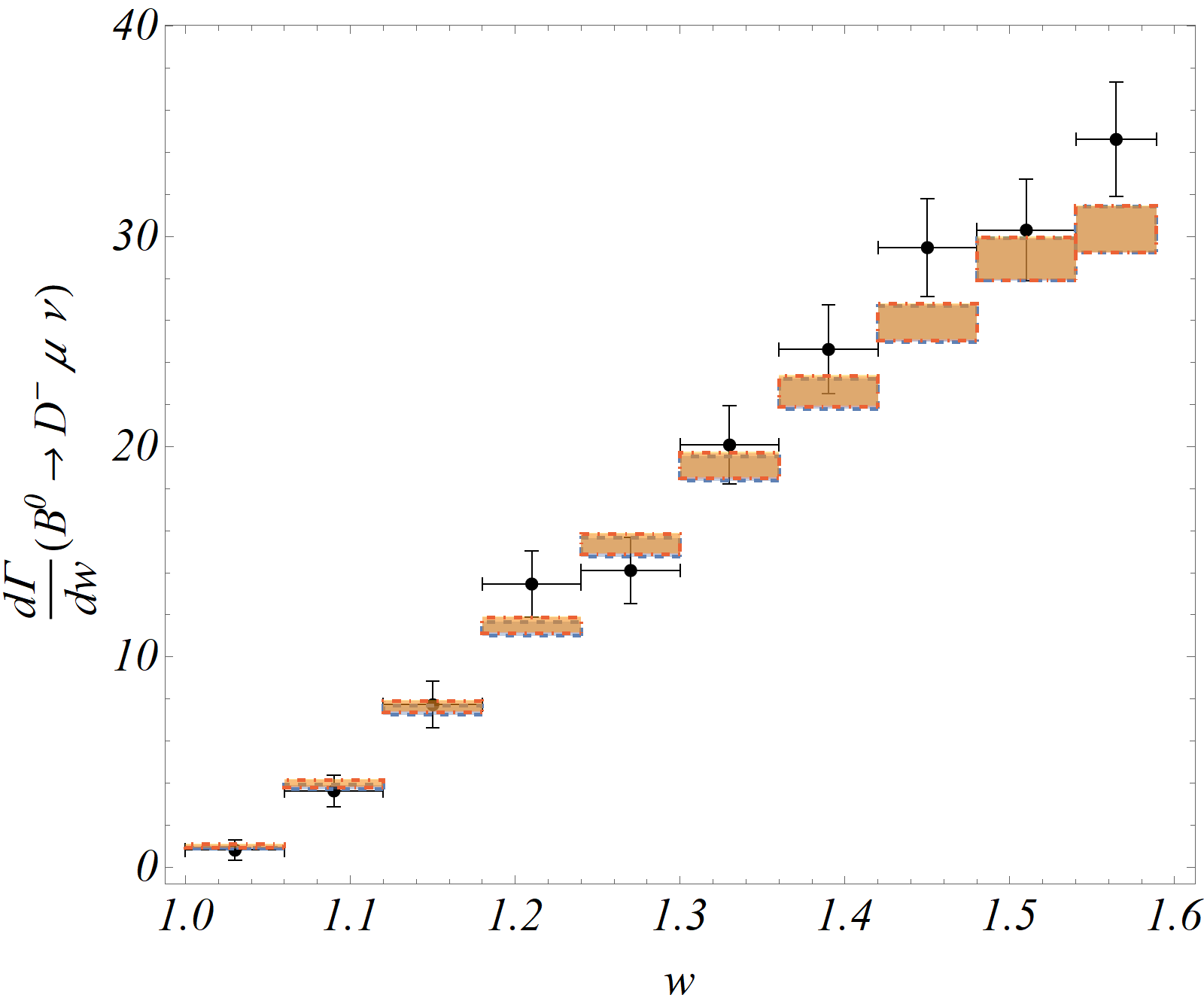}\label{fig:gamupl}}~~~
	\subfloat[]{\includegraphics[width=0.32\textwidth]{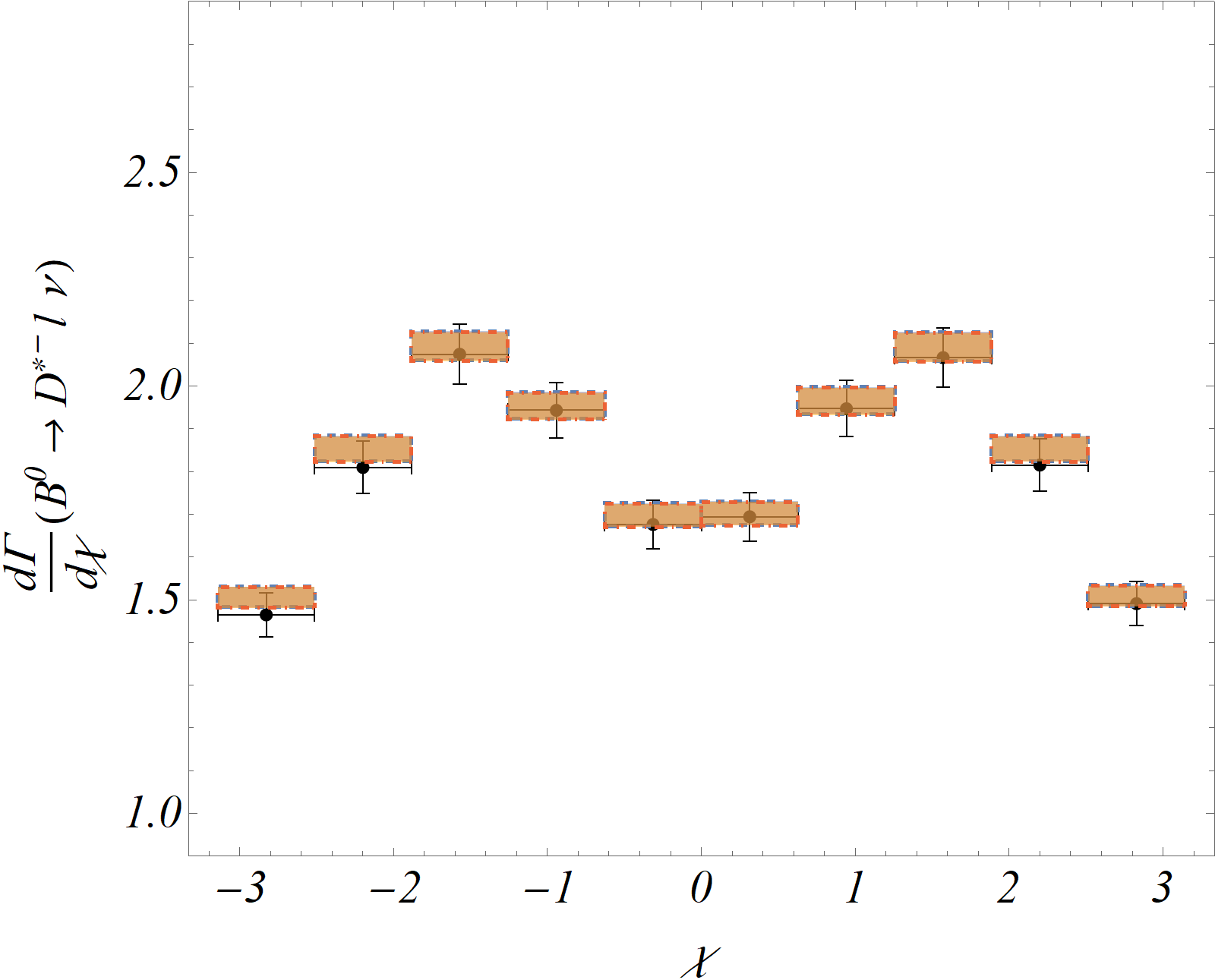}\label{fig:gadstchi}}\\
	\subfloat[]{\includegraphics[width=0.32\textwidth]{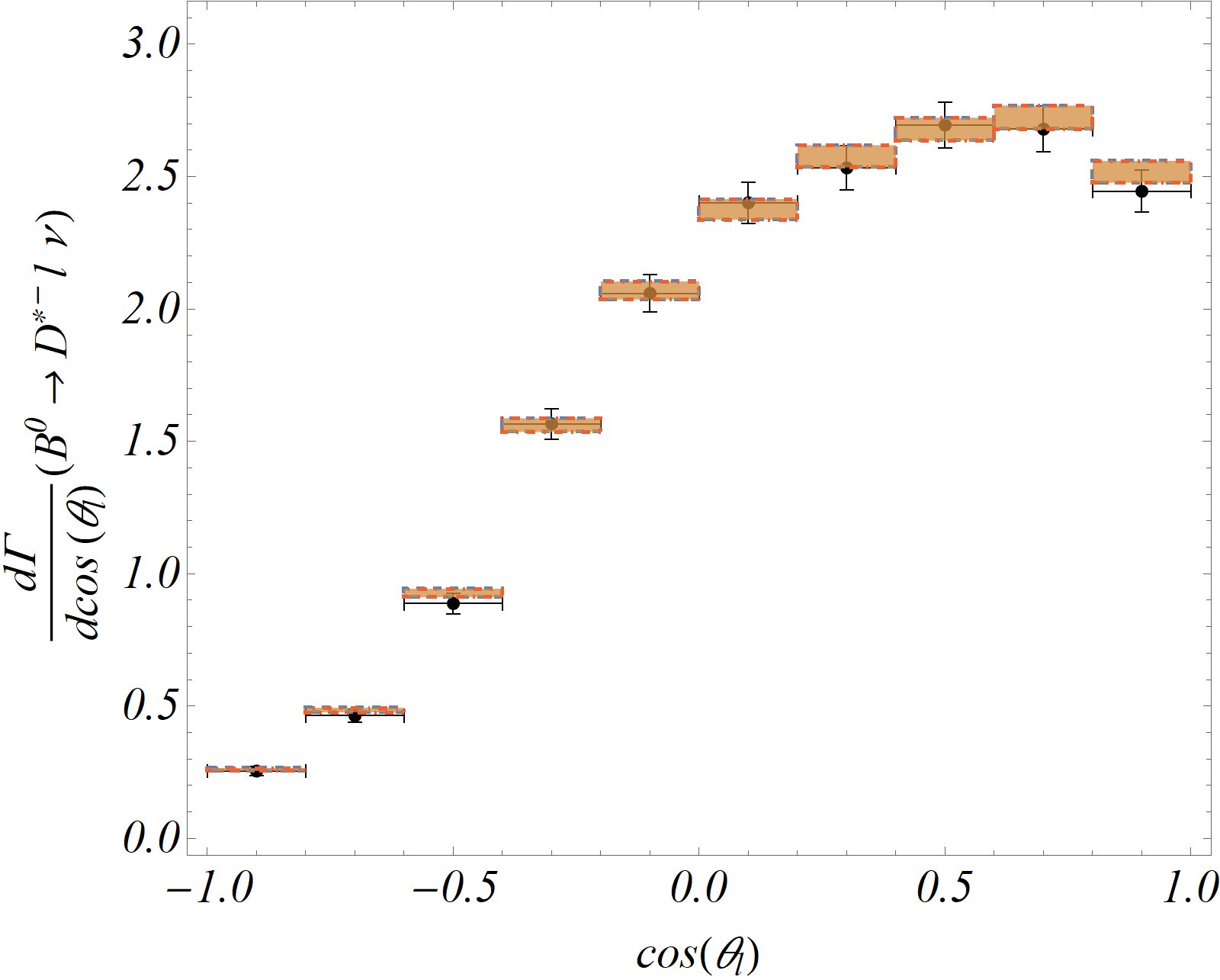}\label{fig:gadstcthl}}~~~	\subfloat[]{\includegraphics[width=0.32\textwidth]{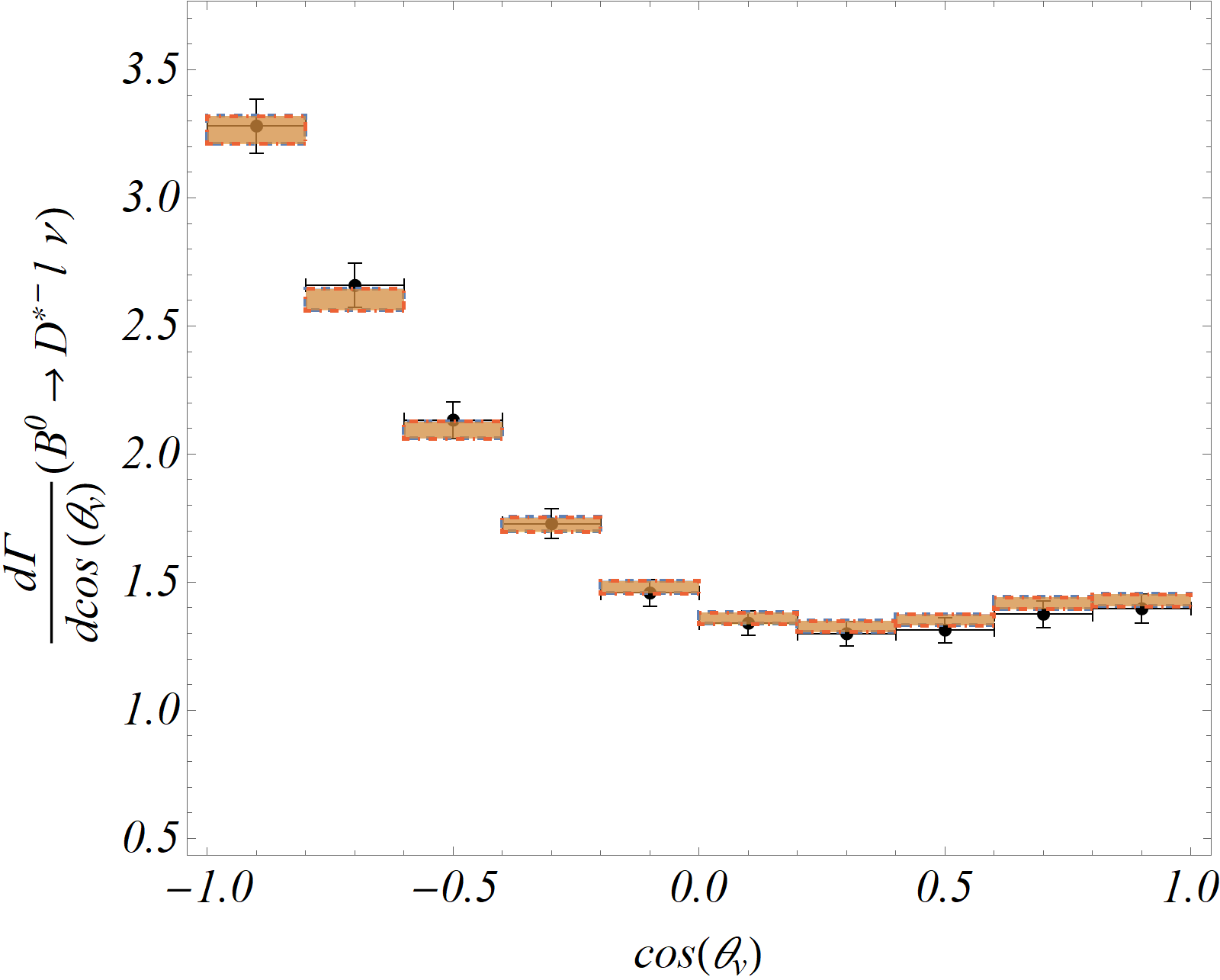}\label{fig:gadstcthv}}~~~
	\subfloat[]{\includegraphics[width=0.32\textwidth]{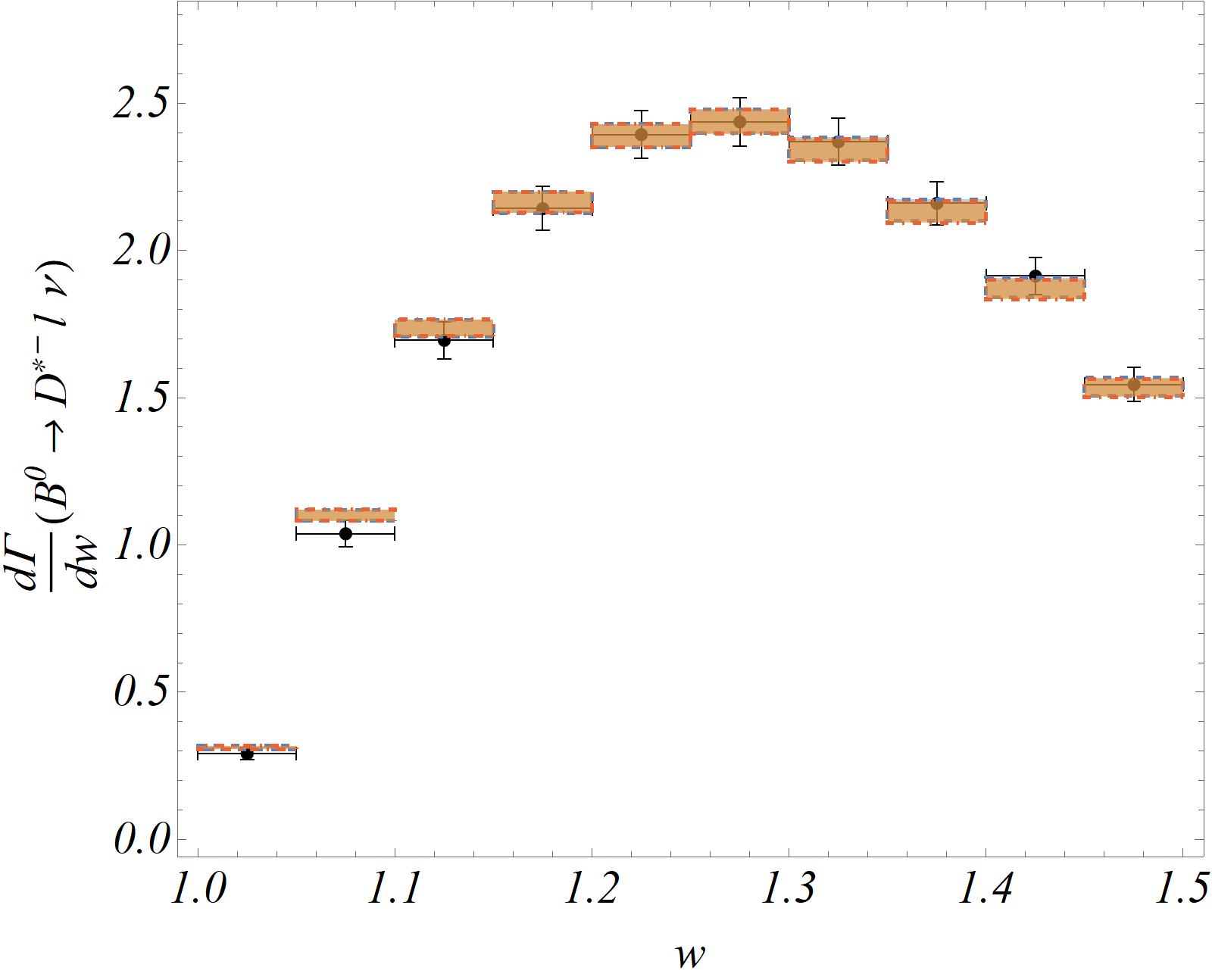}\label{fig:gadstw}}\\
	\subfloat[]{\includegraphics[width=0.35\textwidth]{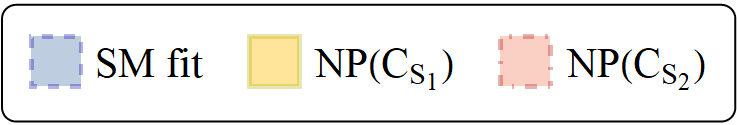}\label{fig:legend}}
	\caption{The predicted angular and $q^2$ distributions of the rates in $\BtoDDstellnu$ decays obtained in the scenarios without NP and with $C_{S_1}$ and $C_{S_2}$, which have been compared with the measured one.}
	\label{fig:gamma}
\end{figure*}

The predictions of the different observables related to $\bar{B}\to D^{(*)}(\tau^-,\mu^-,e^-)\bar{\nu}$ and $\bar{B}\to \pi\mu^-\bar{\nu}$ decays in the SM and different NP scenarios in small $q^2$-bins are presented in this section. In the one operator scenarios, the predictions are presented in tables \ref{tab:predB2D1OP} and \ref{tab:predB2Dst1OP} for $B\to D$ and $B\to D^*$ decays, respectively. The respective predictions in the two operator scenarios are given in tables \ref{tab:predB2DOP2a}, \ref{tab:predB2DOP2b} for $B\to D$, and in tables \ref{tab:predB2Dst2OPa} and \ref{tab:predB2DstOP2b} for $B\to D^*$ decays, respectively. These predictions are obtained using the fit results of tables \ref{tab:fitNPnonMFV} and \ref{tab:NPfit2opr}. 

\begin{table}[htbp]
	\begin{center}
		\small
		\renewcommand*{\arraystretch}{1.6}
\begin{tabular}{|*{8}{c|}}
	\hline
		\hline
	Obs  & $q^2$ bins & SM value & $C_{V_1}$ & $C_{V_2}$ & $C_T$  & $C_{S_1}$ & $C_{S_2}$  \\
\hline
$A_{FB}^{\bar{B} \to D \mu^- \bar{\nu}}$ & 
	$\{0.01,2\}$  &  $\text{0.03319(6)}$  &  $\text{0.03325(5)}$  &  $\text{0.03324(5)}$  &  $\text{0.03324(6)}$  &  $\text{0.033(2)}$  &  $\text{0.033(2)}$  \\
	\hline
	& $\{2,4\}$  &  $\text{0.00649(1)}$  &  $\text{0.00649(1)}$  &  $\text{0.00649(1)}$  &  $\text{0.00649(4)}$  &  $\text{0.006(3)}$  &  $\text{0.006(3)}$  \\
	\hline
	& $\{4,6\}$  &  $\text{0.00423(1)}$  &  $\text{0.00423(1)}$  &  $\text{0.00423(1)}$  &  $\text{0.00423(5)}$  &  $\text{0.004(3)}$  &  $\text{0.004(3)}$  \\
	\hline
 &	$\{6,8\}$  &  $\text{0.00350(1)}$  &  $\text{0.00350(1)}$  &  $\text{0.00350(1)}$  &  $\text{0.00350(5)}$  &  $\text{0.003(4)}$  &  $\text{0.003(4)}$  \\
	\hline
 &	$\{8,10\}$  &  $\text{0.00348(1)}$  &  $\text{0.00348(1)}$  &  $\text{0.00348(1)}$  &  $\text{0.00348(6)}$  &  $\text{0.003(5)}$  &  $\text{0.003(5)}$  \\
	\hline
 &	$\{10,11.628\}$  &  $\text{0.00458(2)}$  &  $\text{0.00458(2)}$  &  $\text{0.00459(2)}$  &  $\text{0.0046(1)}$  &  $\text{0.004(7)}$  &  $\text{0.004(7)}$  \\
	\hline
$A_{\lambda_\ell}^{\bar{B} \to D \mu^- \bar{\nu}}$ &	$\{0.01,2\}$  &  $\text{0.9107(2)}$  &  $\text{0.9106(1)}$  &  $\text{0.9106(1)}$  &  $\text{0.9106(1)}$  &  $\text{0.911(10)}$  &  $\text{0.911(10)}$  \\
	\hline
  &	$\{2,4\}$  &  $\text{0.98165(5)}$  &  $\text{0.98165(5)}$  &  $\text{0.98165(5)}$  &  $\text{0.98165(6)}$  &  $\text{0.98(1)}$  &  $\text{0.98(1)}$  \\
	\hline
  &	$\{4,6\}$  &  $\text{0.98721(6)}$  &  $\text{0.98722(5)}$  &  $\text{0.98722(6)}$  &  $\text{0.98721(6)}$  &  $\text{0.99(1)}$  &  $\text{0.99(1)}$  \\
	\hline
  &	$\{6,8\}$  &  $\text{0.98825(6)}$  &  $\text{0.98826(6)}$  &  $\text{0.98826(6)}$  &  $\text{0.98826(8)}$  &  $\text{0.99(2)}$  &  $\text{0.99(2)}$  \\
	\hline
  &	$\{8,10\}$  &  $\text{0.98580(10)}$  &  $\text{0.98581(9)}$  &  $\text{0.98580(9)}$  &  $\text{0.9858(1)}$  &  $\text{0.99(3)}$  &  $\text{0.99(3)}$  \\
	\hline
   &	$\{10,11.628\}$  &  $\text{0.9706(2)}$  &  $\text{0.9706(2)}$  &  $\text{0.9706(2)}$  &  $\text{0.9706(2)}$  &  $\text{0.98(8)}$  &  $\text{0.98(8)}$  \\
	\hline
   $A_{FB}^{\bar{B} \to D \tau^- \bar{\nu}}$ &	$\{3.157,5\}$  &  $\text{0.44311(8)}$  &  $\text{0.44315(8)}$  &  $\text{0.44314(8)}$  &  $\text{0.436(2)}$  &  $\text{0.436(3)}$  &  $\text{-0.13(3)}$  \\
	\hline
  &	$\{5,7\}$  &  $\text{0.39400(8)}$  &  $\text{0.39404(6)}$  &  $\text{0.39403(8)}$  &  $\text{0.382(4)}$  &  $\text{0.385(3)}$  &  $\text{-0.34(2)}$  \\
	\hline
   &	$\{7,9\}$  &  $\text{0.3494(2)}$  &  $\text{0.3494(2)}$  &  $\text{0.3494(2)}$  &  $\text{0.330(6)}$  &  $\text{0.333(6)}$  &  $\text{-0.339(4)}$  \\
	\hline
   &	$\{9,11.628\}$  &  $\text{0.2848(5)}$  &  $\text{0.2849(5)}$  &  $\text{0.2848(5)}$  &  $\text{0.259(8)}$  &  $\text{0.249(9)}$  &  $\text{-0.209(8)}$  \\
	\hline
  $A_{\lambda_\ell}^{\bar{B} \to D \tau^- \bar{\nu}}$ &	$\{3.157,5\}$  &  $\text{-0.314(2)}$  &  $\text{-0.314(2)}$  &  $\text{-0.314(2)}$  &  $\text{-0.339(7)}$  &  $\text{-0.40(2)}$  &  $\text{0.42(2)}$  \\
	\hline
   &	$\{5,7\}$  &  $\text{-0.255(2)}$  &  $\text{-0.255(2)}$  &  $\text{-0.255(2)}$  &  $\text{-0.280(7)}$  &  $\text{-0.38(3)}$  &  $\text{0.15(6)}$  \\
	\hline
&	$\{7,9\}$  &  $\text{-0.278(3)}$  &  $\text{-0.277(3)}$  &  $\text{-0.277(3)}$  &  $\text{-0.303(8)}$  &  $\text{-0.44(4)}$  &  $\text{-0.33(5)}$  \\
	\hline
  &	$\{9,11.628\}$  &  $\text{-0.496(3)}$  &  $\text{-0.496(3)}$  &  $\text{-0.496(3)}$  &  $\text{-0.520(7)}$  &  $\text{-0.66(3)}$  &  $\text{-0.76(2)}$  \\
	\hline
\end{tabular}
\caption{Predictions of various observables in the SM and different 1-operator scenarios in the $\bar{B} \to D l \bar{\nu}$ channel. }
\label{tab:predB2D1OP}
\end{center}
\end{table} 

\begin{table}[t]
	\begin{center}
 \small
		\renewcommand*{\arraystretch}{1.6}
		\begin{tabular}{|*{8}{c|}}
			\hline
			Obs  & $q^2$ bins & SM value & $C_{V_1}$ & $C_{V_2}$ & $C_T$  & $C_{S_1}$ & $C_{S_2}$  \\
			\hline
$A_{FB}^{\bar{B} \to D^* \mu^- \bar{\nu}}$ &
	$\{0.01,2\}$  &  $\text{-0.08(1)}$  &  $\text{-0.059(5)}$  &  $\text{-0.059(5)}$  &  $\text{-0.059(5)}$  &  $\text{-0.059(5)}$  &  $\text{-0.059(5)}$  \\
	\hline
&	$\{2,4\}$  &  $\text{-0.23(1)}$  &  $\text{-0.198(5)}$  &  $\text{-0.198(5)}$  &  $\text{-0.198(5)}$  &  $\text{-0.198(5)}$  &  $\text{-0.198(5)}$  \\
	\hline
&	$\{4,6\}$  &  $\text{-0.29(1)}$  &  $\text{-0.259(4)}$  &  $\text{-0.259(4)}$  &  $\text{-0.259(4)}$  &  $\text{-0.259(4)}$  &  $\text{-0.259(4)}$  \\
	\hline
&	$\{6,8\}$  &  $\text{-0.290(9)}$  &  $\text{-0.271(5)}$  &  $\text{-0.271(5)}$  &  $\text{-0.271(5)}$  &  $\text{-0.271(5)}$  &  $\text{-0.271(5)}$  \\
	\hline
&	$\{8,10\}$  &  $\text{-0.235(7)}$  &  $\text{-0.224(5)}$  &  $\text{-0.223(5)}$  &  $\text{-0.224(5)}$  &  $\text{-0.224(5)}$  &  $\text{-0.224(5)}$  \\
	\hline
&	$\{10,10.68\}$  &  $\text{-0.128(4)}$  &  $\text{-0.124(3)}$  &  $\text{-0.123(3)}$  &  $\text{-0.124(3)}$  &  $\text{-0.124(3)}$  &  $\text{-0.124(3)}$  \\
	\hline
  $A_{\lambda_\ell}^{\bar{B} \to D^* \mu^- \bar{\nu}}$ &	$\{0.01,2\}$  &  $\text{0.930(2)}$  &  $\text{0.9276(5)}$  &  $\text{0.9276(5)}$  &  $\text{0.9276(9)}$  &  $\text{0.928(4)}$  &  $\text{0.927(4)}$  \\
	\hline
&	$\{2,4\}$  &  $\text{0.9905(2)}$  &  $\text{0.9902(1)}$  &  $\text{0.9902(1)}$  &  $\text{0.9902(2)}$  &  $\text{0.990(3)}$  &  $\text{0.990(3)}$  \\
	\hline
&	$\{4,6\}$  &  $\text{0.99564(6)}$  &  $\text{0.99552(5)}$  &  $\text{0.99551(5)}$  &  $\text{0.9955(1)}$  &  $\text{0.996(2)}$  &  $\text{0.995(2)}$  \\
	\hline
&	$\{6,8\}$  &  $\text{0.99750(3)}$  &  $\text{0.99746(2)}$  &  $\text{0.99746(2)}$  &  $\text{0.99746(8)}$  &  $\text{0.9975(10)}$  &  $\text{0.997(1)}$  \\
	\hline
&	$\{8,10\}$  &  $\text{0.99842(1)}$  &  $\text{0.998410(10)}$  &  $\text{0.998409(10)}$  &  $\text{0.99841(8)}$  &  $\text{0.9984(5)}$  &  $\text{0.9984(5)}$  \\
	\hline
&	$\{10,10.68\}$  &  $\text{0.998844(2)}$  &  $\text{0.998843(2)}$  &  $\text{0.998843(2)}$  &  $\text{0.99885(8)}$  &  $\text{0.9989(1)}$  &  $\text{0.9988(1)}$  \\
	\hline
$F_L^{\bar{B} \to D^* \mu^- \bar{\nu}}$ &	$\{0.01,2\}$  &  $\text{0.82(1)}$  &  $\text{0.848(5)}$  &  $\text{0.848(5)}$  &  $\text{0.848(5)}$  &  $\text{0.848(5)}$  &  $\text{0.848(5)}$  \\
	\hline
 &	$\{2,4\}$  &  $\text{0.60(2)}$  &  $\text{0.636(5)}$  &  $\text{0.636(5)}$  &  $\text{0.636(5)}$  &  $\text{0.636(5)}$  &  $\text{0.636(5)}$  \\
	\hline
&	$\{4,6\}$  &  $\text{0.47(1)}$  &  $\text{0.501(3)}$  &  $\text{0.501(3)}$  &  $\text{0.501(3)}$  &  $\text{0.501(3)}$  &  $\text{0.501(3)}$  \\
	\hline
&	$\{6,8\}$  &  $\text{0.399(6)}$  &  $\text{0.415(3)}$  &  $\text{0.415(3)}$  &  $\text{0.415(3)}$  &  $\text{0.415(3)}$  &  $\text{0.415(3)}$  \\
	\hline
&	$\{8,10\}$  &  $\text{0.355(3)}$  &  $\text{0.363(2)}$  &  $\text{0.363(2)}$  &  $\text{0.363(2)}$  &  $\text{0.363(2)}$  &  $\text{0.362(2)}$  \\
	\hline
&	$\{10,10.68\}$  &  $\text{0.3371(8)}$  &  $\text{0.3388(5)}$  &  $\text{0.3388(5)}$  &  $\text{0.3388(5)}$  &  $\text{0.3388(5)}$  &  $\text{0.3388(5)}$  \\
	\hline
 $A_{FB}^{\bar{B} \to D^* \tau^- \bar{\nu}}$ &	$\{3.157,5\}$  &  $\text{0.09(1)}$  &  $\text{0.126(4)}$  &  $\text{0.143(9)}$  &  $\text{0.17(1)}$  &  $\text{0.140(5)}$  &  $\text{0.213(3)}$  \\
	\hline
&	$\{5,7\}$  &  $\text{-0.034(10)}$  &  $\text{-0.009(4)}$  &  $\text{0.01(1)}$  &  $\text{0.05(2)}$  &  $\text{0.008(6)}$  &  $\text{0.106(4)}$  \\
	\hline
&	$\{7,9\}$  &  $\text{-0.110(7)}$  &  $\text{-0.095(4)}$  &  $\text{-0.07(1)}$  &  $\text{-0.03(2)}$  &  $\text{-0.080(6)}$  &  $\text{0.016(5)}$  \\
	\hline
&	$\{9,10.68\}$  &  $\text{-0.103(5)}$  &  $\text{-0.095(3)}$  &  $\text{-0.080(8)}$  &  $\text{-0.03(2)}$  &  $\text{-0.087(4)}$  &  $\text{-0.025(4)}$  \\
	\hline
$A_{\lambda_\ell}^{\bar{B} \to D^* \tau^- \bar{\nu}}$ &	$\{3.157,5\}$  &  $\text{0.159(8)}$  &  $\text{0.144(6)}$  &  $\text{0.140(6)}$  &  $\text{0.165(8)}$  &  $\text{0.11(1)}$  &  $\text{-0.15(1)}$  \\
	\hline
&	$\{5,7\}$  &  $\text{0.385(6)}$  &  $\text{0.376(5)}$  &  $\text{0.373(5)}$  &  $\text{0.364(6)}$  &  $\text{0.34(1)}$  &  $\text{0.05(1)}$  \\
	\hline
&	$\{7,9\}$  &  $\text{0.568(3)}$  &  $\text{0.564(3)}$  &  $\text{0.563(3)}$  &  $\text{0.53(1)}$  &  $\text{0.539(8)}$  &  $\text{0.29(1)}$  \\
	\hline
&	$\{9,10.68\}$  &  $\text{0.686(1)}$  &  $\text{0.685(1)}$  &  $\text{0.685(1)}$  &  $\text{0.63(2)}$  &  $\text{0.674(3)}$  &  $\text{0.551(7)}$  \\
	\hline
 $F_L^{\bar{B} \to D^* \tau^- \bar{\nu}}$ &	$\{3.157,5\}$  &  $\text{0.61(1)}$  &  $\text{0.636(4)}$  &  $\text{0.643(5)}$  &  $\text{0.60(1)}$  &  $\text{0.647(5)}$  &  $\text{0.731(5)}$  \\
	\hline
&	$\{5,7\}$  &  $\text{0.500(9)}$  &  $\text{0.522(3)}$  &  $\text{0.528(4)}$  &  $\text{0.496(10)}$  &  $\text{0.534(5)}$  &  $\text{0.634(5)}$  \\
	\hline
&	$\{7,9\}$  &  $\text{0.411(5)}$  &  $\text{0.423(3)}$  &  $\text{0.427(3)}$  &  $\text{0.416(4)}$  &  $\text{0.432(4)}$  &  $\text{0.522(5)}$  \\
	\hline
&	$\{9,10.68\}$  &  $\text{0.357(2)}$  &  $\text{0.361(1)}$  &  $\text{0.362(2)}$  &  $\text{0.371(3)}$  &  $\text{0.365(2)}$  &  $\text{0.412(3)}$  \\
	\hline
		\end{tabular}
		\caption{Predictions of various observables in the SM and different 1-operator scenarios for the $\bar{B} \to D^* l \bar{\nu}$ channel. }
		\label{tab:predB2Dst1OP}
	\end{center}
\end{table}

\begin{table}
	\begin{center}
		\small
		\renewcommand*{\arraystretch}{1.6}
		\begin{tabular}{|*{7}{c|}}
			\hline
			Obs  & $q^2$ bins &  $C_{V_1}$, $C_{V_2}$ & $C_{V_1}$, $C_{S_1}$ & $C_{V_1}$, $C_{S_2}$ & $C_{V_1}$, $C_T$ &  $C_{V_2}$, $C_{S_1}$   \\
			\hline
			
 $A_{FB}^{\bar{B} \to D \mu^- \bar{\nu}}$ & 
 	$\{0.01,2\}$  &  $\text{0.03325(5)}$  &  $\text{0.033(2)}$  &  $\text{0.033(2)}$  &  $\text{0.03325(6)}$  &  $\text{0.033(2)}$  \\
 	\hline
 &	$\{2,4\}$  &  $\text{0.00649(1)}$  &  $\text{0.006(3)}$  &  $\text{0.006(3)}$  &  $\text{0.00649(4)}$  &  $\text{0.006(3)}$  \\
 	\hline
 &	$\{4,6\}$  &  $\text{0.00423(1)}$  &  $\text{0.004(3)}$  &  $\text{0.004(3)}$  &  $\text{0.00423(5)}$  &  $\text{0.004(3)}$  \\
 	\hline
 &	$\{6,8\}$  &  $\text{0.00350(1)}$  &  $\text{0.003(4)}$  &  $\text{0.003(4)}$  &  $\text{0.00350(5)}$  &  $\text{0.003(4)}$  \\
 	\hline
 &	$\{8,10\}$  &  $\text{0.00348(1)}$  &  $\text{0.003(5)}$  &  $\text{0.003(5)}$  &  $\text{0.00348(6)}$  &  $\text{0.003(5)}$  \\
 	\hline
 &	$\{10,11.628\}$  &  $\text{0.00458(2)}$  &  $\text{0.004(7)}$  &  $\text{0.004(7)}$  &  $\text{0.0046(1)}$  &  $\text{0.004(7)}$  \\
 	\hline
  &	$\bf\{0.01,11.628\}$  &  $\text{0.01381(8)}$  &  $\text{0.014(3)}$  &  $\text{0.014(3)}$  &  $\text{0.01380(9)}$  &  $\text{0.014(3)}$  \\
 	\hline
 $A_{\lambda_\ell}^{\bar{B} \to D \mu^- \bar{\nu}}$ & $\{0.01,2\}$   &  $\text{0.9106(1)}$  &  $\text{0.911(10)}$  &  $\text{0.911(10)}$  &  $\text{0.9106(1)}$  &  $\text{0.91(1)}$  \\
 	\hline
 &	$\{2,4\}$  &  $\text{0.98166(5)}$  &  $\text{0.98(1)}$  &  $\text{0.98(1)}$  &  $\text{0.98165(6)}$  &  $\text{0.98(1)}$  \\
 	\hline
 &	$\{4,6\}$  &  $\text{0.98722(5)}$  &  $\text{0.99(1)}$  &  $\text{0.99(2)}$  &  $\text{0.98722(6)}$  &  $\text{0.99(2)}$  \\
 	\hline
 &	$\{6,8\}$  &  $\text{0.98827(6)}$  &  $\text{0.99(2)}$  &  $\text{0.99(2)}$  &  $\text{0.98826(8)}$  &  $\text{0.99(2)}$  \\
 	\hline
 &	$\{8,10\}$  &  $\text{0.98581(9)}$  &  $\text{0.99(3)}$  &  $\text{0.99(3)}$  &  $\text{0.9858(1)}$  &  $\text{0.99(3)}$  \\
 	\hline
 &	$\{10,11.628\}$  &  $\text{0.9706(2)}$  &  $\text{0.98(8)}$  &  $\text{0.97(8)}$  &  $\text{0.9706(2)}$  &  $\text{0.98(8)}$  \\
 	\hline
 &	$\bf\{0.01,11.628\}$  &  $\text{0.9615(2)}$  &  $\text{0.96(1)}$  &  $\text{0.96(2)}$  &  $\text{0.9615(2)}$  &  $\text{0.96(2)}$  \\
 	\hline
 $A_{FB}^{\bar{B} \to D \tau^- \bar{\nu}}$ &	$\{3.157,5\}$  &  $\text{0.44315(8)}$  &  $\text{0.442(3)}$  &  $\text{-0.20(5)}$  &  $\text{0.445(4)}$  &  $\text{0.434(3)}$  \\
 	\hline
 &	$\{5,7\}$  &  $\text{0.39405(8)}$  &  $\text{0.393(3)}$  &  $\text{-0.36(1)}$  &  $\text{0.397(7)}$  &  $\text{0.383(4)}$  \\
 	\hline
 &	$\{7,9\}$  &  $\text{0.3494(2)}$  &  $\text{0.346(7)}$  &  $\text{-0.327(10)}$  &  $\text{0.35(1)}$  &  $\text{0.329(7)}$  \\
 	\hline
 &	$\{9,11.628\}$  &  $\text{0.2849(5)}$  &  $\text{0.28(2)}$  &  $\text{-0.19(1)}$  &  $\text{0.29(1)}$  &  $\text{0.24(1)}$  \\
 	\hline
 &	$\bf\{3.157,11.628\}$  &  $\text{0.3600(2)}$  &  $\text{0.355(9)}$  &  $\text{-0.268(5)}$  &  $\text{0.365(10)}$  &  $\text{0.333(8)}$  \\
 	\hline
 $A_{\lambda_\ell}^{\bar{B} \to D \tau^- \bar{\nu}}$ &	$\{3.157,5\}$  &  $\text{-0.314(2)}$  &  $\text{-0.34(4)}$  &  $\text{0.37(5)}$  &  $\text{-0.31(1)}$  &  $\text{-0.42(2)}$  \\
 	\hline
 &	$\{5,7\}$  &  $\text{-0.254(2)}$  &  $\text{-0.29(6)}$  &  $\text{0.02(9)}$  &  $\text{-0.25(1)}$  &  $\text{-0.40(3)}$  \\
 	\hline
 &	$\{7,9\}$  &  $\text{-0.277(3)}$  &  $\text{-0.32(8)}$  &  $\text{-0.44(7)}$  &  $\text{-0.27(1)}$  &  $\text{-0.47(4)}$  \\
 	\hline
 &	$\{9,11.628\}$  &  $\text{-0.496(3)}$  &  $\text{-0.54(8)}$  &  $\text{-0.81(3)}$  &  $\text{-0.49(1)}$  &  $\text{-0.68(3)}$  \\
 	\hline
 &	$\bf\{3.157,11.628\}$  &  $\text{-0.323(3)}$  &  $\text{-0.36(7)}$  &  $\text{-0.48(6)}$  &  $\text{-0.32(2)}$  &  $\text{-0.50(4)}$  \\
 	\hline
 \end{tabular}
		\caption{Predictions of various observables in different 2-operator scenarios in the $\bar{B} \to D l \bar{\nu}$ channel. The rows with the $q^2$-bins written in bold font represent the predictions for the $q^2$ integrated observables.  }
		\label{tab:predB2DOP2a}
	\end{center}
\end{table} 

\begin{table}
	\begin{center}
		\small
		\renewcommand*{\arraystretch}{1.6}
		\begin{tabular}{|*{7}{c|}}
			\hline
			Obs  & $q^2$ bins  & $C_{V_2}$, $C_{S_2}$ & $C_{V_2}$, $C_T$ & $C_{S_1}$, $C_T$ & $C_{S_2}$, $C_T$ & $C_{S_1}$, $C_{S_2}$   \\
			\hline
$A_{FB}^{\bar{B} \to D \mu^- \bar{\nu}}$ & 
	$\{0.01,2\}$  &  $\text{0.033(2)}$  &  $\text{0.03325(6)}$  &  $\text{0.033(2)}$  &  $\text{0.033(2)}$  &  $\text{0.033(2)}$  \\
	\hline
&	$\{2,4\}$  &  $\text{0.006(3)}$  &  $\text{0.00649(4)}$  &  $\text{0.006(3)}$  &  $\text{0.006(3)}$  &  $\text{0.006(3)}$  \\
	\hline
&	$\{4,6\}$  &  $\text{0.004(3)}$  &  $\text{0.00423(5)}$  &  $\text{0.004(3)}$  &  $\text{0.004(3)}$  &  $\text{0.004(3)}$  \\
	\hline
&	$\{6,8\}$  &  $\text{0.003(4)}$  &  $\text{0.00350(5)}$  &  $\text{0.003(4)}$  &  $\text{0.003(4)}$  &  $\text{0.003(4)}$  \\
	\hline
&	$\{8,10\}$  &  $\text{0.003(5)}$  &  $\text{0.00348(6)}$  &  $\text{0.003(5)}$  &  $\text{0.003(5)}$  &  $\text{0.003(5)}$  \\
	\hline
&	$\{10,11.628\}$  &  $\text{0.005(7)}$  &  $\text{0.0046(1)}$  &  $\text{0.004(7)}$  &  $\text{0.004(7)}$  &  $\text{0.004(7)}$  \\
	\hline
&	$\bf\{0.01,11.628\}$  &  $\text{0.014(3)}$  &  $\text{0.01381(9)}$  &  $\text{0.014(3)}$  &  $\text{0.014(3)}$  &  $\text{0.014(3)}$  \\
	\hline
$A_{\lambda_\ell}^{\bar{B} \to D \mu^- \bar{\nu}}$ &	$\{0.01,2\}$  &  $\text{0.91(1)}$  &  $\text{0.9106(1)}$  &  $\text{0.911(10)}$  &  $\text{0.911(10)}$  &  $\text{0.911(10)}$  \\
	\hline
&	$\{2,4\}$  &  $\text{0.98(1)}$  &  $\text{0.98165(6)}$  &  $\text{0.98(1)}$  &  $\text{0.98(1)}$  &  $\text{0.98(1)}$  \\
	\hline
&	$\{4,6\}$  &  $\text{0.99(2)}$  &  $\text{0.98722(6)}$  &  $\text{0.99(1)}$  &  $\text{0.99(2)}$  &  $\text{0.99(1)}$  \\
	\hline
&	$\{6,8\}$  &  $\text{0.99(2)}$  &  $\text{0.98826(8)}$  &  $\text{0.99(2)}$  &  $\text{0.99(2)}$  &  $\text{0.99(2)}$  \\
	\hline
&	$\{8,10\}$  &  $\text{0.99(3)}$  &  $\text{0.9858(1)}$  &  $\text{0.99(3)}$  &  $\text{0.99(3)}$  &  $\text{0.99(3)}$  \\
	\hline
&	$\{10,11.628\}$  &  $\text{0.97(9)}$  &  $\text{0.9706(2)}$  &  $\text{0.98(8)}$  &  $\text{0.97(8)}$  &  $\text{0.98(8)}$  \\
	\hline
&	$\bf\{0.01,11.628\}$  &  $\text{0.96(2)}$  &  $\text{0.9615(2)}$  &  $\text{0.96(1)}$  &  $\text{0.96(2)}$  &  $\text{0.96(2)}$  \\
	\hline
$A_{FB}^{\bar{B} \to D \tau^- \bar{\nu}}$ &	$\{3.157,5\}$  &  $\text{-0.12(3)}$  &  $\text{0.430(3)}$  &  $\text{0.432(3)}$  &  $\text{-0.15(3)}$  &  $\text{-0.16(4)}$  \\
	\hline
&	$\{5,7\}$  &  $\text{-0.33(2)}$  &  $\text{0.369(6)}$  &  $\text{0.377(5)}$  &  $\text{-0.35(1)}$  &  $\text{-0.35(1)}$  \\
	\hline
&	$\{7,9\}$  &  $\text{-0.340(3)}$  &  $\text{0.311(9)}$  &  $\text{0.322(8)}$  &  $\text{-0.351(9)}$  &  $\text{-0.335(5)}$  \\
	\hline
&	$\{9,11.628\}$  &  $\text{-0.212(8)}$  &  $\text{0.23(1)}$  &  $\text{0.24(1)}$  &  $\text{-0.22(1)}$  &  $\text{-0.202(8)}$  \\
	\hline
&	$\bf\{3.157,11.628\}$  &  $\text{-0.2725(2)}$  &  $\text{0.326(8)}$  &  $\text{0.329(8)}$  &  $\text{-0.285(8)}$  &  $\text{-0.271(2)}$  \\
	\hline
$A_{\lambda_\ell}^{\bar{B} \to D \tau^- \bar{\nu}}$ &	$\{3.157,5\}$  &  $\text{0.43(2)}$  &  $\text{-0.36(1)}$  &  $\text{-0.40(2)}$  &  $\text{0.41(2)}$  &  $\text{0.41(2)}$  \\
	\hline
&	$\{5,7\}$  &  $\text{0.17(5)}$  &  $\text{-0.30(1)}$  &  $\text{-0.37(3)}$  &  $\text{0.12(5)}$  &  $\text{0.10(6)}$  \\
	\hline
&	$\{7,9\}$  &  $\text{-0.31(5)}$  &  $\text{-0.33(1)}$  &  $\text{-0.43(4)}$  &  $\text{-0.34(5)}$  &  $\text{-0.37(5)}$  \\
	\hline
&	$\{9,11.628\}$  &  $\text{-0.75(2)}$  &  $\text{-0.540(10)}$  &  $\text{-0.64(4)}$  &  $\text{-0.76(2)}$  &  $\text{-0.78(2)}$  \\
	\hline
&	$\bf\{3.157,11.628\}$  &  $\text{-0.37(5)}$  &  $\text{-0.37(1)}$  &  $\text{-0.46(4)}$  &  $\text{-0.40(4)}$  &  $\text{-0.43(5)}$  \\
	\hline
\end{tabular}
		\caption{Predictions of various observables in different 2-operator scenarios in the $\bar{B} \to D l \bar{\nu}$ channel. The rows with the $q^2$-bins written in bold font represent the predictions for the $q^2$ integrated observables.  }
		\label{tab:predB2DOP2b}
	\end{center}
\end{table}

\begin{table}
	\begin{center}
		\scriptsize
		\renewcommand*{\arraystretch}{1.6}
		\begin{tabular}{|*{7}{c|}}
			\hline
			Obs  & $q^2$ bins &  $C_{V_1}$, $C_{V_2}$ & $C_{V_1}$, $C_{S_1}$ & $C_{V_1}$, $C_{S_2}$ & $C_{V_1}$, $C_T$ &  $C_{V_2}$, $C_{S_1}$   \\
			\hline
		
 $A_{FB}^{\bar{B} \to D^* \mu^- \bar{\nu}}$ &
 	$\{0.01,2\}$  &  $\text{-0.059(5)}$  &  $\text{-0.059(5)}$  &  $\text{-0.059(5)}$  &  $\text{-0.059(5)}$  &  $\text{-0.059(5)}$  \\
 	\hline
 &	$\{2,4\}$  &  $\text{-0.198(5)}$  &  $\text{-0.198(5)}$  &  $\text{-0.198(5)}$  &  $\text{-0.198(5)}$  &  $\text{-0.198(5)}$  \\
 	\hline
 &	$\{4,6\}$  &  $\text{-0.259(4)}$  &  $\text{-0.259(4)}$  &  $\text{-0.259(4)}$  &  $\text{-0.259(4)}$  &  $\text{-0.259(4)}$  \\
 	\hline
 &	$\{6,8\}$  &  $\text{-0.270(5)}$  &  $\text{-0.271(5)}$  &  $\text{-0.271(5)}$  &  $\text{-0.271(5)}$  &  $\text{-0.270(5)}$  \\
 	\hline
 &	$\{8,10\}$  &  $\text{-0.223(5)}$  &  $\text{-0.224(5)}$  &  $\text{-0.224(5)}$  &  $\text{-0.224(5)}$  &  $\text{-0.223(5)}$  \\
 	\hline
 &	$\{10,10.68\}$  &  $\text{-0.123(3)}$  &  $\text{-0.124(3)}$  &  $\text{-0.124(3)}$  &  $\text{-0.124(3)}$  &  $\text{-0.123(3)}$  \\
 	\hline
 &	$\bf\{0.01,10.68\}$  &  $\text{-0.209(3)}$  &  $\text{-0.209(3)}$  &  $\text{-0.209(3)}$  &  $\text{-0.209(3)}$  &  $\text{-0.209(3)}$  \\
 	\hline
 $A_{\lambda_\ell}^{\bar{B} \to D^* \mu^- \bar{\nu}}$ &	$\{0.01,2\}$  &  $\text{0.9276(5)}$  &  $\text{0.928(4)}$  &  $\text{0.927(4)}$  &  $\text{0.9276(9)}$  &  $\text{0.928(4)}$  \\
 	\hline
 &	$\{2,4\}$  &  $\text{0.9902(1)}$  &  $\text{0.990(3)}$  &  $\text{0.990(3)}$  &  $\text{0.9902(2)}$  &  $\text{0.990(3)}$  \\
 	\hline
 &	$\{4,6\}$  &  $\text{0.99552(5)}$  &  $\text{0.996(2)}$  &  $\text{0.995(2)}$  &  $\text{0.9955(1)}$  &  $\text{0.996(2)}$  \\
 	\hline
 &	$\{6,8\}$  &  $\text{0.99746(2)}$  &  $\text{0.9975(10)}$  &  $\text{0.997(1)}$  &  $\text{0.99746(8)}$  &  $\text{0.998(1)}$  \\
 	\hline
 &	$\{8,10\}$  &  $\text{0.998409(10)}$  &  $\text{0.9984(4)}$  &  $\text{0.9984(5)}$  &  $\text{0.99841(8)}$  &  $\text{0.9984(5)}$  \\
 	\hline
 &	$\{10,10.68\}$  &  $\text{0.998843(2)}$  &  $\text{0.9989(1)}$  &  $\text{0.9988(1)}$  &  $\text{0.99885(8)}$  &  $\text{0.9988(1)}$  \\
 	\hline
 &	$\bf\{0.01,10.68\}$  &  $\text{0.9852(2)}$  &  $\text{0.985(2)}$  &  $\text{0.985(2)}$  &  $\text{0.9852(2)}$  &  $\text{0.985(2)}$  \\
 	\hline
 $F_L^{\bar{B} \to D^* \mu^- \bar{\nu}}$ &	$\{0.01,2\}$  &  $\text{0.847(5)}$  &  $\text{0.848(5)}$  &  $\text{0.848(5)}$  &  $\text{0.848(5)}$  &  $\text{0.847(5)}$  \\
 	\hline
 &	$\{2,4\}$  &  $\text{0.636(5)}$  &  $\text{0.636(5)}$  &  $\text{0.636(5)}$  &  $\text{0.636(5)}$  &  $\text{0.636(5)}$  \\
 	\hline
 &	$\{4,6\}$  &  $\text{0.501(3)}$  &  $\text{0.501(3)}$  &  $\text{0.501(3)}$  &  $\text{0.501(3)}$  &  $\text{0.501(3)}$  \\
 	\hline
 &	$\{6,8\}$  &  $\text{0.415(3)}$  &  $\text{0.415(3)}$  &  $\text{0.415(3)}$  &  $\text{0.415(3)}$  &  $\text{0.415(3)}$  \\
 	\hline
 &	$\{8,10\}$  &  $\text{0.363(2)}$  &  $\text{0.363(2)}$  &  $\text{0.363(2)}$  &  $\text{0.363(2)}$  &  $\text{0.363(2)}$  \\
 	\hline
 &	$\{10,10.68\}$  &  $\text{0.3388(5)}$  &  $\text{0.3388(5)}$  &  $\text{0.3388(5)}$  &  $\text{0.3388(5)}$  &  $\text{0.3388(5)}$  \\
 	\hline
 &	$\bf\{0.01,10.68\}$  &  $\text{0.530(3)}$  &  $\text{0.531(3)}$  &  $\text{0.531(3)}$  &  $\text{0.531(3)}$  &  $\text{0.530(3)}$  \\
 	\hline
 $A_{FB}^{\bar{B} \to D^* \tau^- \bar{\nu}}$ &	$\{3.157,5\}$  &  $\text{0.122(10)}$  &  $\text{0.130(7)}$  &  $\text{0.218(4)}$  &  $\text{0.11(2)}$  &  $\text{0.158(9)}$  \\
 	\hline
 &	$\{5,7\}$  &  $\text{-0.01(1)}$  &  $\text{-0.005(9)}$  &  $\text{0.112(6)}$  &  $\text{-0.03(4)}$  &  $\text{0.03(1)}$  \\
 	\hline
 &	$\{7,9\}$  &  $\text{-0.10(1)}$  &  $\text{-0.091(8)}$  &  $\text{0.024(7)}$  &  $\text{-0.12(4)}$  &  $\text{-0.06(1)}$  \\
 	\hline
 &	$\{9,10.68\}$  &  $\text{-0.098(9)}$  &  $\text{-0.093(5)}$  &  $\text{-0.019(5)}$  &  $\text{-0.12(4)}$  &  $\text{-0.070(8)}$  \\
 	\hline
 &	$\bf\{3.157,10.68\}$  &  $\text{-0.06(1)}$  &  $\text{-0.049(7)}$  &  $\text{0.059(6)}$  &  $\text{-0.07(4)}$  &  $\text{-0.018(10)}$  \\
 	\hline
 $A_{\lambda_\ell}^{\bar{B} \to D^* \tau^- \bar{\nu}}$ &	$\{3.157,5\}$  &  $\text{0.145(6)}$  &  $\text{0.14(2)}$  &  $\text{-0.18(2)}$  &  $\text{0.14(2)}$  &  $\text{0.10(1)}$  \\
 	\hline
 &	$\{5,7\}$  &  $\text{0.376(5)}$  &  $\text{0.37(2)}$  &  $\text{0.02(2)}$  &  $\text{0.377(5)}$  &  $\text{0.34(1)}$  \\
 	\hline
 &	$\{7,9\}$  &  $\text{0.564(3)}$  &  $\text{0.56(1)}$  &  $\text{0.27(2)}$  &  $\text{0.57(2)}$  &  $\text{0.537(8)}$  \\
 	\hline
 &	$\{9,10.68\}$  &  $\text{0.685(1)}$  &  $\text{0.683(5)}$  &  $\text{0.54(1)}$  &  $\text{0.70(3)}$  &  $\text{0.674(3)}$  \\
 	\hline
 &	$\bf\{3.157,10.68\}$  &  $\text{0.506(3)}$  &  $\text{0.50(1)}$  &  $\text{0.21(2)}$  &  $\text{0.51(1)}$  &  $\text{0.477(8)}$  \\
 	\hline
 $F_L^{\bar{B} \to D^* \tau^- \bar{\nu}}$ &	$\{3.157,5\}$  &  $\text{0.634(5)}$  &  $\text{0.638(6)}$  &  $\text{0.738(6)}$  &  $\text{0.64(1)}$  &  $\text{0.655(5)}$  \\
 	\hline
 &	$\{5,7\}$  &  $\text{0.520(5)}$  &  $\text{0.525(6)}$  &  $\text{0.644(8)}$  &  $\text{0.527(9)}$  &  $\text{0.541(5)}$  \\
 	\hline
 &	$\{7,9\}$  &  $\text{0.422(4)}$  &  $\text{0.425(5)}$  &  $\text{0.532(8)}$  &  $\text{0.423(3)}$  &  $\text{0.437(4)}$  \\
 	\hline
 &	$\{9,10.68\}$  &  $\text{0.360(2)}$  &  $\text{0.362(2)}$  &  $\text{0.417(5)}$  &  $\text{0.356(10)}$  &  $\text{0.367(2)}$  \\
 	\hline
 &	$\bf\{3.157,10.68\}$  &  $\text{0.452(4)}$  &  $\text{0.455(5)}$  &  $\text{0.561(7)}$  &  $\text{0.455(3)}$  &  $\text{0.468(4)}$  \\
 	\hline
 \end{tabular}
		\caption{Predictions of various observables in different 2-operator scenarios for the $\bar{B} \to D^* l \bar{\nu}$ channel. The rows with the $q^2$-bins written in bold font represent the predictions for the $q^2$ integrated observables.  }
		\label{tab:predB2Dst2OPa}
	\end{center}
\end{table}

\begin{table}
	\begin{center}
		\scriptsize
		\renewcommand*{\arraystretch}{1.6}
		\begin{tabular}{|*{7}{c|}}
			\hline
			Obs  & $q^2$ bins  & $C_{V_2}$, $C_{S_2}$ & $C_{V_2}$, $C_T$ & $C_{S_1}$, $C_T$ & $C_{S_2}$, $C_T$ & $C_{S_1}$, $C_{S_2}$   \\
			\hline
	
	$A_{FB}^{\bar{B} \to D^* \mu^- \bar{\nu}}$ &
		$\{0.01,2\}$  &  $\text{-0.059(5)}$  &  $\text{-0.059(5)}$  &  $\text{-0.059(5)}$  &  $\text{-0.059(5)}$  &  $\text{-0.060(10)}$  \\
		\hline
	&	$\{2,4\}$  &  $\text{-0.198(5)}$  &  $\text{-0.198(5)}$  &  $\text{-0.198(5)}$  &  $\text{-0.198(5)}$  &  $\text{-0.199(8)}$  \\
		\hline
	&	$\{4,6\}$  &  $\text{-0.259(4)}$  &  $\text{-0.259(4)}$  &  $\text{-0.259(4)}$  &  $\text{-0.259(4)}$  &  $\text{-0.260(5)}$  \\
		\hline
	&	$\{6,8\}$  &  $\text{-0.270(5)}$  &  $\text{-0.270(5)}$  &  $\text{-0.271(5)}$  &  $\text{-0.271(5)}$  &  $\text{-0.271(5)}$  \\
		\hline
	&	$\{8,10\}$  &  $\text{-0.223(5)}$  &  $\text{-0.223(5)}$  &  $\text{-0.224(5)}$  &  $\text{-0.224(5)}$  &  $\text{-0.224(5)}$  \\
		\hline
	&	$\{10,10.68\}$  &  $\text{-0.123(3)}$  &  $\text{-0.123(3)}$  &  $\text{-0.124(3)}$  &  $\text{-0.124(3)}$  &  $\text{-0.124(3)}$  \\
		\hline
	&	$\bf\{0.01,10.68\}$  &  $\text{-0.209(3)}$  &  $\text{-0.209(3)}$  &  $\text{-0.209(3)}$  &  $\text{-0.209(3)}$  &  $\text{-0.210(5)}$  \\
		\hline
	$A_{\lambda_\ell}^{\bar{B} \to D^* \mu^- \bar{\nu}}$ &	$\{0.01,2\}$  &  $\text{0.928(4)}$  &  $\text{0.9276(9)}$  &  $\text{0.928(4)}$  &  $\text{0.927(4)}$  &  $\text{0.93(3)}$  \\
		\hline
	&	$\{2,4\}$  &  $\text{0.990(3)}$  &  $\text{0.9902(2)}$  &  $\text{0.990(3)}$  &  $\text{0.990(3)}$  &  $\text{0.99(2)}$  \\
		\hline
	&	$\{4,6\}$  &  $\text{0.996(2)}$  &  $\text{0.9955(1)}$  &  $\text{0.996(2)}$  &  $\text{0.995(2)}$  &  $\text{0.997(8)}$  \\
		\hline
	&	$\{6,8\}$  &  $\text{0.997(1)}$  &  $\text{0.99746(8)}$  &  $\text{0.9975(10)}$  &  $\text{0.997(1)}$  &  $\text{0.998(4)}$  \\
		\hline
	&	$\{8,10\}$  &  $\text{0.9984(5)}$  &  $\text{0.99841(8)}$  &  $\text{0.9984(5)}$  &  $\text{0.9984(5)}$  &  $\text{0.999(1)}$  \\
		\hline
	&	$\{10,10.68\}$  &  $\text{0.9988(1)}$  &  $\text{0.99885(8)}$  &  $\text{0.9989(1)}$  &  $\text{0.9988(1)}$  &  $\text{0.9989(2)}$  \\
		\hline
	&	$\bf\{0.01,10.68\}$  &  $\text{0.985(2)}$  &  $\text{0.9852(2)}$  &  $\text{0.985(2)}$  &  $\text{0.985(2)}$  &  $\text{0.99(1)}$  \\
		\hline
	$F_L^{\bar{B} \to D^* \mu^- \bar{\nu}}$ &	$\{0.01,2\}$  &  $\text{0.847(5)}$  &  $\text{0.847(5)}$  &  $\text{0.848(5)}$  &  $\text{0.848(5)}$  &  $\text{0.847(5)}$  \\
		\hline
	&	$\{2,4\}$  &  $\text{0.636(5)}$  &  $\text{0.636(5)}$  &  $\text{0.636(5)}$  &  $\text{0.636(5)}$  &  $\text{0.636(5)}$  \\
		\hline
	&	$\{4,6\}$  &  $\text{0.501(3)}$  &  $\text{0.501(3)}$  &  $\text{0.501(3)}$  &  $\text{0.501(3)}$  &  $\text{0.501(3)}$  \\
		\hline
	&	$\{6,8\}$  &  $\text{0.415(3)}$  &  $\text{0.415(3)}$  &  $\text{0.415(3)}$  &  $\text{0.415(3)}$  &  $\text{0.415(3)}$  \\
		\hline
	&	$\{8,10\}$  &  $\text{0.363(2)}$  &  $\text{0.363(2)}$  &  $\text{0.363(2)}$  &  $\text{0.363(2)}$  &  $\text{0.362(2)}$  \\
		\hline
	&	$\{10,10.68\}$  &  $\text{0.3388(5)}$  &  $\text{0.3388(5)}$  &  $\text{0.3388(5)}$  &  $\text{0.3388(5)}$  &  $\text{0.3388(5)}$  \\
		\hline
	&	$\bf\{0.01,10.68\}$  &  $\text{0.530(3)}$  &  $\text{0.530(3)}$  &  $\text{0.531(3)}$  &  $\text{0.531(3)}$  &  $\text{0.530(3)}$  \\
		\hline
	$A_{FB}^{\bar{B} \to D^* \tau^- \bar{\nu}}$ &	$\{3.157,5\}$  &  $\text{0.205(6)}$  &  $\text{0.19(2)}$  &  $\text{0.17(1)}$  &  $\text{0.19(1)}$  &  $\text{0.20(1)}$  \\
		\hline
	&	$\{5,7\}$  &  $\text{0.097(7)}$  &  $\text{0.08(2)}$  &  $\text{0.05(2)}$  &  $\text{0.07(2)}$  &  $\text{0.08(2)}$  \\
		\hline
	&	$\{7,9\}$  &  $\text{0.007(8)}$  &  $\text{0.007(21)}$  &  $\text{-0.03(2)}$  &  $\text{-0.03(3)}$  &  $\text{-0.01(2)}$  \\
		\hline
	&	$\{9,10.68\}$  &  $\text{-0.033(7)}$  &  $\text{0.004(20)}$  &  $\text{-0.04(2)}$  &  $\text{-0.08(4)}$  &  $\text{-0.04(1)}$  \\
		\hline
	&	$\bf\{3.157,10.68\}$  &  $\text{0.044(7)}$  &  $\text{0.04(2)}$  &  $\text{0.003(18)}$  &  $\text{0.01(3)}$  &  $\text{0.03(2)}$  \\
		\hline
	$A_{\lambda_\ell}^{\bar{B} \to D^* \tau^- \bar{\nu}}$ &	$\{3.157,5\}$  &  $\text{-0.18(2)}$  &  $\text{0.19(1)}$  &  $\text{0.13(2)}$  &  $\text{-0.20(3)}$  &  $\text{-0.07(6)}$  \\
		\hline
	&	$\{5,7\}$  &  $\text{0.03(2)}$  &  $\text{0.34(2)}$  &  $\text{0.34(1)}$  &  $\text{0.02(3)}$  &  $\text{0.15(7)}$  \\
		\hline
	&	$\{7,9\}$  &  $\text{0.27(2)}$  &  $\text{0.47(3)}$  &  $\text{0.52(1)}$  &  $\text{0.28(2)}$  &  $\text{0.38(6)}$  \\
		\hline
	&	$\{9,10.68\}$  &  $\text{0.53(1)}$  &  $\text{0.55(4)}$  &  $\text{0.64(2)}$  &  $\text{0.564(9)}$  &  $\text{0.60(3)}$  \\
		\hline
	&	$\bf\{3.157,10.68\}$  &  $\text{0.21(2)}$  &  $\text{0.43(3)}$  &  $\text{0.47(1)}$  &  $\text{0.21(2)}$  &  $\text{0.32(6)}$  \\
		\hline
	$F_L^{\bar{B} \to D^* \tau^- \bar{\nu}}$ &	$\{3.157,5\}$  &  $\text{0.732(5)}$  &  $\text{0.52(4)}$  &  $\text{0.62(2)}$  &  $\text{0.750(8)}$  &  $\text{0.70(2)}$  \\
		\hline
	&	$\{5,7\}$  &  $\text{0.638(7)}$  &  $\text{0.45(3)}$  &  $\text{0.51(1)}$  &  $\text{0.653(9)}$  &  $\text{0.60(2)}$  \\
		\hline
	&	$\{7,9\}$  &  $\text{0.528(7)}$  &  $\text{0.40(1)}$  &  $\text{0.425(6)}$  &  $\text{0.534(7)}$  &  $\text{0.49(2)}$  \\
		\hline
	&	$\{9,10.68\}$  &  $\text{0.416(5)}$  &  $\text{0.377(3)}$  &  $\text{0.372(3)}$  &  $\text{0.408(5)}$  &  $\text{0.39(1)}$  \\
		\hline
	&	$\bf\{3.157,10.68\}$  &  $\text{0.557(7)}$  &  $\text{0.42(1)}$  &  $\text{0.453(7)}$  &  $\text{0.564(8)}$  &  $\text{0.52(2)}$  \\
		\hline
	\end{tabular}
		\caption{Predictions of various observables in different 2-operator scenarios for the $\bar{B} \to D^* l \bar{\nu}$ channel. The rows with the $q^2$-bins written in bold font represent the predictions for the $q^2$ integrated observables.   }
		\label{tab:predB2DstOP2b}
	\end{center}
\end{table}
\begin{table}
	\begin{center}
		\small
		\renewcommand*{\arraystretch}{1.6}
		\begin{tabular}{|*{8}{c|}}
			\hline
			\hline
			Obs  & $q^2$ bins & SM value & $C_{V_1}$ & $C_{V_2}$ & $C_T$  & $C_{S_1}$ & $C_{S_2}$  \\
			\hline
			\hline
			$A_{FB}^{\bar{B} \to \pi \mu^- \bar{\nu}}$ &
			$\{0.01,2\}$  &  $\text{0.0317(2)}$  &  $\text{0.0315(1)}$  &  $\text{0.0315(1)}$  &  $\text{0.04(1)}$  &  $\text{0.031(8)}$  &  $\text{0.031(8)}$  \\
			\hline
			& $\{2,4\}$  &  $\text{0.00588(4)}$  &  $\text{0.00585(3)}$  &  $\text{0.00585(3)}$  &  $\text{0.02(1)}$  &  $\text{0.005(9)}$  &  $\text{0.005(9)}$  \\
			\hline
			& $\{4,6\}$  &  $\text{0.00351(4)}$  &  $\text{0.00348(3)}$  &  $\text{0.00348(3)}$  &  $\text{0.02(1)}$  &  $\text{0.003(9)}$  &  $\text{0.003(9)}$  \\
			\hline
			& $\{6,8\}$  &  $\text{0.00254(4)}$  &  $\text{0.00252(3)}$  &  $\text{0.00252(3)}$  &  $\text{0.02(1)}$  &  $\text{0.002(9)}$  &  $\text{0.002(9)}$  \\
			\hline
			& $\{8,10\}$  &  $\text{0.00201(3)}$  &  $\text{0.00199(3)}$  &  $\text{0.00199(3)}$  &  $\text{0.02(1)}$  &  $\text{0.001(10)}$  &  $\text{0.001(10)}$  \\
			\hline
			& $\{10,12\}$  &  $\text{0.00169(3)}$  &  $\text{0.00167(3)}$  &  $\text{0.00167(3)}$  &  $\text{0.02(1)}$  &  $\text{0.0007(98)}$  &  $\text{0.0007(98)}$  \\
			\hline
			& $\{12,14\}$  &  $\text{0.00147(3)}$  &  $\text{0.00146(3)}$  &  $\text{0.00146(3)}$  &  $\text{0.02(1)}$  &  $\text{0.0005(101)}$  &  $\text{0.0005(101)}$  \\
			\hline
			& $\{14,16\}$  &  $\text{0.00133(3)}$  &  $\text{0.00132(3)}$  &  $\text{0.00132(3)}$  &  $\text{0.02(1)}$  &  $\text{0.0003(105)}$  &  $\text{0.0003(105)}$  \\
			\hline
			& $\{16,18\}$  &  $\text{0.00124(3)}$  &  $\text{0.00123(3)}$  &  $\text{0.00123(3)}$  &  $\text{0.02(1)}$  &  $\text{0.0001(111)}$  &  $\text{0.0001(111)}$  \\
			\hline
			& $\{18,20\}$  &  $\text{0.00118(3)}$  &  $\text{0.00118(3)}$  &  $\text{0.00118(3)}$  &  $\text{0.02(1)}$  &  $\text{   0.000009(11934)}$  &  $\text{ 0.000009(11934)}$  \\
			\hline
			
			& $\{20,22\}$  &  $\text{0.00118(3)}$  &  $\text{0.00118(3)}$  &  $\text{0.00118(3)}$  &  $\text{0.02(2)}$  &  $\text{-0.0001(132)}$  &  $\text{-0.0001(132)}$  \\
			\hline
			& $\{22,24\}$  &  $\text{0.00125(3)}$  &  $\text{0.00125(3)}$  &  $\text{0.00125(3)}$  &  $\text{0.02(2)}$  &  $\text{-0.0002(152)}$  &  $\text{-0.0002(152)}$  \\
			\hline
			& $\{24,26.4\}$  &  $\text{0.00154(4)}$  &  $\text{0.00154(4)}$  &  $\text{0.00154(4)}$  &  $\text{0.03(2)}$  &  $\text{-0.0005(204)}$  &  $\text{-0.0005(204)}$  \\
			\hline
			$A_{\lambda_\ell}^{\bar{B} \to \pi \mu^- \bar{\nu}}$ & $\{0.01,2\}$  &  $\text{0.9151(5)}$  &  $\text{0.9157(2)}$  &  $\text{0.9157(2)}$  &  $\text{0.89(3)}$  &  $\text{0.92(3)}$  &  $\text{0.92(3)}$  \\
			\hline
			& $\{2,4\}$  &  $\text{0.9842(2)}$  &  $\text{0.9843(1)}$  &  $\text{0.9843(1)}$  &  $\text{0.94(7)}$  &  $\text{0.99(3)}$  &  $\text{0.99(3)}$  \\
			\hline
			& $\{4,6\}$  &  $\text{0.9905(2)}$  &  $\text{0.9906(1)}$  &  $\text{0.9906(1)}$  &  $\text{0.93(10)}$  &  $\text{0.99(3)}$  &  $\text{0.99(3)}$  \\
			\hline
			& $\{6,8\}$  &  $\text{0.9930(2)}$  &  $\text{0.9931(1)}$  &  $\text{0.9931(1)}$  &  $\text{0.9(1)}$  &  $\text{1.00(2)}$  &  $\text{1.00(2)}$  \\
			\hline
			& $\{8,10\}$  &  $\text{0.9944(1)}$  &  $\text{0.9945(1)}$  &  $\text{0.9945(1)}$  &  $\text{0.9(2)}$  &  $\text{1.00(2)}$  &  $\text{1.00(2)}$  \\
			\hline
			& $\{10,12\}$  &  $\text{0.9952(1)}$  &  $\text{0.9953(1)}$  &  $\text{0.9953(1)}$  &  $\text{0.9(2)}$  &  $\text{1.00(2)}$  &  $\text{1.00(2)}$  \\
			\hline
			& $\{12,14\}$  &  $\text{0.9958(1)}$  &  $\text{0.9958(1)}$  &  $\text{0.9958(1)}$  &  $\text{0.9(2)}$  &  $\text{1.00(1)}$  &  $\text{1.00(1)}$  \\
			\hline
			& $\{14,16\}$  &  $\text{0.9961(1)}$  &  $\text{0.9961(1)}$  &  $\text{0.9961(1)}$  &  $\text{0.9(2)}$  &  $\text{1.00(1)}$  &  $\text{1.00(1)}$  \\
			\hline
			& $\{16,18\}$  &  $\text{0.9962(1)}$  &  $\text{0.9963(1)}$  &  $\text{0.9963(1)}$  &  $\text{0.8(3)}$  &  $\text{0.999(6)}$  &  $\text{0.999(6)}$  \\
			\hline
			& $\{18,20\}$  &  $\text{0.9962(2)}$  &  $\text{0.9962(2)}$  &  $\text{0.9962(2)}$  &  $\text{0.8(3)}$  &  $\text{0.9994(4)}$  &  $\text{0.9994(4)}$  \\
			\hline
			& $\{20,22\}$  &  $\text{0.9960(2)}$  &  $\text{0.9960(2)}$  &  $\text{0.9960(2)}$  &  $\text{0.8(3)}$  &  $\text{0.999(8)}$  &  $\text{0.999(8)}$  \\
			\hline
			& $\{22,24\}$  &  $\text{0.9952(2)}$  &  $\text{0.9952(2)}$  &  $\text{0.9952(2)}$  &  $\text{0.8(3)}$  &  $\text{1.00(2)}$  &  $\text{1.00(2)}$  \\
			\hline
			& $\{24,26.4\}$  &  $\text{0.9922(4)}$  &  $\text{0.9922(4)}$  &  $\text{0.9922(4)}$  &  $\text{0.8(4)}$  &  $\text{1.00(6)}$  &  $\text{1.00(6)}$  \\
			\hline
		\end{tabular}
		\caption{Predictions of various observables in the SM and different new physics scenarios for the $\bar{B} \to \pi l \bar{\nu}$ channel using the fit results of the left column of table \ref{tab:fitNPBtopi} for the scenarios with $\mathcal{O}_T$, $\mathcal{O}_{S_1}$, $\mathcal{O}_{S_2}$ and the right column for the scenarios with $\mathcal{O}_{V_1}$, $\mathcal{O}_{V_2}$. }
		\label{tab:predB2Pi}
	\end{center}
\end{table}
	
\bibliographystyle{JHEP}	
\bibliography{ref_ASSI.bib}	
	
\end{document}